\documentclass[11pt,letterpaper]{article}
\usepackage[margin=1in]{geometry}
\usepackage[T1]{fontenc}
\usepackage[utf8]{inputenc}
\usepackage{newtxtext}
\usepackage{amsmath,amssymb,amsthm}
\usepackage{newtxmath}
\usepackage{bm}
\usepackage[mathlines]{lineno}
\usepackage{microtype}
\usepackage{booktabs,longtable,array,calc}
\usepackage{graphicx}
\usepackage{tocloft}
\usepackage{fvextra}
\usepackage{xurl}
\usepackage{xcolor} 
\usepackage{hyperref}
\usepackage{aliascnt}
\usepackage[capitalise,nameinlink,noabbrev]{cleveref}

\newtheorem{theorem}{Theorem}[section]
\newaliascnt{lemma}{theorem}
\newtheorem{lemma}[lemma]{Lemma}
\aliascntresetthe{lemma}
\newaliascnt{proposition}{theorem}
\newtheorem{proposition}[proposition]{Proposition}
\aliascntresetthe{proposition}
\newaliascnt{corollary}{theorem}
\newtheorem{corollary}[corollary]{Corollary}
\aliascntresetthe{corollary}
\theoremstyle{definition}
\newaliascnt{definition}{theorem}
\newtheorem{definition}[definition]{Definition}
\aliascntresetthe{definition}
\newaliascnt{example}{theorem}
\newtheorem{example}[example]{Example}
\aliascntresetthe{example}
\theoremstyle{remark}
\newaliascnt{remark}{theorem}
\newtheorem{remark}[remark]{Remark}
\aliascntresetthe{remark}
\theoremstyle{plain}
\newcommand{\restatementheading}{}
\newtheorem*{restatement}{\restatementheading}
\newcommand{\restatestatement}[2]{%
  \begingroup
  \renewcommand{\restatementheading}{\Cref{#1} (restated)}%
  \expandafter\let\expandafter\equation\csname equation*\endcsname
  \expandafter\let\expandafter\endequation\csname endequation*\endcsname
  \begin{restatement}#2\end{restatement}%
  \endgroup
}
\numberwithin{equation}{section}
\crefname{subsection}{Section}{Sections}
\Crefname{subsection}{Section}{Sections}
\crefname{subsubsection}{Section}{Sections}
\Crefname{subsubsection}{Section}{Sections}
\crefname{appendix}{Appendix}{Appendices}
\Crefname{appendix}{Appendix}{Appendices}

\providecommand{\tightlist}{\setlength{\itemsep}{0pt}\setlength{\parskip}{0pt}}
\title{Search and Verification for Strategyproof Facility Location on Metric Graphs}
\author{%
  Krzysztof Rogowski\\
  Institute of Informatics, University of Warsaw\\
  \href{mailto:kr418382@mimuw.edu.pl}{kr418382@mimuw.edu.pl}
  \and
  Marcin Dziubiński\\
  Institute of Informatics, University of Warsaw\\
  \href{mailto:m.dziubinski@mimuw.edu.pl}{m.dziubinski@mimuw.edu.pl}
}
\date{}
\hypersetup{
  colorlinks=false,
  pdfborder={0 0 1},
  allbordercolors={0 0.6 0},
  pdftitle={Search and Verification for Strategyproof Facility Location on Metric Graphs},
  pdfauthor={Krzysztof Rogowski and Marcin Dziubiński},
  pdfkeywords={Facility location, Strategyproof mechanisms, Piecewise-linear attributions}
}

\begin{document}
\begin{titlepage}
\maketitle
\thispagestyle{empty}
\begin{abstract}
\normalsize
We study automated design of strategyproof mechanisms for randomized facility location on continuous metric
graphs, with the objective of minimizing total distance. We develop methods for computing upper and lower bounds
on the optimal approximation ratio achievable by strategyproof mechanisms for a given graph and number of agents.

For the lower bound, we propose a cost-preserving reduction to a finite domain
that yields lower bounds for unrestricted strategyproof mechanisms through
linear programming dual certificates.

For the upper bound, we introduce finite representations of mechanisms, called piecewise-linear attributions (PL attributions),
that mix piecewise-linear deterministic mechanisms using affine weights on a triangulation of the report space.
We show that for mechanisms represented by PL attributions, checking strategyproofness and an approximation guarantee reduces to
finitely many quadratic nonnegativity tests over compact polytopes. With exact arithmetic and comparisons, each verification certifies the proposed property or returns locations where it fails.
In addition, we introduce a strengthening of the incentive notion called strong strategyproofness.
We show that for fixed component mechanisms and regions of a PL attribution, the strong strategyproofness requirement
can be expressed by a finite number of linear constraints on the attribution coefficients.
Using these theoretical results, we propose a heuristic procedure for computing an upper bound on the approximation ratio.

We compute certified upper and lower bounds on several metric graphs and numerically evaluate the proposed search procedure.

Lastly, we show that deterministic strategyproof onto mechanisms on graphs are strongly strategyproof and admit PL attributions.
Thus restricting attention to strongly strategyproof mechanisms admitting PL attributions loses no deterministic strategyproof onto mechanisms.

\end{abstract}
\end{titlepage}

\pagenumbering{roman}
\tableofcontents
\clearpage
\pagenumbering{arabic}

\section{Introduction}

\label{simp-new:introduction}
The problem of strategyproof facility location concerns choosing a location for a facility in a metric space for a group of agents, each
characterized by its own location in that space. Each agent seeks to minimize the distance from its own location to the facility,
while the social objective is to minimize the sum of these distances. A solution to this problem is a function, called a mechanism,
that maps the agents' reported locations to a lottery over facility locations.
If, for every report profile, this lottery chooses a single location with probability $1$, the mechanism is called deterministic.
A mechanism is strategyproof (SP) if no agent can reduce its expected
distance to the facility by misreporting its own location. The aim is to minimize the expected sum of distances from the facility to the agents' true locations, under the constraint of strategyproofness.

In this paper we study strategyproof facility location on continuous metric graphs, where
agents and the facility may occupy both vertices and edge interiors,
as in Alon, Feldman, Procaccia, and Tennenholtz~\cite{alon2009strategyproof}.
We use the approximation ratio to quantify the trade-off between social
efficiency and strategyproofness, following the agenda of approximate
mechanism design without money introduced by Procaccia and Tennenholtz
\cite{procaccia2013approximate}. A mechanism's approximation ratio is the
worst-case ratio, over all location profiles, of its expected total distance
to the minimum possible total distance. Our goal is to design strategyproof
mechanisms with the best possible approximation ratio on a given graph,
for a fixed number of agents. Solving this design problem directly in the
continuous setting is difficult: it requires optimizing over functions on
a continuous profile space, with no prescribed finite parameterization,
subject to infinitely many strategyproofness and approximation constraints.
We therefore approach the problem by establishing lower and upper bounds
on the optimal approximation ratio.

The two bounds require different arguments. A lower bound must apply to
every strategyproof mechanism. Building on the finite-domain linear
programming approach of Meir \cite[Section 4.1]{meir2019strategyproof}, we use
a cost-preserving reduction under which every continuous-domain
strategyproof mechanism induces a mechanism with reports and outputs
in suitable finite sets, with no worse approximation ratio. The finite
problem, therefore, provides a lower bound on the continuous problem. Any feasible solution of its
linear programming dual certifies such a bound, connecting this formulation
to the weighted certificates for the continuous circle of Chan, Lin, and Wang
\cite[Section 4]{chan2026randomized}.

\phantomsection\label{simp-new:main-results}

An upper bound requires finding only one mechanism with the desired guarantees.
The difficulty is to describe such a mechanism and verify its properties
over the continuous domain. To address this difficulty, we introduce
\emph{attributions}, which organize the definition of a mechanism into three
ingredients: component mechanisms, regions covering the profile space,
and mixing functions on each region. Within a region, the mixing functions
assign probabilities to the component mechanisms according to the reported
locations. The resulting weighted mixture of their outcomes gives the
lottery selected by the represented mechanism at the given reported locations.
The representation of mechanisms by attributions is without loss of generality, as every
facility location mechanism admits such a representation.

To facilitate the computation of mechanisms and the verification of their approximation
ratio and strategyproofness guarantees, we restrict attention to
\emph{piecewise-linear attributions} (PL attributions). In these attributions the
component mechanisms are deterministic and piecewise linear, the regions
are the maximal simplices of a finite triangulation of the profile space,
and the mixing functions are affine on each region.
We first investigate how the structure of PL attributions
enables \emph{global verification} of strategyproofness and approximation
ratio guarantees for the mechanisms they represent. Their expected costs are
piecewise quadratic in location coordinates, and this structure carries
over to the strategyproofness and approximation comparisons. Checking
these properties reduces to finitely many quadratic nonnegativity tests
over compact polytopes. Using exact arithmetic and comparisons, a finite
procedure certifies nonnegativity or returns a negative witness. Each
verification therefore either certifies the proposed
property throughout the continuous domain or returns locations where it
fails.

Verification lets us assess a proposed mechanism, but we still need to find
good candidates. We therefore investigate restrictions of the attribution
class, seeking structure for efficient optimization. This leads to a
heuristic search approach in which optimization within a chosen subclass
supplies candidates for global verification.

We first fix the component mechanisms and regions of a PL attribution,
allowing only the mixing functions to vary. These attributions
form a compact convex polytope in a finite-dimensional vector space.
At fixed locations, the strategyproofness and approximation comparisons
are linear in the attribution coefficients and the proposed approximation
factor. These properties give a linear programming formulation with
finitely many variables, although the incentive and approximation
constraints remain infinite.

To obtain finitely many incentive constraints, we introduce a stronger
incentive notion, \emph{strong strategyproofness (SSP)}, and restrict the search
to mechanisms satisfying it. This requires an agent's expected cost to be
nondecreasing as its reported location moves along any shortest path away
from its true location, with the other agents' reports fixed. Checking this
condition reduces to checking monotonicity of quadratic polynomials over
polytopes. This can be done by checking nonnegativity of affine directional
derivatives at the vertices of those polytopes.
This gives an explicit finite system of linear incentive constraints.

The approximation constraints still range over all location profiles.
We retain comparisons on a finite sample and minimize the resulting
sampled worst-case approximation factor. The resulting finite linear
program completes our heuristic search formulation: an optimal solution
supplies a candidate that is best within the chosen SSP class for the
sampled criterion. The sampled factor may underestimate the global
approximation ratio, so the candidate's proposed guarantee must then be
checked by global verification.

To evaluate the approach to mechanism design developed in this paper,
we conducted a computational study. We implemented exact-arithmetic verifiers based on our global verification
procedure and finite-domain lower-bound reduction. They check strategyproofness and approximation ratio guarantees
for mechanisms represented by PL attributions, as well as lower-bound certificates for all SP
mechanisms.

With these verifiers in place, we gave large language models the mathematical
framework. The models developed heuristic search procedures to find
candidate mechanisms and lower-bound certificates. We then used the
upper-bound verifier to certify that the reported PL attributions represent
SP mechanisms with the stated approximation guarantees, and the lower-bound
verifier to certify lower bounds for all SP mechanisms.

For three agents, these results place the optimal SP approximation ratio in
$[1.0591,1.10747]$ on the continuous cycle and $[1.0782,1.2028]$ on the
equal-path theta graph, leaving gaps below $0.049$ and $0.125$, respectively.
Following Guo and Conitzer \cite{guo2010computationally}, these comparisons
bound the possible cost of the restrictions imposed during search:
allowing arbitrary SP mechanisms can improve on the certified factors by
less than the corresponding gaps.

We also implemented the preceding heuristic-search construction on cycle graphs,
giving a numerical realization of the representation by PL attributions and finite SSP
characterization. We examined its performance under mesh refinement, using
floating-point arithmetic for linear optimization and continuous
approximation checking. On the finest mesh, with $128$ equally spaced points,
we obtain a numerical approximation upper bound of $1.072288$.
Rational reconstructions of selected candidates on smaller meshes are also
accepted by the general exact verifier, providing a cross-check between
the two implementations.

To evaluate the expressive power of strongly strategyproof mechanisms that admit
representation by PL attributions, we show that every deterministic strategyproof
and onto mechanism on metric graphs is strongly strategyproof and admits a PL attribution.


\paragraph{Related literature.}
Moulin~\cite{moulin1980strategy} characterized deterministic SP rules for
single-peaked preferences on the line. Schummer and Vohra
\cite{schummer2002strategy} studied deterministic SP mechanisms on metric
graphs, showing that every deterministic SP onto mechanism is dictatorial
on profiles confined to a cycle.
Dokow, Feldman, Meir, and Nehama~\cite{dokow2012mechanism} showed that deterministic SP onto
mechanisms on sufficiently large discrete cycles are nearly dictatorial.

To address the loss of social efficiency caused by the SP requirement,
Procaccia and Tennenholtz \cite{procaccia2013approximate} introduced approximate
mechanism design without money. They studied worst-case approximation ratios
for deterministic and randomized SP mechanisms on the real line.
Alon, Feldman, Procaccia, and Tennenholtz~\cite{alon2009strategyproof,alon2010strategyproof} extended this
approach to metric networks, studying SP mechanisms for
social-cost and minimax objectives.
For total distance and $n$ agents, random dictatorship guarantees a factor
of $2-2/n$ on general metric graphs \cite{alon2009strategyproof}.
On a cycle, Meir \cite{meir2019strategyproof} studied the three-agent case,
Rogowski and Dziubi\'{n}ski \cite{rogowski2025improved} improved bounds for odd
numbers of agents, and Chan, Lin, and Wang \cite{chan2026randomized} established
a factor of $3/2$ for every $n\geq3$.
Despite these improvements, the optimal ratio remains unknown in general,
even on a cycle.

The difficulty of finding optimal mechanisms analytically motivates
computational search. Conitzer and Sandholm \cite{conitzer2002complexity}
introduced automated mechanism design, showing that optimal randomized SP
mechanisms for finite type and outcome spaces can be computed using linear
programming.
Revenue-maximizing auction design has been a major application of this approach
\cite{likhodedov2005approximating,sandholm2015automated,albert2015assesing},
including recent deep-learning approaches
\cite{dutting2024optimal,wang2024gemnet,wang2026duality}.
%

To our knowledge, this paper presents the first framework for automated search
for randomized SP facility-location mechanisms on general continuous metric graphs.
Earlier automated facility-location approaches considered line segments,
multidimensional domains, and discrete graphs. Sui, Boutilier, and Sandholm
\cite{sui2013analysis} optimized SP percentile mechanisms for a distribution
of preferences in multidimensional domains. Narasimhan, Agarwal, and Parkes
\cite{narasimhan2016automated} learned SP mechanisms from a parameterized
family, with facility-location experiments on a finite set of points of a
line segment. Golowich, Narasimhan, and Parkes
\cite{golowich2018deep} used deep learning for multiple facilities on a
continuous line segment, targeting expected social cost under a distribution
of profiles.
In the related setting of false-name-proof facility location, Okada, Todo, and
Yokoo \cite{okada2019sat} used SAT solvers to obtain mechanisms and impossibility
results on discrete graphs.

Computational methods also help bound what any SP mechanism can achieve.
Meir \cite[Section~4.1]{meir2019strategyproof} formulated an LP for optimizing
randomized SP mechanisms for three agents on finite graphs and proposed a
support reduction to obtain lower bounds for the continuous cycle.\footnote{In \cref{lower-exp:app-correction}, we give a counterexample
showing that Meir's construction may not preserve SP and provide a corrected reduction using a fixed output set.}
Chan, Lin, and Wang \cite{chan2026randomized} subsequently developed weighted
lower-bound certificates for the continuous cycle.

Feldman and Wilf \cite{feldman2013strategyproof} provide a precedent for
composing randomized SP mechanisms from deterministic components. Their
construction on trees mixes boomerang mechanisms and a weighted-average
component with probabilities independent of reports, guaranteeing SP.
Our PL attributions allow the probabilities to depend affinely on reports
within each region, with SP checked separately for the mixture.

Like us, Br\^{a}nzei and Procaccia \cite{branzei2015verifiably} link a finite
representation of mechanisms to verification of their incentives. For their
comparison decision trees on the real line, they verify deterministic SP and
randomized universal truthfulness in time polynomial in the representation
size and number of agents. Our PL attributions support verification of SP
in expectation and approximation guarantees on continuous metric graphs
through finitely many quadratic nonnegativity tests under the stated
assumptions on exact arithmetic and comparisons.

Like Haeringer and Ha\l{}aburda's \cite{haeringer2016monotone} monotone
strategyproofness, SSP requires that a more truthful report be weakly
preferable. Their ordering uses inclusion of the sets of pairwise
comparisons reversed relative to the truth; ours orders reports by their
position along shortest paths from the true location, with lotteries
evaluated by expected distance.
Carroll \cite{carroll2012when} studied when local incentive constraints imply
global incentive compatibility. Our locality theorem gives a corresponding
reduction for SSP, a strengthening of SP. His local constraints compare
truthful reporting with nearby deviations; ours compare nearby reports
evaluated at a fixed true location that may be far from both.

\section{Model}

\label{simp-new:model}

Our network model follows Schummer and Vohra \cite[Section 2]{schummer2002strategy}.
The randomized mechanisms, expected-distance costs, and approximation
objective follow Alon, Feldman, Procaccia, and Tennenholtz
\cite[Sections 2 and 3]{alon2009strategyproof}. We restrict lotteries to finite
support, so all expectations below are finite sums.

\subsection{Metric graphs}
\label{simp-new:metric-graphs}

A \emph{graph} is a connected compact subset of Euclidean space $\mathbb R^k$,
for some positive integer $k$, that is the union of finitely many compact
curves of positive finite length. These curves are called \emph{edges},
and their endpoints are the \emph{vertices}. Each edge has two distinct endpoints and no
self-intersections, and distinct edges meet only at common endpoints.
We assume that every graph has at least one edge.

We denote graphs by $G,H$, their edge and vertex sets by $E(G)$ and
$V(G)$, edges by $e$, and vertices by $u,v,w$. The length of an edge $e$
is denoted by $\ell_e$. Points $x,y,z\in G$ may be vertices or lie in
edge interiors.

\phantomsection\label{simp-new:eq-distance}
A \emph{path} between $x,y\in G$ is a connected subset of $G$ containing
both points that is minimal under inclusion. The \emph{graph distance}
is $d_G(x,y)=\min_{\gamma}\operatorname{len}(\gamma)$, where $\gamma$
ranges over paths between $x$ and $y$ and $\operatorname{len}(\gamma)$
denotes the length of $\gamma$. After inserting $x$ and $y$ as vertices,
there are only finitely many such paths, so the minimum exists.
We write $d$ when $G$ is fixed.

We denote the set of all points lying on a shortest path between $x$ and
$y$ by $[x,y]_G=\{z\in G:d_G(x,z)+d_G(z,y)=d_G(x,y)\}$.
This set need not be a single path.

\subsection{Agents, lotteries, and mechanisms}

\label{simp-new:mechanisms}

Fix an integer $n\geq2$ and let $N=\{1,\ldots,n\}$ be the set of agents.
Agent $i$ has an
ideal location, or \emph{peak}, $x_i\in G$. The tuple
$\bm{x}=(x_1,\ldots,x_n)\in G^n$ is a \emph{location profile}.
We denote profiles by $\bm{x},\bm{y}$, and write
$(z,\bm{x}_{-i})$ for the profile obtained by replacing $x_i$ with
$z\in G$.

A \emph{lottery} $\mu$ on $G$ assigns a probability $\mu(y)\geq 0$ to each
$y\in G$, with finite support $\operatorname{supp}(\mu)=\{y\in G:\mu(y)>0\}$
and $\sum_{y\in\operatorname{supp}(\mu)}\mu(y)=1$. Let $\Delta(G)$ be the set
of these lotteries. A \emph{degenerate lottery}, or \emph{point mass},
$\delta_y$ selects $y$ with probability $1$. Throughout, we identify each
point $y\in G$ with the degenerate lottery $\delta_y$. Finite support does not
restrict outcomes to graph vertices or to any fixed finite set of locations.

A \emph{mechanism} is a function $M:G^n\to\Delta(G)$ mapping reported
location profiles to lotteries. For the fixed number of agents $n$, write
$\mathcal M_G$ for the class of all such mechanisms.
A mechanism is \emph{deterministic} if $M(\bm{x})$
is a degenerate lottery for every profile $\bm{x}\in G^n$.

For an agent at $x\in G$, the cost of an outcome lottery $\mu\in\Delta(G)$
is its expected distance from $x$:

\begin{equation}
\label{simp-new:eq-cost}
\operatorname{cost}_G(\mu,x)
=\sum_{y\in\operatorname{supp}(\mu)}\mu(y)d_G(y,x).
\end{equation}

We omit the graph subscript when it is clear.

\subsection{Strategyproofness}

\label{simp-new:strategyproofness}

A mechanism is \emph{strategyproof in expectation} if no agent can lower its
expected cost by misreporting, whatever the other agents report. Throughout,
we refer to this property simply as \emph{strategyproofness (SP)}. Formally,

\begin{equation}
\label{simp-new:eq-sp}
\operatorname{cost}(M(\bm{x}),x_i)
\leq
\operatorname{cost}(M(z,\bm{x}_{-i}),x_i),
\quad\text{for all }\bm{x}\in G^n, i\in N, z\in G.
\end{equation}

The true location in both costs is $x_i$; only the report changes.
Write $\mathcal M_G^{\mathrm{SP}}$ for the class of SP mechanisms on $G$.

\subsection{Social cost and approximation}

\label{simp-new:social-cost}

We use the utilitarian objective: the \emph{social cost} of a lottery
$\mu\in\Delta(G)$ at a profile $\bm{x}\in G^n$ is the sum of the
agents' expected costs:

\begin{equation}
\label{simp-new:eq-social-cost}
\operatorname{SC}_G(\mu,\bm{x})
=\sum_{i\in N}\operatorname{cost}_G(\mu,x_i).
\end{equation}

The \emph{optimal social cost} is the value

\begin{equation}
\label{simp-new:eq-opt}
\operatorname{OPT}_G(\bm{x})
=\min_{\mu\in\Delta(G)}\operatorname{SC}_G(\mu,\bm{x}).
\end{equation}

This minimum is attained by a point mass. Indeed, the social cost of a
lottery is an average of the social costs of its support points, so one
of those points is at least as good as the lottery. A minimizing point
exists because $G$ is compact and distance is continuous.

For a real number $\alpha\geq 1$, a mechanism is an
\emph{$\alpha$-approximation} on $G$ if

\begin{equation}
\label{simp-new:eq-approximation}
\operatorname{SC}_G(M(\bm{x}),\bm{x})
\leq\alpha\operatorname{OPT}_G(\bm{x}),
\quad\text{for all }\bm{x}\in G^n.
\end{equation}

Its \emph{approximation ratio} $\operatorname{apx}(M)$ is the infimum of such
factors, with value $+\infty$ if no finite factor exists.

Using an inequality avoids division by $0$ at unanimous profiles:
$\operatorname{OPT}_G(x,\ldots,x)=0$. In particular, every mechanism with
a finite approximation ratio must satisfy $M(x,\ldots,x)=\delta_x$.

For a class $\mathcal C$ of mechanisms, its \emph{optimal approximation ratio} is

\begin{equation}
\operatorname{infapx}(\mathcal C)
=\inf_{M\in\mathcal C}\operatorname{apx}(M).
\end{equation}

\subsection{Reference mechanisms}

\label{simp-new:preliminaries}
\label{simp-new:reference-mechanisms}

We recall the following standard SP mechanisms and their
exact approximation ratios.

{
\begin{longtable}[]{@{}
  >{\raggedright\arraybackslash}p{(\linewidth - 6\tabcolsep) * \real{0.2500}}
  >{\raggedright\arraybackslash}p{(\linewidth - 6\tabcolsep) * \real{0.2500}}
  >{\raggedright\arraybackslash}p{(\linewidth - 6\tabcolsep) * \real{0.2500}}
  >{\raggedright\arraybackslash}p{(\linewidth - 6\tabcolsep) * \real{0.2500}}@{}}
\toprule\noalign{}
\begin{minipage}[b]{\linewidth}\raggedright
Mechanism
\end{minipage} & \begin{minipage}[b]{\linewidth}\raggedright
Outcome at $\bm{x}$
\end{minipage} & \begin{minipage}[b]{\linewidth}\raggedright
Approximation ratio
\end{minipage} & \begin{minipage}[b]{\linewidth}\raggedright
Remarks
\end{minipage} \\
\midrule\noalign{}
\endhead
\bottomrule\noalign{}
\endlastfoot
Constant $\kappa_y$ & $y$ & $+\infty$ & Fixed $y\in G$. \\
Dictatorship $\pi_i$ & $x_i$ & $n-1$ & Fixed $i\in N$. \\
Median & Middle report in increasing order & $1$ & Interval; odd $n$. \\
Random dictatorship (RD) & $\displaystyle\frac1n\sum_{i\in N}\delta_{x_i}$ & $\displaystyle 2-\frac2n$ & \cite[Theorem 3.1]{alon2009strategyproof}. \\
Proportional circle distance (PCD) & Defined below. & $5/4$ for $n=3$ & Cycle; odd $n$. \cite[Section 3.1]{meir2019strategyproof}. \\
\end{longtable}
}

PCD selects each report with probability proportional to the length of its
opposite arc. For $n=2k+1$, this arc lies between the $k$th and $(k+1)$st
reports encountered after the selected report clockwise around the cycle. Tied
reports may be ordered arbitrarily. Selection probabilities at the same
location are added.

Throughout the paper, it will be useful to consider the following families
of mechanisms:

\begin{equation}
\Pi=\{\pi_i:i\in N\},
\qquad
\Pi(W)=\Pi\cup\{\kappa_y:y\in W\},
\end{equation}

where $W\subseteq G$ is a finite set. The family $\Pi$ consists of all
dictatorships, which we also call \emph{report selectors}, and $\Pi(W)$
additionally includes the constant mechanisms at locations in $W$.

\section{Representation and global verification}

\label{simp-new:analysis}

For a fixed graph $G$ and number of agents $n$, our objective is to
determine $\operatorname{infapx}(\mathcal M_G^{\mathrm{SP}})$. We approach
this objective by finding an SP mechanism $M$ and bounds $L,U$
such that

\begin{equation}
L\leq \operatorname{infapx}(\mathcal M_G^{\mathrm{SP}})
\leq \operatorname{apx}(M)\leq U.
\end{equation}

In this section, we focus on establishing upper bounds through automated
verification. We seek an expressive subclass of mechanisms that admit
finite descriptions and whose guarantees can be verified over the continuous
domain. To develop such a subclass, we adopt the piecewise-linear framework of
Rourke and Sanderson \cite{rourke1982introduction}.
Every graph in our model admits an isometric realization to which this
framework applies, allowing us to work in this setting without loss of
generality. We then introduce attributions, which describe mechanisms
through component mechanisms, regions, and mixing functions. Finally,
imposing PL structure on
these ingredients gives finite representations and reduces verification
to finitely many quadratic nonnegativity tests under the stated assumptions
on exact arithmetic and comparisons.
\Cref{simp-new:heuristic-search} develops a heuristic search approach for finding
candidates within this representation.

\subsection{Piecewise-linear preliminaries}

\label{simp-new:pl-preliminaries}

A \emph{convex polytope} is the convex hull of finitely many points.
A \emph{$k$-simplex} is the convex hull of $k+1$ affinely independent
points. Line segments and triangles are examples. The compact
\emph{polyhedra} used as domains here are finite unions of simplices
in a Euclidean space and need not be convex. Finite products of these
domains are again polyhedra.

A \emph{subdivision} divides a domain into finitely many closed convex
polytopes, called \emph{cells}, meeting in common faces. A
\emph{triangulation} is a subdivision into simplices. Its maximal
simplices are those not contained in larger ones. A \emph{refinement}
subdivides the cells further. Any finitely many subdivisions of the
same domain have a common refinement, which can be triangulated.

A map $f$ is \emph{affine} if it preserves affine combinations:
$f(\sum_{j=0}^k\lambda_jx_j)=\sum_{j=0}^k\lambda_jf(x_j)$, for all
points $x_0,\ldots,x_k$ in its domain and
real coefficients $\lambda_0,\ldots,\lambda_k$ with
$\sum_{j=0}^k\lambda_j=1$, provided that
$\sum_{j=0}^k\lambda_jx_j$ also belongs to the domain.
A map
$f:P\to\mathbb R^m$ on a polyhedron $P$ is \emph{piecewise linear
(PL)} if it is affine on every cell of some subdivision of $P$.
For maps into a polyhedron $Q\subseteq\mathbb R^m$, we use the same
definition, viewing their values in $\mathbb R^m$.
\phantomsection\label{simp-new:pq-preliminaries}
A real-valued function on a polyhedron is \emph{piecewise quadratic
(PQ)} if, on every cell of some subdivision, it is a finite sum of
products of two real-valued affine functions.
When defining a function by such formulas, we require agreement on
overlapping cells. The finite closed cover then also ensures continuity.

The following rules let us combine piecewise formulas and substitute
maps while retaining PL or PQ structure.

\newcommand{\claimPLPQCalculus}{%
On compact polyhedra, combining finitely many PL maps $f_1,\ldots,f_k$
with a common domain into the map
$x\mapsto(f_1(x),\ldots,f_k(x))$ gives a PL map.
For real-valued functions on a common domain, finite sums with fixed
real coefficients preserve both PL and PQ structure.
Finite sums of products of two real-valued PL functions are PQ.
In particular, every real-valued PL function is PQ.
Composing PL maps gives a PL map, and substituting a PL map into a
PQ function gives a PQ function.
}

\begin{lemma}[PL and PQ calculus]\label{simp-new:lem-pl-pq-calculus}

\claimPLPQCalculus

\end{lemma}

\Cref{simp-new:app-piecewise-calculus} gives the precise geometric
conventions and a proof of \cref{simp-new:lem-pl-pq-calculus}, using
Rourke and Sanderson's PL theory
\cite{rourke1982introduction}.

\subsection{Standing assumptions}

\label{simp-new:restrictions}

For the remainder of the paper, we assume that every edge of $G$ is a
straight line segment.
This is without loss of generality: every graph in our model admits such
an isometric realization after subdivision, and transporting mechanisms
preserves costs, SP, and approximation ratios.
\Cref{simp-new:app-metric} gives the construction and transport argument.

\phantomsection\label{simp-new:graphs-polyhedra}

With this realization, $G$ and its finite Cartesian powers are polyhedra,
so the piecewise-linear framework applies to these domains. In particular,
we have the following.

\newcommand{\claimPLGraphDistance}{%
For such a graph $G$, the distance function $d_G:G\times G\to\mathbb R$
is PL.
}

\begin{lemma}[PL graph distance]\label{simp-new:lem-pl-distance}

\claimPLGraphDistance

\end{lemma}

Related formulations of this fact appear as auxiliary results in earlier
work \cite[Proposition 4.1]{deeley2011configuration} and
\cite[Lemma 3.7]{garijo2023continuous}. For completeness and ease of
exposition, \cref{simp-new:app-distance} gives a proof in our
setting, together with the more precise distance-branch formulas used in
the later analysis.

\subsection{Attributions}

\label{simp-new:framework}

We now give a formal definition of the attribution representation
introduced above.

\begin{definition}[Attribution]

Let $F$ be a finite nonempty family of mechanisms $f:G^n\to\Delta(G)$,
called \emph{component mechanisms}, and let $\mathcal T$ be a
cover of $G^n$ by regions $T$. An \emph{$(F,\mathcal T)$-attribution of a
mechanism} $M:G^n\to\Delta(G)$ is a family
$A=(a_{T,f})_{T\in\mathcal T,\;f\in F}$
of \emph{mixing functions} $a_{T,f}:T\to[0,1]$ such that

\begin{equation}
\label{simp-new:eq-attribution}
\sum_{f\in F}a_{T,f}(\bm{x})=1,
\qquad
M(\bm{x})=\sum_{f\in F}a_{T,f}(\bm{x})\,f(\bm{x}),
\quad\text{for all }T\in\mathcal T, \bm{x}\in T.
\end{equation}

The second sum is a mixture of lotteries. A mechanism admitting such an
attribution is \emph{$(F,\mathcal T)$-attributable}. Since $\mathcal T$
covers the profile space, an attribution determines its represented
mechanism uniquely, which we denote by $M_A$.

\end{definition}

An attribution describes how a mechanism $M$ is assembled from the
component mechanisms in $F$. On each region $T\in\mathcal T$, it expresses
$M$ as a weighted mixture of these components. At a reported profile
$\bm{x}\in T$, the mixing function $a_{T,f}$ specifies the probability
of using component $f$, whose outcome is the lottery $f(\bm{x})$.
Thus the attribution gives a regional description of how the component
mechanisms combine to produce the lottery $M(\bm{x})$.

Attributions by themselves impose no restrictions: every mechanism $M$
admits a trivial attribution with $F=\{M\}$, $\mathcal T=\{G^n\}$, and
the sole mixing function identically equal to $1$. Their usefulness lies
in separating the component mechanisms, the cover of profile space, and
the mixing functions, allowing us to choose or restrict these ingredients
independently. Following Guo and Conitzer's approach
\cite{guo2010computationally}, we seek restrictions that yield expressive
classes of mechanisms with finite representations and constraints suitable
for optimization by linear programming.

We first illustrate this representation with the median mechanism
from \cref{simp-new:reference-mechanisms}. The
example shows how these ingredients work together and motivates the
particular restrictions we study below.

\begin{example}[Attribution for the median mechanism]\label{simp-new:ex-median-attribution}

For three agents on $G=[0,1]$, consider the median mechanism. Take the global
family of \emph{report selectors} $F=\{\pi_1,\pi_2,\pi_3\}$, regarded as
deterministic mechanisms through our identification of points with degenerate
lotteries. For each permutation $\rho$ of $\{1,2,3\}$, take the region $T_\rho
= \{\bm{x}\in[0,1]^3: x_{\rho(1)}\le x_{\rho(2)}\le x_{\rho(3)}\}$. These
six regions form the cover $\mathcal T$ of the profile space. On each
$T_\rho$, assign constant mixing weight $1$ to $\pi_{\rho(2)}$ and
weight $0$ to the other selectors.

This is an $(F,\mathcal T)$-attribution of the
median mechanism: on each $T_\rho$, the mixture assigns probability $1$
to $x_{\rho(2)}$, the middle report.

\end{example}

We have represented the median using the same three dictatorship mechanisms
throughout the profile space. On each ordering region, the mixing functions
assign weight $1$ to the dictatorship of the agent whose report occupies
the middle position.

At ties, more than one regional description applies. For example, when
$x_1=x_2<x_3$, the region defined by $x_1\leq x_2\leq x_3$ selects
$\pi_2$, while the region defined by $x_2\leq x_1\leq x_3$ selects
$\pi_1$. Both assign probability $1$ to the same location. We therefore
do not need a rule choosing which tied agent counts as the median.
Mixing functions on overlapping regions may assign different weights to
the selectors, provided that they define the same lottery at every shared
profile.

The ingredients of this attribution have a particularly simple form.
Viewed as point-valued maps through our identification of points with
degenerate lotteries, the component mechanisms are affine coordinate
projections. The ordering regions are the maximal simplices of a finite
triangulation, and the mixing functions are constant, hence affine.

\paragraph{PL attributions.}\label{simp-new:pl-attributions}

Building on the median example, we impose restrictions on the three
ingredients of an attribution.

\begin{definition}[PL attributions]\label{simp-new:def-pl-attribution}

An $(F,\mathcal T)$-attribution is \emph{piecewise-linear} (PL) if it
satisfies the following requirements:

\phantomsection\label{simp-new:selector-assumptions}

\begin{description}
\tightlist
\item[Component mechanisms.]
$F$ is a finite nonempty family of deterministic component mechanisms
$f:G^n\to G$ that are PL maps.
Throughout the paper, we refer to these deterministic component mechanisms
as \emph{selectors}.
\item[Regions.]
\phantomsection\label{simp-new:region-assumptions}
$\mathcal T$ consists of the maximal simplices of a finite triangulation
of $G^n$, each contained in a product $e_1\times\cdots\times e_n$ of
graph edges.
\item[Mixing functions.]
Each $a_{T,f}:T\to[0,1]$ is affine on $T$.
\end{description}

\end{definition}

The requirements defining PL attributions give each candidate a finite description through its
component mechanisms, triangulation, and regional affine weights. The next subsection
shows how this description supports global verification of SP
and approximation guarantees.

\subsection{Strategyproofness and approximation verification}

\label{simp-new:pl-properties}

Our verification procedure rests on expressing SP and
approximation bounds as nonnegativity conditions for finitely many
\emph{comparison functions}. These comparison functions are differences between the two sides of
the required cost inequalities, written so that nonnegativity means the
inequality holds. We first state that these functions are PQ and,
postponing the proof, explain how this reduces verification to quadratic
nonnegativity tests. We then establish their PQ structure in two steps: showing
that expected costs are PQ, and deriving the comparison functions from
those costs.

\begin{lemma}[PQ comparisons]\label{quad-search:presentation}

For a PL attribution $A$ and a fixed $\alpha\geq1$, SP of
$M_A$ and the bound $\operatorname{apx}(M_A)\leq\alpha$ are each equivalent
to nonnegativity of finitely many PQ functions.

\end{lemma}

\phantomsection\label{quad-search:verification}

Given \cref{quad-search:presentation}, verifying either property reduces to
checking nonnegativity of the associated PQ comparison
functions. Since each such function is PQ, its domain
admits a finite subdivision into polytopes on which it is given by a
polynomial of degree at most $2$. Checking nonnegativity reduces, therefore,
to finitely many polynomial nonnegativity tests on these polytopes. If a
comparison is negative somewhere, the test returns a point identifying
locations where it fails.
\Cref{quad-search:proof-verification} gives the construction, and
\cref{simp-new:app-quadratic-minimization} gives the computational details,
including the required assumptions on exact arithmetic and comparisons.

\phantomsection\label{quad-search:attributions}

We now return to the proof of \cref{quad-search:presentation}, beginning
with the PQ structure of expected costs.

\phantomsection\label{simp-new:quadratic-costs}

For a mechanism $M$, define its expected-cost function by

\begin{equation}
C_M(\bm{x},t)=\operatorname{cost}(M(\bm{x}),t),
\quad\text{for all }\bm{x}\in G^n, t\in G,
\end{equation}

where $\bm{x}$ is the reported profile and $t$ is the location at which
cost is evaluated.

\newcommand{\claimPQExpectedCosts}{%
For every PL attribution $A$, the expected-cost function
$C_{M_A}:G^n\times G\to\mathbb R$ is PQ.
}

\begin{lemma}[PQ expected costs]\label{quad-search:lem-quadratic-cost}

\claimPQExpectedCosts

\end{lemma}

The attribution formula gives

\begin{equation}
C_{M_A}(\bm{x},t)
=\sum_{f\in F}a_{T,f}(\bm{x})\,d_G(f(\bm{x}),t),
\quad\text{for all }T\in\mathcal T, \bm{x}\in T, t\in G.
\end{equation}

The weights $a_{T,f}$ are affine on $T$, and the distance factors
$d_G(f(\bm{x}),t)$ are PL on $T\times G$ by
\cref{simp-new:lem-pl-distance} and the composition rule in
\cref{simp-new:lem-pl-pq-calculus}.
Subdivide the products $T\times G$ compatibly so that every distance
factor is affine on each cell. Each summand is then a product of two
affine functions. The regional cost formulas agree on overlaps because
they give the expected cost of the same mechanism, and so they define
a PQ function on $G^n\times G$ by the definition in
\cref{simp-new:pq-preliminaries}. The proof in
\cref{quad-search:proof-quadratic-cost} gives the details.

\phantomsection\label{quad-search:comparisons}

Strategyproofness is equivalent to

\begin{equation}
\label{simp-new:algebraic-constraints}
C_{M_A}((z,\bm{x}_{-i}),x_i)
-C_{M_A}(\bm{x},x_i)
\geq0,\quad\text{for all }i\in N, \bm{x}\in G^n, z\in G.
\end{equation}

The two cost expressions are compositions of $C_{M_A}$ with the maps
$(\bm{x},z)\mapsto ((z,\bm{x}_{-i}),x_i)$ and
$(\bm{x},z)\mapsto (\bm{x},x_i)$, respectively.
These maps are PL, so substitution and subtraction preserve PQ
structure by \cref{simp-new:lem-pl-pq-calculus}.
Their cost difference is therefore PQ on $G^n\times G$,
giving one comparison function for each agent.

\phantomsection\label{simp-new:approximation-verification}

For approximation verification, we use the following consequence of
Hakimi's network-median theorem
\cite[p. 456, theorem in “Absolute Median”]{hakimi1964optimum}: an optimum is
attained at a peak or graph vertex. These are exactly the outputs
$g(\bm{x})$ for $g\in\Pi(V(G))$, the previously defined family of
report selectors and constants at graph vertices. A direct proof and a
strengthening appear in
\cref{quad-search:proof-benchmarks}.

\newcommand{\claimFiniteBenchmarks}{%
The optimal social cost satisfies

\begin{equation}
\operatorname{OPT}(\bm{x})
=\min_{g\in\Pi(V(G))}\operatorname{SC}(g(\bm{x}),\bm{x}),
\quad\text{for all }\bm{x}\in G^n.
\end{equation}
}

\begin{lemma}[Finite benchmarks for optimal social cost]\label{quad-search:lem-benchmarks}

\claimFiniteBenchmarks

\end{lemma}

By definition and the lemma, $\operatorname{apx}(M_A)\leq\alpha$ is
equivalent to

\begin{equation}
\operatorname{SC}(M_A(\bm{x}),\bm{x})
\leq\alpha\,\operatorname{OPT}(\bm{x})
=\alpha\min_{g\in\Pi(V(G))}
\operatorname{SC}(g(\bm{x}),\bm{x}),\quad\text{for all }\bm{x}\in G^n.
\end{equation}

A bound by the minimum is equivalent to a bound by every benchmark, so

\begin{equation}
\sum_{i\in N}
\bigl(\alpha C_g(\bm{x},x_i)-C_{M_A}(\bm{x},x_i)\bigr)
\geq0,\quad\text{for all }g\in\Pi(V(G)), \bm{x}\in G^n.
\end{equation}

Thus approximation verification reduces to checking the nonnegativity of finitely many
comparison functions on $G^n$, one for each benchmark $g$.
Each benchmark is a deterministic PL mechanism, so
$C_g(\bm{x},t)=d_G(g(\bm{x}),t)$ is PL by
\cref{simp-new:lem-pl-distance} and composition. The candidate's expected
cost function is PQ. Substituting $t=x_i$, multiplying by the fixed
factor $\alpha$, and taking sums and differences preserve PQ structure
by \cref{simp-new:lem-pl-pq-calculus}.
Thus every approximation comparison is PQ, completing the proof of
\cref{quad-search:presentation}.

\section{Heuristic search within attribution classes}

\label{simp-new:heuristic-search}

The preceding section provides a procedure for verifying a candidate
mechanism. Our approach to finding candidates is to optimize within smaller
classes of PL attributions. We fix a selector family $F$ and a triangulation
of $G^n$ whose maximal simplices form $\mathcal T$, leaving the affine
mixing functions to vary.

Let $\mathcal A_{\mathcal T,F}$ denote the class of PL
attributions with these selectors and regions. Its represented mechanism
class and SP subclass are

\begin{equation}
\mathcal M_{\mathcal T,F}
=\{M_A:A\in\mathcal A_{\mathcal T,F}\},
\qquad
\mathcal M_{\mathcal T,F}^{\mathrm{SP}}
=\mathcal M_{\mathcal T,F}\cap\mathcal M_G^{\mathrm{SP}}.
\end{equation}

The \emph{attribution map}
$\Phi_{\mathcal T,F}:\mathcal A_{\mathcal T,F}\to\mathcal M_{\mathcal T,F}$
sends each attribution $A$ to its represented mechanism $M_A$.

We first discuss the choice of component mechanisms and triangulation and
how it restricts the represented mechanisms.
We then establish the linear structure needed for optimization and introduce
a stronger incentive condition with a finite linear characterization.
Sampling the approximation comparisons produces a finite LP. An exact
optimum minimizes the sampled worst-case approximation factor within the
chosen attribution class subject to this stronger condition.
Global verification then checks the candidate's proposed
approximation guarantee throughout the continuous domain.

\subsection{Choosing the attribution class}

\label{simp-new:class-choice}

Fixing $F$ and $\mathcal T$ restricts the mechanisms available to
optimization. These choices serve different purposes: $F$ specifies the
possible outcomes at each profile, while $\mathcal T$ controls how flexibly
their probabilities can vary with the reports.

Choosing $F$ can itself be part of the design goal. For example, requiring
the facility to be placed at a reported location is expressed by taking
$F=\Pi$. When no such requirement is imposed, known mechanisms can
suggest useful selectors. Random dictatorship
and PCD use report selectors. Our later deterministic
characterization (\cref{simp-new:thm-deterministic-representation}) shows
that every deterministic SP onto mechanism admits a PL
attribution using report selectors and finitely many constants.
These examples suggest starting with report selectors, possibly
supplemented by finitely many constant selectors, and optimizing their
mixing weights.

The choice of triangulation is pragmatic. We can repeat the optimization
with different triangulations, using the resulting approximation guarantees
and computational costs to guide the choice.

\subsection{Structure of a fixed attribution class}

\label{simp-new:optimization}

With $F$ and $\mathcal T$ fixed, we optimize the regional affine mixing
functions together with a proposed approximation factor $\alpha\geq1$.
We show that attributions form a convex polytope in a
finite-dimensional vector space of regional affine functions, and that
SP and approximation requirements are affine in the
optimization variables. This gives a semi-infinite linear program for
minimizing the approximation factor, for which the verification procedures
from \cref{simp-new:pl-properties} provide separation oracles.

A finite-dimensional vector space containing $\mathcal A_{\mathcal T,F}$
is the \emph{coefficient space}

\begin{equation}
\label{simp-new:eq-coefficient-space}
\mathcal V_{\mathcal T,F}
=\prod_{T\in\mathcal T}\operatorname{Aff}(T,\mathbb R^F),
\end{equation}

where $\operatorname{Aff}(T,\mathbb R^F)$ consists of all families of
real-valued affine functions on $T$, indexed by $f\in F$. Its dimension is
$m=\dim\mathcal V_{\mathcal T,F}=|F|\sum_{T\in\mathcal T}(\dim T+1)$.

An element of $\mathcal V_{\mathcal T,F}$ is a family of real-valued
affine functions. Such a family defines an attribution when its weights
are nonnegative, sum to $1$ on each region, and give the same lottery on
overlapping regions. We call these coefficient families \emph{admissible};
they form $\mathcal A_{\mathcal T,F}$.

Every attribution already satisfies these conditions. We sometimes write
\emph{admissible attribution} to emphasize that the coefficient family
represents a mechanism $M_A$.

\newcommand{\claimAttributionPolytope}{%
The attribution class $\mathcal A_{\mathcal T,F}$ is a compact
convex polytope in $\mathcal V_{\mathcal T,F}$, described by finitely many
linear equalities and inequalities.
}

\begin{proposition}[Finite attribution constraints]\label{simp-new:prop-attribution-polytope}

\claimAttributionPolytope

\end{proposition}

\phantomsection\label{simp-new:eq-overlap-consistency}

The constraint description follows from the definition of a PL attribution
(\cref{simp-new:def-pl-attribution}). An attribution requires nonnegative
weights summing to $1$ on each simplex and agreement of the resulting
lotteries on overlaps. Since the weights
are affine, the first two conditions reduce to vertex checks. Agreement on
overlaps gives linear equations in the coefficients, of which only finitely
many can be linearly independent. The explicit finite constraint system and the proof of
compactness are given in \cref{simp-new:app-attribution-constraints}.

\begin{proposition}[Affinity of the attribution map]\label{simp-new:prop-attribution-mixing}\label{simp-new:cor-mechanism-mixing}

The attribution map $\Phi_{\mathcal T,F}$ is affine. Consequently, the
represented mechanism class $\mathcal M_{\mathcal T,F}$ is convex.

\end{proposition}

In each regional mixture, the component lotteries are fixed and the
mixing weights enter linearly. For
$A_0,A_1\in\mathcal A_{\mathcal T,F}$, $\lambda\in[0,1]$, and
$\bm{x}\in T\in\mathcal T$, the regional lottery formula gives

\begin{equation}
M_{\lambda A_0+(1-\lambda)A_1}(\bm{x})
=\sum_{f\in F}
\bigl(\lambda a^0_{T,f}(\bm{x})
 +(1-\lambda)a^1_{T,f}(\bm{x})\bigr)\delta_{f(\bm{x})}
=\lambda M_{A_0}(\bm{x})
 +(1-\lambda)M_{A_1}(\bm{x}),
\end{equation}

where $a^j_{T,f}$ are the weights of $A_j$.

The affinity of $\Phi_{\mathcal T,F}$ implies that the SP constraints are
affine in $A$, and the approximation constraints are affine jointly in
$(A,\alpha)$. Indeed, these constraints are linear in the mechanism's output
lotteries and, for approximation, the factor $\alpha$, as in Meir's
finite-domain LP formulation \cite[Section 4.1]{meir2019strategyproof}.
Composing them with the affine attribution map transfers this structure to
attribution coordinates.
\Cref{simp-new:app-regional-expressions} gives the explicit
coefficient formulas.

\phantomsection\label{simp-new:prop-semi-infinite-program}\label{simp-new:sp-separation}

Since $\mathcal A_{\mathcal T,F}$ is a convex polytope described by finitely
many linear constraints (\cref{simp-new:prop-attribution-polytope}), minimizing
$\alpha$ over this class subject to the SP and approximation inequalities
and $\alpha\geq1$ defines a semi-infinite linear program: finitely many
variables with infinitely many constraints.

Moreover, the verification procedures yield separation oracles for the SP
and approximation constraints of this program.

\newcommand{\claimSeparation}{%
Assume explicit finite input descriptions, exact arithmetic and
comparisons, and access to an oracle that decides nonnegativity of
quadratic polynomials on nonempty compact polytopes and returns a negative
witness when nonnegativity fails. For an admissible attribution
$A$ and a proposed factor $\alpha\geq1$, the SP and
approximation verification procedures from \cref{simp-new:pl-properties}
each either certify the corresponding requirement or return an affine
inequality in $(A,\alpha)$ that the mechanism $M_A$ violates and every
admissible candidate satisfying that requirement obeys.
}

\begin{corollary}[Separation from verification]\label{simp-new:prop-separation-reduction}

\claimSeparation

\end{corollary}

A failed verification supplies locations where a cost comparison is
violated. Holding those locations fixed makes the comparison an affine
inequality in the optimization variables, giving the required separation.
The explicit inequalities and proof appear in
\cref{simp-new:proof-oracle-separation}.

These separation oracles suggest a possible route to optimization through
the ellipsoid method, following Grötschel, Lovász, and Schrijver
\cite{grotschel1981ellipsoid}, subject to further representation and complexity
assumptions. We do not pursue this route here: our incentive separation
procedure requires quadratic nonnegativity tests at each query. We therefore
develop a stronger incentive condition that admits an explicit finite
linear description.

\subsection{Strong strategyproofness}

\label{simp-new:ssp-analysis}

\phantomsection\label{quad-search:ssp}

The preceding verification procedure checks SP through
nonnegativity of the cost differences $C_{M_A}((z,\bm{x}_{-i}),t)
-C_{M_A}((t,\bm{x}_{-i}),t)$. Both terms evaluate the PQ
expected-cost function. A sufficient condition for these differences to be
nonnegative is that, with $t$ and
$\bm{x}_{-i}$ fixed, cost be nondecreasing as the report moves along any shortest
path starting at $t$. We formalize this requirement as strong
strategyproofness.

\begin{definition}[Strong strategyproofness]\label{quad-search:def-ssp}

A mechanism $M$ is \emph{strongly strategyproof (SSP)} if, for every agent
$i$, true location $t$, other-report profile $\bm{x}_{-i}$, and reports
$y,z$ satisfying $d_G(t,z)=d_G(t,y)+d_G(y,z)$,

\begin{equation}
\label{simp-new:eq-ssp}
C_M((z,\bm{x}_{-i}),t)-C_M((y,\bm{x}_{-i}),t)\geq0.
\end{equation}

\end{definition}

\phantomsection\label{simp-new:ssp}

Taking $y=t$ recovers the SP condition. The additional comparisons between
successive reports allow us to assemble long comparisons from the short ones.
To see this, keep the truth and other reports fixed, and choose
$y=y_0,\ldots,y_m=z$ in order along a shortest path from $t$ to $z$. The
cost difference telescopes:

\begin{equation}
C_M((z,\bm{x}_{-i}),t)-C_M((y,\bm{x}_{-i}),t)
=\sum_{j=1}^{m}
\left[
C_M((y_j,\bm{x}_{-i}),t)
-C_M((y_{j-1},\bm{x}_{-i}),t)
\right].
\end{equation}

Each summand is another SSP comparison, evaluated at the same true
location. If these shorter comparisons are nonnegative, so is the
original one.

\newcommand{\claimSSPLocality}{%
Let $\mathcal U$ be any finite subdivision of $G^n\times G$ into closed
convex polytopes. A mechanism $M$ is SSP if and only if it satisfies the
subset of constraints from \cref{quad-search:def-ssp} consisting of those
for which the two points
$((y,\bm{x}_{-i}),t)$ and $((z,\bm{x}_{-i}),t)$ lie in a common
cell $U\in\mathcal U$.
}

\begin{theorem}[Locality of SSP]\label{finite-ssp:thm-locality-ssp}
\label{finite-ssp:presentation}\label{finite-ssp:locality}

\claimSSPLocality

\end{theorem}

Indeed, for an SSP comparison between reports $y$ and $z$, choose a
shortest path in $G$ from the true location $t$ to $z$ passing through $y$.
Moving agent $i$'s report along the portion from $y$ to $z$, while keeping
$t$ and the other reports fixed, traces a path in $G^n\times G$.
This path can be divided into finitely many portions contained in
individual cells. Summing the SSP inequalities between consecutive
endpoints gives the original comparison by telescoping.

The same shortest-path splitting argument used to prove locality lets us
restrict attention to SSP comparisons between reports less than any
prescribed $\varepsilon>0$ apart. Longer comparisons split into shorter
steps whose cost differences telescope. Since $\varepsilon$ can be
arbitrarily small, this suggests an infinitesimal test: divide a short
cost increment by the distance traveled and let that distance tend to
zero. Within a polynomial piece, this connects monotonicity of expected
cost to nonnegativity of a directional derivative.

\phantomsection\label{finite-ssp:directional-test}

Choose a subdivision so that, on each maximal cell, expected cost has a
quadratic formula and each agent's outward report movements have a fixed
direction. Here an outward movement extends a shortest path from the true
location, while that location and the other reports remain fixed.
The earlier SP verification procedure tests nonnegativity of quadratic
comparison functions on cells. For SSP, we can replace these oracle
calls with vertex checks: the derivative of the cell's cost polynomial
in the outward direction is affine, so its nonnegativity throughout the
cell is equivalent to its nonnegativity at every vertex.
The distance factors are fixed and the weight factors depend linearly
on the original attribution coefficients, so each derivative check is
linear in those coefficients. These are the finite incentive constraints
we can impose during search.

\phantomsection\label{finite-ssp:finite-characterization}

At a shared vertex, we test the polynomial from each incident maximal
cell separately; their derivatives need not agree.
\Cref{finite-ssp:proofs} makes this reduction precise,
accounting for the shortest-path restriction and for boundaries where
expected cost need not be differentiable. It constructs a suitable
subdivision $\mathcal U$ depending only on $G$, $F$, and $\mathcal T$, not
on the candidate attribution $A$ or approximation factor $\alpha$. This
yields the following characterization.

\newcommand{\claimFiniteSSP}{%
For fixed $G$, $F$, and $\mathcal T$, the SSP attribution set

\begin{equation}
\mathcal A_{\mathcal T,F}^{\mathrm{SSP}}
=\{A\in\mathcal A_{\mathcal T,F}:M_A\text{ is SSP}\}
\end{equation}

is a compact polytope described by the admissibility constraints together
with finitely many inequalities linear in the original attribution
coefficients.
}

\begin{proposition}[Finite linear characterization of SSP]\label{simp-new:ssp-constraints}

\claimFiniteSSP

\end{proposition}

Once constructed, this finite system can be evaluated to verify SSP or
imposed directly during search.

\subsection{Sampled approximation and global certification}

\label{simp-new:ssp-optimization}

\phantomsection\label{quad-search:search}

Our aim is a finite LP for candidate generation. The SSP characterization
supplies finitely many incentive constraints, but the approximation
requirements still range over all profiles. We relax these requirements by
retaining comparisons only at a finite set $\mathcal P\subseteq G^n$, while
keeping all admissibility and SSP constraints:

\begin{equation}
\label{simp-new:eq-sampled-program}
\begin{aligned}
\operatorname{minimize}\quad &\alpha\\
\operatorname{subject\ to}\quad
&A\in\mathcal A_{\mathcal T,F}^{\mathrm{SSP}},\qquad \alpha\geq1,\\
&\alpha\operatorname{SC}(g(\bm{x}),\bm{x})
 -\operatorname{SC}(M_A(\bm{x}),\bm{x})\geq0,
\quad\text{for all }\bm{x}\in\mathcal P, g\in\Pi(V(G)).
\end{aligned}
\end{equation}

Let $(A,\alpha_{\mathcal P})$ be an exactly feasible optimal solution of
this LP. Then $\alpha_{\mathcal P}$ is the best worst-case approximation
factor on the sampled profiles $\mathcal P$ within the chosen SSP class.
The mechanism $M_A$ is globally SSP, but the sampled optimum can
underestimate its global
approximation ratio (we only have
$\alpha_{\mathcal P}\leq\operatorname{apx}(M_A)$).

\phantomsection\label{simp-new:search-certification}

For any threshold $\beta\geq1$ and tolerance $\varepsilon>0$, repeated calls
to exact global verification from \cref{simp-new:pl-properties} can either certify
that $\operatorname{apx}(M_A)>\beta$ or return $U\leq\beta$ satisfying
$U-\varepsilon\leq\operatorname{apx}(M_A)\leq U$. The procedure checks the
endpoint bounds and, when necessary, bisects $[\alpha_{\mathcal P},\beta]$.
\Cref{quad-search:sampled-bounds} gives the details.
A known mechanism in the class can supply a useful comparison threshold
$\beta$.

Every globally feasible solution in the chosen SSP class is feasible for
the sampled LP. Consequently, whenever verification returns a certified
upper bound $U$,

\begin{equation}
\alpha_{\mathcal P}
\leq
\inf_{A'\in\mathcal A_{\mathcal T,F}^{\mathrm{SSP}}}
\operatorname{apx}(M_{A'})
\leq\operatorname{apx}(M_A)
\leq U.
\end{equation}

Thus $U-\alpha_{\mathcal P}$ bounds the candidate's suboptimality within
this class and quantifies the possible loss from optimizing the sampled
objective. \Cref{quad-search:sampled-bounds} gives the details.

\section{Lower bounds for unrestricted mechanisms}

\label{simp-new:lower-bounds}

\phantomsection\label{lower-exp:lower-bounds}

The preceding section compares the candidate with the optimum within the
chosen SSP class. Following Guo and Conitzer \cite{guo2010computationally},
we also assess its performance relative to the unrestricted design
problem. We therefore seek a lower bound on the optimal approximation
ratio among all SP mechanisms on $G$, against which to compare our
certified upper bound.

Building on Meir's finite lottery LP and endpoint redistribution
\cite{meir2019strategyproof}, we obtain lower bounds through a cost-preserving
finite reduction for general metric graphs, allowing the report and output
sets to differ. The resulting LP dual certificates are related to the weighted
lower-bound certificates of Chan, Lin, and Wang \cite{chan2026randomized}.
The construction and proof of the reduction appear in
\cref{lower-exp:proof-output-reduction}.

Suppose that finite sets $D\subseteq B\subseteq G$ admit the following
transfer: every SP mechanism on $G$ induces an SP mechanism with reports
in $D$ and outputs in $B$, without worsening its approximation factor.
Write

\begin{equation}
\mathcal M_{D,B}=\{M:D^n\to\Delta(B)\},
\end{equation}

using the graph distance $d_G$ and measuring approximation over profiles in
$D^n$ against the optimal social cost attainable in $B$.
We denote by $\mathcal M_{D,B}^{\mathrm{SP}}$ the subclass satisfying
SP for true locations and deviations in $D$.

\phantomsection\label{simp-new:lower-bound-certificates}

Extending Meir's finite-domain LP formulation to reports in $D$ and
outputs in $B$ gives a finite program whose optimum is
$\operatorname{infapx}(\mathcal M_{D,B}^{\mathrm{SP}})$. A feasible dual
solution supplies a certified lower bound $L$ on this optimum. The full
formulation and its dual appear in \cref{lower-exp:app-dual}.

\begin{equation}
\label{lower-exp:prop-lower-certificate}
L
\overset{(1)}{\leq}
\operatorname{infapx}(\mathcal M_{D,B}^{\mathrm{SP}})
\overset{(2)}{\leq}
\operatorname{infapx}(\mathcal M_G^{\mathrm{SP}}),
\end{equation}

where the inequalities follow from:

\begin{enumerate}
\def\labelenumi{\arabic{enumi}.}
\tightlist
\item
  Weak duality for the finite mechanism-design LP.
\item
  The assumed transfer, which makes every approximation factor
  achievable on $G$ also achievable with reports in $D$ and outputs in $B$.
\end{enumerate}

\phantomsection\label{simp-new:output-reduction}

It remains to justify the transfer.

\begin{definition}[Compatible output sets]\label{lower-exp:def-compatible-set}

Let $D\subseteq G$ be finite. A finite set $B\subseteq G$ is
\emph{compatible with $D$} if it contains $D$ and the graph vertices, and,
for every $t\in D$, the function $y\mapsto d_G(t,y)$ is affine on each
edge of the subdivision of $G$ at $B$.

\end{definition}

Such an output set exists for every finite $D$: include the graph
vertices, the points of $D$, and all breakpoints of the finitely many
PL functions $d_G(t,\cdot)$, $t\in D$
(\cref{simp-new:lem-pl-distance,simp-new:lem-pl-pq-calculus}).

\newcommand{\claimFiniteDomainReduction}{%
Let $D\subseteq G$ be finite and let $B$ be compatible with $D$.
Every SP mechanism $M$ on $G$ induces an SP mechanism
$M_{D,B}:D^n\to\Delta(B)$ satisfying
$\operatorname{apx}(M_{D,B})\leq\operatorname{apx}(M)$.
}

\begin{lemma}[Finite-domain reduction]\label{lower-exp:lem-output-reduction}

\claimFiniteDomainReduction

\end{lemma}

The idea is to adapt Meir's endpoint redistribution \cite{meir2019strategyproof} to the fixed
subdivision determined by $B$. Replace an outcome inside a subdivision
edge by a lottery on its endpoints, with probabilities determined by its
position along the edge. Compatibility ensures that this replacement
preserves expected costs at every true location in $D$, and hence
preserves SP within $D$.
The finite-benchmark lemma, \cref{quad-search:lem-benchmarks},
ensures that the optimal social cost is also preserved, so the
approximation guarantee is preserved.

For bipartite graphs with equal edge lengths, this reduction takes the
particularly simple form of restricting both reports and outputs to the
graph vertices, which we use in the computational study in
\cref{simp-new:experiments}.

\newcommand{\claimBipartiteReduction}{%
Let $G$ be a bipartite graph with equal edge lengths. Then
$B=D=V(G)$ is compatible, and every feasible dual solution of the finite
mechanism-design LP on $V(G)$ gives a lower bound for
$\operatorname{infapx}(\mathcal M_G^{\mathrm{SP}})$.
}

\begin{corollary}[Finite domains from bipartite graphs]\label{lower-exp:cor-bipartite}

\claimBipartiteReduction

\end{corollary}

In a bipartite graph, shortest paths from a fixed vertex to the two
endpoints of an edge have opposite parity in their numbers of edges:
one is even and the other odd. These numbers differ by at most one, so
they differ by exactly one. With equal edge lengths, the distances
therefore differ by exactly one edge length, making distance from the
fixed vertex affine along the edge.
\Cref{lower-exp:proof-bipartite} gives the full proof.

\section{Computational study}

\label{simp-new:experiments}

\phantomsection\label{lower-exp:experiments}

We perform two experiments that apply different parts of the theory
introduced in this paper and rely on different computational guarantees.
The first uses exact
verification and finite-domain lower bounds to assess candidates supplied
by LLM-based search. The second gives a numerical realization of our
complete heuristic-search construction on the cycle.

\subsection{Upper- and lower-bound verification with LLM-based search}

\label{simp-new:verified-experiments}

We implemented exact-arithmetic upper- and lower-bound verifiers for
continuous metric graphs represented by finite connected simple bipartite
graphs with unit-length edges.
Upper certificates describe rational PL attributions using report selectors
and constant selectors at rational graph locations.

The upper-bound verifier validates the supplied attribution and comparison
covers, including probability validity and agreement of lotteries across
faces. It then applies \cref{quad-search:prop-verification}, checking
quadratic nonnegativity exactly (\cref{simp-new:lem-quadratic-minimization}), to verify
that the mechanism is SP and has approximation ratio at most the submitted
factor, over the full continuous domain. These checks do not certify SSP.
The lower-bound verifier reconstructs the anonymous finite mechanism-design
LP on the explicit graph, with every vertex allowed as a report and an
output, and checks its dual certificates (\cref{lower-exp:app-dual,lower-exp:app-symmetry}).
The bipartite finite-domain reduction (\cref{lower-exp:cor-bipartite}) makes
their bounds applicable to all SP mechanisms on the continuous graph,
independently of the representation and incentive restrictions used during
search.

We supplied the mathematical framework and verification interfaces to
large language models, which developed procedures for finding mechanisms
and certificates. We verified the resulting candidates on graphs in the
implemented class. The correctness of the established results depends on
the mathematical reductions, the verifier implementations, and their exact
arithmetic. It does not depend on the correctness of the LLM reasoning or
the procedures used to generate the candidates. The detailed search
procedures are outside the scope of this paper.

\phantomsection\label{simp-new:cycle-lower-certificate}\label{simp-new:remark-cycle-lower-bound}

We highlight three results. For three agents on the cycle, we obtain a
verified approximation factor of $1.10747$, improving on the $5/4$
guarantee of PCD established by Meir \cite{meir2019strategyproof}.\footnote{The smaller $7/6$ factor claimed for Meir's
  $\tfrac14$-QCD rule is not used as a benchmark: the stated rule fails
  SP and exceeds that factor, as shown in
  \cref{simp-new:app-qcd}.}
Together with the unrestricted lower bound $1.0591173691$, our verified
upper bound places the optimal approximation ratio for three agents on the
cycle in an interval of width less than $0.049$. Thus, allowing arbitrary
SP mechanisms, beyond the family used in our search, can improve the
certified factor by less than $0.049$.

The experiments also extend beyond the cycle and trees to graphs
containing both cycles and branching. For three agents on the equal-path
theta graph (the complete bipartite graph $K_{2,3}$ with equal edge
lengths), we obtain a verified factor of $1.2028$, improving on random
dictatorship's $4/3$. We also certify an unrestricted lower bound of
$1.078216584909$ for the same graph and number of agents.

Our strongest recorded finite-domain lower certificate concerns six
agents on the equal-edge complete bipartite graph $K_{2,6}$, whose two
hubs are joined by six equal-length paths. It gives the unrestricted lower
bound $1.140204823651$. This illustrates the application of the
finite-domain reduction to larger numbers of agents and graphs beyond the cycle.

The broader inventory includes a four-agent cycle and other branching
graphs. All $13$ selected upper bounds improve on the corresponding
random-dictatorship guarantee.
The complete bounds are listed in \cref{lower-exp:table-verified-bounds}
in Appendix~\ref{lower-exp:app-certificates}, alongside the verification
records. The accompanying repository provides
the exact verifiers, selected certificates, verification records, and
numerical cycle study with reproduction instructions:
\url{https://github.com/baktrius/sp-facility-location-on-metric-graphs/tree/arxiv}.
The procedures used to generate the LLM-search certificates or recover
the rational candidates are excluded.

\subsection{Numerical realization of the heuristic search}

\label{simp-new:cycle-search-class}

\phantomsection\label{lower-exp:cycle-search}

We implemented our heuristic-search construction for three agents on the
cycle, using report selectors and a uniform grid of $q$ equally spaced
points around the cycle. Each cell of the corresponding product grid is
subdivided into simplices according to the ordering of the agents' local
coordinates within their respective grid intervals. Increasing $q$ gives
a finer grid on which to represent the PL probability assignments.

The finite attribution constraints (\cref{simp-new:prop-attribution-polytope})
and SSP characterization (\cref{simp-new:ssp-constraints}), together with
sampled approximation comparisons, give the LP for candidate generation
in \cref{simp-new:eq-sampled-program}. The implementation also imposes
the symmetry constraints specified in \cref{lower-exp:app-legacy}.
After optimization, we estimate the candidate's global approximation ratio
using bisection. Optimization and ratio estimation use floating-point
arithmetic, so these results provide numerical evidence rather than exact
certificates.

\phantomsection\label{simp-new:experimental-results}

Across the $17$ tested mesh sizes from $q=4$ to $q=128$, the numerical
continuous approximation upper endpoints decrease, reaching $1.072288$
at the largest mesh (\cref{lower-exp:fig-cycle-refinement} in
Appendix~\ref{lower-exp:app-legacy}). Compared with
the certified unrestricted lower bound $1.0591173691$, this suggests a
remaining gap below $0.0132$.

For four smaller meshes, rational reconstruction yields mechanisms that
pass exact verification of strategyproofness and approximation guarantees
closely matching the numerical estimates. This cross-check increases
confidence in the numerical implementation, although it does not certify
the candidate at $q=128$. Details are given in \cref{lower-exp:app-legacy}.

\section{Expressiveness of deterministic mechanisms}

\label{simp-new:deterministic-expressiveness}

In \cref{simp-new:experiments}, we compared certified approximation
guarantees for SP mechanisms represented by PL attributions with
unrestricted lower bounds, following the approach advocated by Guo and
Conitzer \cite{guo2010computationally}. Here we study deterministic
mechanisms, for which we obtain a full characterization.
Building on Schummer and Vohra's characterization
\cite[Theorems 1 and 4]{schummer2002strategy}, we show that every deterministic
SP onto mechanism admits a PL attribution and is SSP.

A deterministic mechanism is onto if every location in $G$ occurs as its
output at some report profile. Finite approximation ratio implies this
property: at a unanimous profile $(t,\ldots,t)$, the optimal social cost
is $0$, so the mechanism must output $t$.

We first specialize PL attributions by requiring each region to select
a single component.

\begin{definition}[Deterministic attributions]\label{simp-new:def-deterministic-attribution-class}

A PL $(F,\mathcal T)$-attribution is \emph{deterministic} if, on every
region $T\in\mathcal T$, one fixed selector $f_T\in F$ has weight $1$
throughout $T$ and every other selector has weight $0$.

\end{definition}

\newcommand{\claimDeterministicCharacterization}{%
Let $M:G^n\to G$ be a deterministic onto mechanism. The following are
equivalent:

\begin{enumerate}
\def\labelenumi{\arabic{enumi}.}
\tightlist
\item
  $M$ is SP.
\item
  $M$ is SSP.
\item
  $M$ admits a deterministic PL $(\Pi(W),\mathcal T)$-attribution for some finite
  $W\subseteq G$ and some $\mathcal T$ satisfying the PL
  attribution assumptions in \cref{simp-new:framework}.
\end{enumerate}
}

\begin{theorem}[Characterization of deterministic mechanisms]\label{simp-new:thm-deterministic-representation}

\claimDeterministicCharacterization

\end{theorem}

The implication $(2)\Rightarrow(1)$ follows directly from the definition
of SSP. For $(3)\Rightarrow(2)$, fix the other agents' reports. Within each
region, the selected outcome either follows the varying report or remains
fixed, so its distance from the agent's true location cannot decrease as
the report moves outward along a shortest path. The locality result,
\cref{finite-ssp:thm-locality-ssp}, then gives SSP globally. The main construction
is needed for $(1)\Rightarrow(3)$: using Schummer and Vohra's
characterization, we construct selectors and a finite triangulation such
that the mechanism agrees with one report or constant selector throughout
each region. This construction and the complete proof are given in
\cref{simp-new:app-deterministic}.

\section{Discussion and future directions}

\label{simp-new:discussion}

Attributions are the central organizing idea of this paper. They provide
a common representation through which we study facility location mechanisms
on continuous metric graphs, verify their guarantees, and search for
improved approximation ratios. PL attributions give finite
descriptions suitable for global verification, while requiring SSP yields
finite linear incentive constraints within each fixed attribution class.
Our deterministic characterization and computational study provide two
forms of support for this approach: every deterministic SP onto
mechanism admits a PL attribution and satisfies SSP, and the experiments
provide verified examples of randomized mechanisms and a numerical
realization of the SSP search procedure.

These results suggest three directions for further work. The first concerns
the search itself: how richer mechanism representations and larger profile
samples affect approximation guarantees and computational cost. The second
uses search and verification to investigate what facility-location
mechanisms can achieve. The third explores extensions to other
facility-location objectives and other mechanism-design domains.

\begin{enumerate}
\def\labelenumi{\arabic{enumi}.}
\item
  \textbf{Further theoretical and algorithmic directions.}

  \begin{itemize}
  \item
    \textbf{Convergence and approximation.} One direction is to establish
    convergence guarantees for sampled optimization within a fixed SSP
    attribution class. A broader question is whether selector enlargement
    and triangulation refinement can approach the unrestricted SSP optimum.
  \item
    \textbf{Complexity.} The finite formulations invite analysis of construction,
    optimization, and verification costs. For SP, the separation
    oracles suggest a route through the ellipsoid method, subject to
    suitable bounds on representation size and oracle complexity.
  \item
    \textbf{Locality and decomposition.} Can the locality of SSP be used to
    decompose optimization into smaller regional problems? Attributions
    provide a representation suited to exploring this question: mixing
    weights are specified region by region, subject to agreement of the
    resulting lotteries on overlaps. The challenge is to identify the
    constraints coupling these regional problems and to determine when
    solving them separately is computationally useful.
  \end{itemize}
\item
  \textbf{Structural questions in facility location.}

  \begin{itemize}
  \item
    \textbf{SSP versus SP.} For a fixed metric graph and number of agents,
    whenever the optimal approximation ratio among randomized SP mechanisms
    is attained, is it attained by an SSP mechanism? For deterministic onto
    mechanisms, we proved the stronger statement that every SP mechanism is SSP.
  \item
    \textbf{Expressiveness.} Does requiring a PL attribution change the optimal
    approximation ratio within the SP or SSP class? When do report selectors
    suffice, and when can additional selectors improve the achievable ratio?
  \end{itemize}

  Automated search and verification can help investigate these questions.
  More broadly, computationally discovered mechanisms and dual certificates
  may reveal recurring patterns that lead to explicit rules and analytic
  proofs.
\item
  \textbf{Extensions to other settings.}

  The maximum-cost objective is a natural first extension: incentive
  requirements remain the same, while the approximation analysis would need
  adaptation. Multiple facilities would require representations of facility
  configurations and an analysis of the resulting costs and computational
  complexity. Beyond facility location, a more exploratory direction is to
  investigate whether analogues of SSP and regional representations can
  support automated mechanism design in domains such as auctions.
\end{enumerate}

\subsection*{Acknowledgments}

This work was supported by the Polish National Science Centre through grant
2021/42/E/HS4/00196.

\clearpage
\appendix
\crefalias{section}{appendix}
\crefalias{subsection}{appendix}
\crefalias{subsubsection}{appendix}

\section*{Appendices}\phantomsection\label{simp-new:appendix}\addcontentsline{toc}{section}{Appendices}

Statements repeated from the main text retain their original numbers.

\section{Glossary}

\label{simp-new:glossary}

The two lists distinguish the notation used in the main text from
the additional notation used in the appendix. Within each list, entries follow
the order in which their meanings are introduced. Reused symbols are scoped
by the indicated construction. Local indices, scalar multipliers, and generic
points, polytopes, and maps are defined where they occur.

\subsection*{Notation introduced in the main text}

\begin{itemize}
\tightlist
\item
  $G,H$; $V(G),E(G)$; $e,\ell_e$ --- Metric graphs, their vertex and edge sets, an edge, and its length.
\item
  $d_G,[x,y]_G$ --- Graph distance and the union of shortest paths from $x$ to $y$.
\item
  $n,N$ --- The number of agents and the set $\{1,\ldots,n\}$.
\item
  $\bm{x},(z,\bm{x}_{-i})$ --- A report profile and the profile after agent $i$ reports $z$.
\item
  $\Delta(G),\mu,\delta_y$ --- Finite-support lotteries, a lottery, and a point mass at $y$.
\item
  $M,\mathcal M_G$ --- A mechanism on $G$ and the class of all such mechanisms, with $n$ fixed.
\item
  $\operatorname{cost}_G$ --- Expected individual distance from a lottery's outcome to the true location.
\item
  SP, $\mathcal M_G^{\mathrm{SP}}$ --- Strategyproofness and the SP subclass of mechanisms on $G$.
\item
  $\operatorname{SC}_G,\operatorname{OPT}_G$ --- Social cost and its optimal value at a profile.
\item
  $\alpha,\operatorname{apx}(M)$ --- A proposed approximation factor and a mechanism's approximation ratio.
\item
  $\operatorname{infapx}(\mathcal C)$ --- The optimal approximation ratio of a mechanism class: $\inf_{M\in\mathcal C}\operatorname{apx}(M)$.
\item
  $\pi_i,\kappa_y,\Pi,\Pi(W)$ --- Report selectors, constant selectors, the report-selector family, and its extension by constants in a finite set $W$.
\item
  $L,U$ --- Lower and upper bounds on an optimal approximation ratio; $U$ may also bound the ratio of a particular candidate.
\item
  PL, PQ --- Piecewise linear and piecewise quadratic.
\item
  $P,Q$ --- Compact polyhedra, with convexity or a role as subdivision cells specified locally; $P$ denotes a cell of the product graph complex in the SSP construction.
\item
  $F$ --- A finite nonempty family of component mechanisms; in a PL attribution, deterministic PL selectors.
\item
  $\mathcal T,T$ --- A cover of profile space and an attribution region; in a PL attribution, the maximal simplices of a triangulation and one such simplex.
\item
  $A,a_{T,f}$ --- An attribution and its regional mixing functions, affine in the PL setting; the same notation denotes coefficient families before admissibility is imposed.
\item
  $M_A$ --- The mechanism represented by an admissible attribution $A$.
\item
  $C_M(\bm{x},t)$ --- Expected cost at true location $t$ for reported profile $\bm{x}$.
\item
  $\Pi(V(G)),g$ --- The finite benchmark family of report selectors and constants at graph vertices, and one benchmark mechanism in this family.
\item
  $\mathcal A_{\mathcal T,F}$ --- The class of admissible attributions for fixed $F,\mathcal T$, subsequently shown to be a polytope.
\item
  $\mathcal M_{\mathcal T,F},\mathcal M_{\mathcal T,F}^{\mathrm{SP}}$ --- The represented mechanism class and its SP subclass.
\item
  $\Phi_{\mathcal T,F}$ --- The attribution map $A\mapsto M_A$.
\item
  $\mathcal V_{\mathcal T,F}$ --- The vector space of regional affine coefficient families for fixed $F,\mathcal T$.
\item
  $\operatorname{Aff}(T,\mathbb R^F)$ --- The vector space of families of real-valued affine functions on $T$, indexed by $F$; distinct from the affine-hull notation $\operatorname{aff}$.
\item
  SSP --- Strong strategyproofness.
\item
  $\mathcal U,U$ --- A finite subdivision of $G^n\times G$ and one of its cells, used for local SSP comparisons.
\item
  $\mathcal A_{\mathcal T,F}^{\mathrm{SSP}}$ --- The polytope of admissible attributions representing SSP mechanisms.
\item
  $\mathcal P,\alpha_{\mathcal P}$ --- The sampled profiles and the exact optimum of the sampled approximation LP within the fixed SSP attribution class.
\item
  $\beta,\varepsilon$ --- A comparison threshold and positive error tolerance for certification of a candidate mechanism's approximation ratio.
\item
  $D,B$ --- A finite report set and an output set compatible with it.
\item
  $\mathcal M_{D,B},\mathcal M_{D,B}^{\mathrm{SP}}; M_{D,B}$ --- Mechanisms with reports in $D$ and outputs in $B$, their SP subclass, and the mechanism obtained from $M$ by restricting reports to $D$ and redistributing outputs into compatible $B$.
\item
  $q$ --- The number of intervals in the uniform cycle grid used to represent a mechanism.
\end{itemize}

\subsection*{Notation introduced in the appendix}

\begin{itemize}
\tightlist
\item
  $K,|K|$ --- A geometric simplicial complex and its underlying polyhedron.
\item
  $\operatorname{aff},\operatorname{relint},\operatorname{Vert}$ --- Affine hull, relative interior, and vertex set.
\item
  $s_e$ --- An oriented arclength coordinate on edge $e$.
\item
  $\mathcal D,b_Q,\varepsilon_Q,\sigma_Q$ --- A subdivision of $G\times G$ with affine distance branches, and the intercept and two signed unit arclength slopes of the branch on cell $Q$.
\item
  $\varphi,\mu^\varphi,M^\varphi$ --- A graph isometry and the transported lottery and mechanism.
\item
  $h_e,\bm{b}_e$ --- The detour height and unit vector in the coordinate reserved for edge $e$ in the straight-edge realization.
\item
  $Z$ --- The reported locations together with graph vertices of degree at least $3$, used in the strengthened finite-benchmark proof.
\item
  $B,\lambda,I$ --- A symmetric quadratic matrix, barycentric coordinates, and a nonempty coordinate support in the nonnegativity test; $B$ here is distinct from a finite output set.
\item
  $V$ --- The matrix whose columns are the vertices of a test simplex in the quadratic nonnegativity reduction; distinct from the graph vertex set $V(G)$.
\item
  $\lambda_{T,v}$ --- The barycentric coordinate associated with vertex $v$ of attribution simplex $T$.
\item
  $\sim_Q,J$ --- Equality of selectors throughout an overlap polytope $Q$, and an equivalence class for this relation.
\item
  $c_T(A;\bm{x},t)$ --- A regional cost expression on the full coefficient space.
\item
  $X$ --- The report and true-location space $G^n\times G$ used in the SSP proofs.
\item
  $\operatorname{lin}(H),D_dq$ --- The direction space of an affine space $H$ and the directional derivative of a polynomial $q$ in direction $d$.
\item
  $K_G$ --- The geometric simplicial complex formed by the graph's straight edges and vertices.
\item
  $C_f$ --- The expected-cost function for a deterministic selector $f$, namely $C_f(\bm{x},t)=d_G(f(\bm{x}),t)$.
\item
  $T(U),\eta_{i,U}$ --- An original attribution region containing the report coordinates of cell $U$, and a direction in which agent $i$'s report extends a shortest path from the true location.
\item
  $\mathcal U^{\max},H_U$ --- The maximal simplices of a compatible subdivision and the affine hull of a simplex $U$.
\item
  $w_{U,f,A},h_{U,f};\widehat w_{U,f,A},\widehat h_{U,f}$ --- The regional weight and selector-distance factors on $U$, and their unique affine extensions to $H_U$.
\item
  $q_{U,A},L_{i,U,v}$ --- The cell cost polynomial and its outward derivative test at vertex $v$.
\item
  $\mathcal R^{\mathrm{attr}},\mathcal R^{\mathrm{cost}},\mathcal R^{\mathrm{ord}}$ --- Subdivisions resolving attribution regions, selector distances, and outward report directions.
\item
  $\psi_i,\rho_i,\widehat\rho_{i,U},\bm{e}_{i,U}$ --- The coordinate projection $\psi_i(\bm{x},t)=(t,x_i)$, the distance $\rho_i=d_G\circ\psi_i$, its affine extension to $H_U$, and the direction increasing only report $i$'s arclength coordinate at unit speed.
\item
  $\alpha^*_{\mathrm{SSP}}$ --- The infimum of global approximation ratios within the fixed SSP attribution class.
\item
  $\operatorname{OPT}_B$ --- Optimal social cost with outputs restricted to $B$.
\item
  $p_{\bm{x},b}$ --- The finite-domain mechanism's probability of output $b$ at profile $\bm{x}$.
\item
  $\mathcal I,r$ --- The finite LP's index set of incentive inequalities and an index $r=(\bm{x},i,z)$.
\item
  $\bm{x}^-(r),\bm{x}^+(r),t_r$ --- Source profile, deviated profile, and original true location for incentive comparison $r$.
\item
  $\lambda_{\bm{x}},\beta_r,\gamma_{\bm{x}}$ --- Dual multipliers for approximation inequalities, incentive inequalities, and normalization equalities.
\item
  $\omega,D_{\bm{x}}(b)$ --- Total weighted optimal cost and the weighted cost expression in the dual constraint for output $b$ at profile $\bm{x}$.
\item
  $s$ --- Approximation-sample scale for the numerical cycle experiments, giving $sq$ locations.
\item
  $p_i,\widehat p_i,\delta$ --- Saved and recovered report-selector coefficients, and their maximum summed absolute discrepancy over regional vertices, in the exact-recovery comparison.
\item
  $t_S$ --- The threshold associated with an agent coalition $S\subseteq N$ in the generalized-median representation.
\item
  $\operatorname{proj},M_\gamma,\widehat M_\gamma$ --- Nearest-point projection onto a path or attached tree, a path's generalized-median mechanism, and its composition with coordinatewise projection from the graph.
\item
  $C_G,H_j$ --- The subgraph containing and connecting all cycles, and its attached trees, in the deterministic proof.
\end{itemize}

\section{PL and PQ functions}

\label{simp-new:app-piecewise-calculus}

\Cref{simp-new:pl-preliminaries} introduces PL and PQ functions through
formulas on finitely many pieces. Here we connect these definitions with
the geometric language of Rourke and Sanderson
\cite[Chapters 1 and 2]{rourke1982introduction} and use it to prove
\cref{simp-new:lem-pl-pq-calculus} through common-refinement and
substitution rules. These rules justify the subdivisions on which
expected costs and their comparisons have polynomial formulas.
We also prove that graph distance is PL.
All polyhedra in this section are compact, and all subdivisions are finite.

\subsection{Geometric conventions}

\label{simp-new:app-geometric-conventions}

We write \emph{affine} where Rourke and Sanderson use \emph{linear}
in the affine sense.
For a nonnegative integer $k$, a $k$-simplex is
$\sigma=\operatorname{conv}\{v_0,\ldots,v_k\}$, where
$v_1-v_0,\ldots,v_k-v_0$ are linearly independent. Here $\operatorname{conv}$
denotes convex hull. A face of a simplex is the convex hull of a subset
of its vertices. We write $[u,v]$ for the closed Euclidean line segment
joining $u$ and $v$.

A finite geometric simplicial complex $K$ is a finite collection of
simplices containing all their faces, such that the intersection of any
two simplices is a face of both or is empty. Its underlying polyhedron
is $|K|=\bigcup_{\sigma\in K}\sigma$, and $\dim K$ is the largest
dimension of its simplices. We allow the empty face. A triangulation
of a compact polyhedron $P$ is such a complex with $|K|=P$. The description of a compact
polyhedron as a finite union of simplices agrees with the book's
geometric definition, and every such polyhedron admits a triangulation
\cite[Example 1.3(5), Theorems 2.2 and 2.11]{rourke1982introduction}.
A \emph{subpolyhedron} of a polyhedron $P$ is a subset $Q\subseteq P$ that is itself
a polyhedron.

A convex polytope can also be described as a bounded set defined by finitely
many weak affine inequalities. Its faces are its intersections with
supporting hyperplanes, together with the polytope itself and the empty
face. Its vertices are its extreme points. We write
$\operatorname{aff}(Q)$ for its affine hull, the smallest affine space
containing $Q$,
$\operatorname{relint}(Q)$ for its interior within that affine hull,
and $\operatorname{Vert}(Q)$ for its vertex set.
A finite polytopal complex satisfies the same face and intersection
conditions as a simplicial complex, with convex polytopes as cells.
Such a complex admits a simplicial subdivision
\cite[Proposition 2.9]{rourke1982introduction}. Products of polytopal
complexes are polytopal complexes
\cite[Example 2.8(5)]{rourke1982introduction}. In particular, finite
products of compact polyhedra are compact polyhedra.

A complex $L$ subdivides $K$ if $|L|=|K|$ and every cell of $L$ lies
in a cell of $K$. When a triangulation must respect several given
subdivisions, we call it \emph{compatible} with them if every cell of each
given subdivision is a union of its simplices.
The PL definition in
\cref{simp-new:pl-preliminaries} agrees with the book's definition:
affine pieces define PL maps
\cite[Example 1.5(4)]{rourke1982introduction}, while every PL map admits
a compatible triangulation on which it is affine
\cite[Corollary 2.3, Addendum 2.12]{rourke1982introduction}.

\subsection{Common subdivisions and the calculus lemma}

We first state the subdivision facts used in the proof of
\cref{simp-new:lem-pl-pq-calculus} and in later constructions.

\begin{lemma}[Common triangulations and PL substitutions]\label{simp-new:lem-pl-calculus}

Let $P$ and $Q$ be compact polyhedra.

\begin{enumerate}
\def\labelenumi{\arabic{enumi}.}
\tightlist
\item
  Finitely many triangulations of compact subpolyhedra of $P$ admit a
  compatible triangulation of $P$: each subpolyhedron is a union of
  simplices, and the induced triangulation refines its given one. The
  triangulation can also refine any fixed finite polytopal subdivision
  of $P$.
\item
  For a PL map
  $\phi:P\to Q$ and a finite polytopal subdivision of $Q$, there is a
  triangulation of $P$ on which $\phi$ is affine and each simplex maps
  into one target cell.
\end{enumerate}

\end{lemma}

\begin{proof}

The first assertion follows from Rourke and Sanderson's
common-triangulation result \cite[Addendum 2.12]{rourke1982introduction},
applied to a triangulation of $P$ and the given complexes on its
subpolyhedra. A prescribed polytopal subdivision can be triangulated
first \cite[Proposition 2.9]{rourke1982introduction} and included among
those complexes.

For the second assertion, take a triangulation of $P$ on which $\phi$
is affine. Its intersections with the inverse images of the target
cells form a polytopal complex
\cite[Example 2.8(6)]{rourke1982introduction}. A triangulation of that
complex \cite[Proposition 2.9]{rourke1982introduction} gives the required
source triangulation.

\end{proof}

\restatestatement{simp-new:lem-pl-pq-calculus}{\claimPLPQCalculus}

\begin{proof}

For finitely many PL maps on a common domain, choose a common
triangulation of their affine pieces by
point 1 of \cref{simp-new:lem-pl-calculus}. The tuple map is affine on
every resulting simplex, as is any fixed linear combination of
real-valued PL functions. Thus these maps are PL.

The same refinement argument applies to the polynomial pieces of
finitely many PQ functions. On each simplex, a fixed linear combination
of their formulas remains a finite sum of affine products, so the resulting
function is PQ. For a finite sum of products of two real-valued PL
functions, choose a common triangulation of all the factors' affine
pieces. Each summand is a product of two affine functions on
each simplex. Every real-valued PL function is included by multiplying
it by the constant function $1$.

Finally, let $\phi:P\to Q$ be PL, and let $g$ on $Q$ be either a PL
map or a PQ function. By point 2 of \cref{simp-new:lem-pl-calculus},
there is a triangulation of $P$ on which $\phi$ is affine and each
simplex maps into one cell of a subdivision defining $g$. On each
source simplex, $g\circ\phi$ is therefore affine in the PL case and
a finite sum of affine products in the PQ case, obtained by composing
each factor with $\phi$. The formulas agree on overlaps
because they are restrictions of the same composition. This proves
both substitution rules.

\end{proof}

\subsection{Graph distance and its affine branches}

\label{simp-new:app-distance}

\Cref{simp-new:lem-pl-distance} makes selector distances available as PL
factors in expected costs. We prove it by giving an affine formula for
distance on each cell of a subdivision of $G\times G$. We call these
formulas \emph{affine branches}. Their slopes in arclength coordinates
determine, in particular, which report movements extend a shortest path from
the true location in the finite SSP test. We use the straight-edge
assumption from \cref{simp-new:restrictions}.
\Cref{simp-new:app-metric} justifies it by constructing an isometric realization.
Orient each edge $e$ from $u_e$ to $v_e$
and let $s_e:e\to[0,\ell_e]$ be its arclength coordinate. For an endpoint
$u$ of $e$ and $x\in e$, the distance $d_G(x,u)$ is the length of the
subsegment from $x$ to $u$. For $u=u_e$ and $u=v_e$, these are the
affine functions $s_e(x)$ and $\ell_e-s_e(x)$, respectively.

\begin{lemma}[Affine distance branches]\label{simp-new:lem-distance-branches}

There is a finite polyhedral subdivision
$\mathcal D$ of $G\times G$ refining the edge products such that each
cell $Q\in\mathcal D$ lies in a product of edges $e\times e'$ and has a formula

\begin{equation}
d_G(x,y)=b_Q+\varepsilon_Q s_e(x)+\sigma_Q s_{e'}(y),\quad\text{for all }(x,y)\in Q,
\end{equation}

where $b_Q\in\mathbb R$ and $\varepsilon_Q,\sigma_Q\in\{-1,1\}$.

\end{lemma}

\begin{proof}

For distinct edges $e,e'$ and points $x\in e$, $y\in e'$, we have

\begin{equation}
\label{simp-new:eq-distance-branches}
d_G(x,y)=
\min_{\substack{u\in\{u_e,v_e\}\\v\in\{u_{e'},v_{e'}\}}}
\bigl(d_G(x,u)+d_G(u,v)+d_G(v,y)\bigr).
\end{equation}

Each candidate is the length of a walk joining $x$ to $y$, so it is at
least $d_G(x,y)$. Conversely, a shortest path leaves the first edge and
enters the second through endpoints. Its subpaths are shortest paths,
so one candidate attains equality. If $x$ or $y$ is already an endpoint,
the corresponding initial or final subsegment can have length~$0$.
Every candidate is affine, with coefficient $+1$ or $-1$ in each of
the two arclength coordinates.

For $e=e'$, the direct subsegment joins $x$ and $y$, and no path can be
shorter than their Euclidean distance. Hence

\begin{equation}
d_G(x,y)=|s_e(x)-s_e(y)|.
\end{equation}

Splitting the edge product at $s_e(x)=s_e(y)$ gives the two affine
branches, each again with coefficients in $\{-1,1\}$.

For each distinct edge pair, cut its product along the nonempty equality
hyperplanes between nonidentical candidate affine functions. On every resulting polytope,
one fixed candidate attains the minimum throughout. Include the diagonal
cuts for equal edges. Triangulate the resulting polytopal subdivision
of each edge product. Point 1 of
\cref{simp-new:lem-pl-calculus} gives a compatible triangulation
of $G\times G$ refining all of these subdivisions. Use this as
$\mathcal D$.

Every cell is contained in an edge product and in a region of a fixed
affine branch. Restricting that branch gives the claimed formula,
including on boundary cells. On a two-dimensional cell, the coefficients
are uniquely determined because its affine hull is the edge-product
plane.

\end{proof}

The unit slopes above refer specifically to arclength coordinates;
in normalized coordinates $s_e/\ell_e$, their magnitudes are the edge
lengths.

\phantomsection\label{simp-new:proof-pl-distance}
\restatestatement{simp-new:lem-pl-distance}{\claimPLGraphDistance}

\begin{proof}

Let $\mathcal D$ be the finite polyhedral subdivision of $G\times G$
from \cref{simp-new:lem-distance-branches}. The restriction of $d_G$
to each cell of $\mathcal D$ is affine, hence continuous. These
restrictions agree on intersections. Since the cells form a finite
closed cover, the continuous restrictions define a continuous function
on their union. Thus $d_G$ is continuous. Choose a compatible
triangulation refining $\mathcal D$. On each simplex, $d_G$ is the
restriction of an affine function on a containing cell of $\mathcal D$
and is therefore affine.

\end{proof}

\section{Metric invariance and straight-edge realizations}

\label{simp-new:app-metric}

The straight-edge assumption in \cref{simp-new:restrictions} lets us
apply PL methods to graph products. We justify it by constructing an
isometric simplicial realization of every graph in the model. First we
show why an isometry preserves the mechanism-design problem, including
the SSP comparisons used later in the search formulation.

\begin{proposition}[Invariance under isometry]\label{simp-new:prop-isometry}

Let $\varphi:G\to H$ be a bijective isometry. It induces a bijection between
mechanisms on $G$ and on $H$ that preserves
individual costs, social costs, SP, and approximation ratios.
It also preserves SSP as defined in
\cref{quad-search:def-ssp}.

\end{proposition}

\begin{proof}

Write $\varphi^n(\bm{x})=(\varphi(x_1),\ldots,\varphi(x_n))$.
Transport a lottery $\mu$ by assigning probability $\mu(y)$ to
$\varphi(y)$, and denote the resulting lottery by $\mu^\varphi$.
For a mechanism $M:G^n\to\Delta(G)$, define its transported mechanism
$M^\varphi$ by

\begin{equation}
M^\varphi(\varphi^n(\bm{x}))=M(\bm{x})^\varphi,
\quad\text{for all }\bm{x}\in G^n.
\end{equation}

Its outputs have finite support. Since $\varphi$ preserves distances,
it preserves expected costs:

\begin{equation}
C_{M^\varphi}(\varphi^n(\bm{x}),\varphi(t))
=C_M(\bm{x},t),
\quad\text{for all }\bm{x}\in G^n, t\in G.
\end{equation}

Transport by $\varphi^{-1}$ is inverse to transport by $\varphi$, so
this construction gives a bijection of mechanisms. To prove that it
preserves SP, SSP, and each $\alpha$-approximation guarantee, it suffices
to transport a violation for $M$ to a violation for $M^\varphi$.
The same argument with $\varphi^{-1}$ then gives the converse.
Preserving every approximation guarantee also preserves the approximation
ratio.

Suppose $M$ violates SP at a profile $\bm{x}$, where agent $i$
benefits from reporting $z$. The cost identity shows that the same agent
benefits by the same amount from reporting $\varphi(z)$ at
$\varphi^n(\bm{x})$ under $M^\varphi$.
If $M$ violates SSP at true location $t$, reports $y,z$, and other
reports $\bm{x}_{-i}$, transport all these locations by $\varphi$.
The isometry preserves the condition
$d_G(t,z)=d_G(t,y)+d_G(y,z)$, and the cost identity preserves the
strict decrease in expected cost from report $y$ to report $z$.
Thus the transported locations witness an SSP violation for $M^\varphi$.

Suppose $M$ violates an $\alpha$-approximation guarantee at
$\bm{x}$, for some $\alpha\geq1$. Let $y$ be an optimal facility
location for $\bm{x}$. The location $\varphi(y)$ has the same social
cost at $\varphi^n(\bm{x})$, so
$\operatorname{OPT}_H(\varphi^n(\bm{x}))\leq
\operatorname{OPT}_G(\bm{x})$.
Summing the cost identity over the agents gives

\begin{equation}
\begin{aligned}
\operatorname{SC}_H(M^\varphi(\varphi^n(\bm{x})),\varphi^n(\bm{x}))
&=\operatorname{SC}_G(M(\bm{x}),\bm{x})\\
&>\alpha\operatorname{OPT}_G(\bm{x})\\
&\geq\alpha\operatorname{OPT}_H(\varphi^n(\bm{x})).
\end{aligned}
\end{equation}

Hence the transported profile violates the same guarantee.

\end{proof}

\begin{lemma}[Isometric simplicial realization]\label{simp-new:lem-realization}

Every metric graph in our model is isometric to the
underlying polyhedron of a finite connected one-dimensional geometric simplicial
complex, equipped with the intrinsic metric given by Euclidean edge lengths.

\end{lemma}

\begin{proof}

Let $k=|V(G)|$ and $m=|E(G)|$.
Work in $\mathbb{R}^{m+1}$, using the first coordinate for the vertices
and one additional coordinate for each edge. Choose the positive spacing
$h_0=(\min_e\ell_e)/(2k)$, and place the vertices at
$v_j=(jh_0,0,\ldots,0)$ for $1\leq j\leq k$.
The Euclidean distance between any two of these vertices is smaller than
every original edge length.

Write $\|\cdot\|_2$ for the Euclidean norm. For an edge $e$ with
endpoints now placed at $u_e,v_e$, let $\bm{b}_e$ be the unit vector
in its reserved coordinate direction and put

\begin{equation}
\label{simp-new:eq-detour-height}
h_e=\frac{\ell_e-\|u_e-v_e\|_2}{2}>0.
\end{equation}

Replace $e$ by the three straight line segments successively joining

\begin{equation}
\label{simp-new:eq-edge-detour}
u_e,\qquad u_e+h_e\bm{b}_e,\qquad
v_e+h_e\bm{b}_e,\qquad v_e.
\end{equation}

Their total Euclidean length is
$h_e+\|u_e-v_e\|_2+h_e=\ell_e$.
Every point of this polygonal path other than its original endpoints has
a positive coordinate in the direction $\bm{b}_e$. All other edge
paths have $0$ in that coordinate. Consequently, different paths,
including those representing parallel edges, meet only at their
prescribed common endpoints. Each path itself is simple.
The line segments and their endpoints therefore form a finite geometric
simplicial complex. Parameterizing each original edge and its polygonal
path by arclength over $[0,\ell_e]$ gives a bijection preserving subarc
lengths. These edge maps agree at common endpoints and preserve all path
lengths, hence the intrinsic shortest-path metric.

\end{proof}

By \cref{simp-new:lem-realization} and \cref{simp-new:prop-isometry}, we may work with this
simplicial realization throughout the analysis. Inserting further
vertices along its line segments also preserves the metric. In particular,
if $x$ and $y$ lie on the same straight edge, then

\begin{equation}
d_G(x,y)=\|x-y\|_2,
\end{equation}

since the line segment between $x$ and $y$ lies in the graph and no path can be
shorter than the Euclidean distance between its endpoints.

\section{PQ expected costs}

\label{quad-search:proof-quadratic-cost}

The verification argument in \cref{simp-new:pl-properties} uses the PQ
expected-cost function of a given PL attribution. Its regional formula
is a sum of products of affine weights and PL distances. We prove that
these formulas define one PQ function even when the individual weights
differ on overlaps.

\restatestatement{quad-search:lem-quadratic-cost}{\claimPQExpectedCosts}

\begin{proof}

Fix a PL attribution $A=(a_{T,f})$. On each product $T\times G$, the
attribution formula gives

\begin{equation}
C_{M_A}(\bm{x},t)
=\sum_{f\in F}a_{T,f}(\bm{x})d_G(f(\bm{x}),t),
\quad\text{for all }\bm{x}\in T, t\in G.
\end{equation}

The function $(\bm{x},t)\mapsto a_{T,f}(\bm{x})$ is affine because it
is affine in $\bm{x}$ and independent of $t$. For each selector $f$,
the tuple $(\bm{x},t)\mapsto(f(\bm{x}),t)$ is PL by
\cref{simp-new:lem-pl-pq-calculus}. Composing it with the PL graph distance
from \cref{simp-new:lem-pl-distance} shows that each distance factor
$d_G(f(\bm{x}),t)$ is PL. Their products and
finite sum are PQ by \cref{simp-new:lem-pl-pq-calculus}. More explicitly,
a common refinement of the distance factors' affine pieces makes every
summand a product of two affine functions on each cell.

The products $T\times G$ form a finite cover of $G^n\times G$ by
compact polyhedra. On their overlaps, the regional lotteries agree, so
the displayed total costs agree. Triangulate each $T\times G$ so that
its cost formula is a finite sum of products of two affine functions on
every simplex. By point 1 of \cref{simp-new:lem-pl-calculus}, these
triangulations admit a compatible triangulation of $G^n\times G$.
On each resulting simplex, $C_{M_A}$ inherits an affine-product formula
from a containing region. The formulas agree on intersections, and
continuity follows from the finite closed cover. Thus $C_{M_A}$ is PQ.
Only the regional lotteries and their costs must agree on overlaps;
individual regional weights need not agree.

\end{proof}

\section{Proof of the finite benchmark lemma}

\label{quad-search:proof-benchmarks}

The finite benchmark lemma is a consequence of Hakimi's network-median
theorem \cite[p. 456, theorem in “Absolute Median”]{hakimi1964optimum}.
We give a direct proof below.

\restatestatement{quad-search:lem-benchmarks}{\claimFiniteBenchmarks}

\begin{proof}

Let $W=\{v\in V(G):\deg_G(v)\geq3\}$, where $\deg_G(v)$ is the number
of edges incident to $v$. We prove the stronger assertion that reports
and vertices in $W$ suffice.
Fix a profile $\bm{x}$ and let $Z=\{x_i:i\in N\}\cup W$.
We show that every location $y\in G$ has a point of $Z$ with no
greater social cost. This is immediate if $y\in Z$. Otherwise, $y$
is an interior point of an edge or an unreported vertex of degree at
most $2$, so there are at most two directions leaving $y$.

Choose a shortest path from $y$ to each $x_i$, and choose an initial
direction followed by $k\geq n/2$ of these paths. Let $z$ be the first
point of $Z$ reached in that direction. Such a point exists because
each selected path ends at a report. Before reaching $z$, the paths
cannot separate at a branching vertex or end at a report, so they
share the entire subpath from $y$ to $z$. Write $s$ for its length.
Subpaths of shortest paths are shortest, so for each of these $k$ agents,

\begin{equation}
d_G(z,x_i)=d_G(y,x_i)-s.
\end{equation}

For every other agent, the triangle inequality gives
$d_G(z,x_i)\leq d_G(y,x_i)+s$. Consequently,

\begin{equation}
\sum_{i\in N}d_G(z,x_i)
\leq\sum_{i\in N}d_G(y,x_i)+(n-2k)s
\leq\sum_{i\in N}d_G(y,x_i).
\end{equation}

The points of $Z$ are exactly the outcomes of the mechanisms in
$\Pi(W)$ at this profile. Thus one of these benchmarks minimizes
social cost over all locations. A lottery averages the social costs
of its outcomes, so allowing lotteries gives the same minimum.
Since $W\subseteq V(G)$, minimizing over $\Pi(V(G))$ also gives
$\operatorname{OPT}(\bm{x})$, as claimed.

\end{proof}

\section{Global verification}

\label{quad-search:proof-verification}

\Cref{quad-search:presentation} expresses SP and approximation as
nonnegativity of finitely many PQ comparisons for a given candidate.
The following proposition turns that description into a finite check,
returning violating locations whenever a comparison fails. Its proof first
constructs the comparison cells and their polynomial formulas from the
candidate's finite representation, then applies the nonnegativity oracle.

\subsection{Verification procedure}

\label{simp-new:proof-global-verification}

\begin{proposition}[Global verification]\label{quad-search:prop-verification}

Let $A$ be a PL attribution on $G$, with the graph, selectors, regions,
and affine weights given by explicit finite descriptions. Assume exact
arithmetic and comparisons and access to an oracle that decides
nonnegativity of a polynomial of degree at most $2$ on a nonempty compact
polytope, returning a negative witness if it fails. There are finite procedures that:

\begin{enumerate}
\def\labelenumi{\arabic{enumi}.}
\tightlist
\item
  verify that $M_A$ is SP, or return an agent, profile, and
  deviating report that violate SP;
\item
  given $\alpha\geq1$, verify that $M_A$ has approximation ratio at most
  $\alpha$, or return a profile and benchmark $g\in\Pi(V(G))$ whose
  approximation comparison fails.
\end{enumerate}

\end{proposition}

\begin{proof}

\noindent\textbf{Finite input and comparison cells.}

\phantomsection\label{simp-new:verification-inputs}

The input for \cref{quad-search:prop-verification} specifies one PL
attribution $A$. It gives the straight-edge graph's incidence data,
vertex coordinates, and edge lengths; the vertices and simplices of
$\mathcal T$; a finite simplicial description of each selector $f\in F$
with its affine formula on each simplex; and the regional affine weights.
An approximation query also gives $\alpha$. Different queries may use
different selectors and regions. Arithmetic operations and equality and
order comparisons on the input data and oracle outputs are exact.

We first construct cells on which expected cost has a polynomial formula.
Distances between graph vertices are
found by comparing lengths of the finitely many simple edge paths.
\Cref{simp-new:lem-distance-branches} then gives the affine distance
branches. Compose these with the selector pieces, and refine the
products $T\times G$ until all distance factors are affine. Multiplying
by the affine weights and summing gives the expected-cost polynomials
from \cref{quad-search:proof-quadratic-cost}.

For an agent $i$, take the inverse images of these cost cells under the
two maps

\begin{equation}
(\bm{x},z)\longmapsto((z,\bm{x}_{-i}),x_i),
\qquad
(\bm{x},z)\longmapsto(\bm{x},x_i),
\end{equation}

which are affine on each product of graph edges. Intersect the inverse
images with these edge products and take a common refinement. On each
resulting cell, both cost terms, and hence their difference, are
polynomials of degree at most $2$. For approximation, the benchmark
costs are PL because every $g\in\Pi(V(G))$ is a projection or a
constant selector. Apply the same inverse-image construction to the
candidate and benchmark cost cells under
$\bm{x}\mapsto(\bm{x},x_i)$, and refine for all agents and benchmarks.
Summing their cost formulas gives the approximation
comparisons. The SP cells have dimension at most $n+1$ and the
approximation cells at most $n$.

All of these constructions use finite intersections, inverse images under affine maps,
and common refinements, as in \cref{simp-new:lem-pl-calculus}. Linear
algebra and inequality checks determine the resulting polytopes and
their faces. They can be triangulated consistently by assigning to each
nonempty face the average of its vertices and taking the simplices
whose vertices correspond to nested chains of faces. Shared faces
receive the same triangulation. The polynomial formulas are
obtained by affine substitution, multiplication, addition, and
subtraction.

\noindent\textbf{Verification.}

For each agent $i$, the SP comparison
$C_{M_A}((z,\bm{x}_{-i}),x_i)-C_{M_A}(\bm{x},x_i)$ is PQ on
$G^n\times G$ by \cref{quad-search:presentation}. The construction above
gives a finite subdivision on which it is a sum of affine products on every closed
cell. Expanding these products in coordinates gives a polynomial of
degree at most $2$, which is the input to the nonnegativity oracle.
Test that polynomial on each nonempty cell. A negative witness gives a
true profile $\bm{x}$ and deviating report $z$ at which agent $i$ reduces
its expected cost. If all tests for all agents certify nonnegativity,
the SP inequalities hold everywhere,
since the cells cover the comparison domains.

For a proposed $\alpha\geq1$ and each $g\in\Pi(V(G))$, apply the
same procedure to the following PQ function on $G^n$:

\begin{equation}
\alpha\operatorname{SC}(g(\bm{x}),\bm{x})
-\sum_{i\in N}C_{M_A}(\bm{x},x_i).
\end{equation}

A negative witness gives a violating profile and benchmark. Otherwise,
all benchmark comparisons hold, and \cref{quad-search:lem-benchmarks}
makes them equivalent to the proposed approximation bound. These
inequalities include unanimous profiles, where optimal social cost is
$0$.

The procedure uses one oracle call per comparison polynomial on each
nonempty cell.
There are finitely many such calls, and
a negative point on a shared face is a valid witness because the
closed-cell formulas agree with the comparison there.

\end{proof}

The verification subdivision may depend on the supplied candidate.
The coefficient-independent subdivision needed to impose finite SSP constraints during search is constructed
separately in \cref{finite-ssp:subdivision-construction}.

Admissibility is a hypothesis of this verification procedure. Its finite
linear constraints, used when searching over coefficients, are given in
\cref{simp-new:app-attribution-constraints}. A finite implementation of
the exact nonnegativity oracle follows in
\cref{simp-new:app-quadratic-minimization}. The experimental checkers are
described in \cref{lower-exp:app-certificates}.

\section{Implementing the quadratic nonnegativity oracle}

\label{simp-new:app-quadratic-minimization}

\Cref{quad-search:prop-verification} reduces each global check to
finitely many nonnegativity tests on compact polytopes. Here we specify the
oracle used in that reduction and give a self-contained finite implementation under
exact arithmetic and comparisons. The implementation covers singular
quadratics and cells of any dimension.

\begin{definition}[Quadratic nonnegativity oracle]\label{simp-new:def-quadratic-oracle}

Given a nonempty compact polytope $P$ described by finitely many affine
equalities and inequalities, and a polynomial $q$ of degree at most $2$
on $P$, the oracle decides whether

\begin{equation}
q(u)\geq0,\quad\text{for all }u\in P.
\end{equation}

If this condition fails, it returns a point $u\in P$ with $q(u)<0$.

\end{definition}

The following finite procedure explains one implementation in
an exact real-arithmetic model. Inputs give the polynomial coefficients
and a finite affine description of the nonempty compact polytope.
The procedure assumes exact arithmetic and exact equality and order
comparisons on the input data and all quantities computed from them.

We separate the implementation into two steps. First we give a
nonnegativity test on the standard simplex, where faces are specified
by the coordinates allowed to be nonzero. We then reduce the general
polytope problem to this test by triangulation and a change to
barycentric coordinates.

\begin{lemma}[Nonnegativity on the standard simplex]\label{simp-new:lem-simplex-nonnegativity}

Assume exact arithmetic and exact equality and order comparisons are
available. Given a symmetric matrix $B\in\mathbb R^{k\times k}$ with
$k\geq1$, a finite procedure decides whether

\begin{equation}
\lambda^{\mathsf T}B\lambda\geq0,
\quad\text{for all }\lambda\in\mathbb R^k
\text{ with }\lambda\geq0\text{ and }\mathbf 1^{\mathsf T}\lambda=1.
\end{equation}

If nonnegativity fails, the procedure returns a feasible $\lambda$ with
$\lambda^{\mathsf T}B\lambda<0$.

\end{lemma}

\begin{proof}

The quadratic attains its minimum because the simplex is compact.
Our goal is to construct a finite list of feasible points containing
at least one minimizer. Checking their values then either finds a
negative witness or establishes nonnegativity. A minimum may occur on
the boundary, so we must consider every face of the simplex.

A face is specified by a nonempty set $I\subseteq\{1,\ldots,k\}$:
coordinates outside $I$ are $0$, and the remaining coordinates sum to
$1$. A point in its relative interior has positive coordinates precisely
on $I$, its \emph{support}. Write $u$ for those coordinates
and $B_{II}$ for the corresponding principal submatrix of $B$.
At a minimum in the relative interior, the directional derivative must
vanish along every direction whose coordinates sum to $0$. Since the
gradient is $2B_{II}u$, this means that all entries of $B_{II}u$ are
equal. Introducing their common value $t$, we obtain the linear system

\begin{equation}
\label{simp-new:eq-face-stationarity}
\begin{pmatrix}
B_{II}&-\mathbf 1\\
\mathbf 1^{\mathsf T}&0
\end{pmatrix}
\begin{pmatrix}u\\t\end{pmatrix}
=\begin{pmatrix}0\\1\end{pmatrix}.
\end{equation}

The last row imposes the sum constraint. Moreover, any solution with
$u\geq0$ gives a feasible point after extending $u$ by zeros outside
$I$, and its quadratic value is
$u^{\mathsf T}B_{II}u=t\mathbf 1^{\mathsf T}u=t$.

To obtain a finite list, we would like to solve only the nonsingular
systems, each of which has a unique solution. This is sufficient because
a singular system at a minimizer allows us to move to a smaller face
without changing the minimum value. To prove this, choose a global
minimizer with the smallest possible support $I$, and let $u$ be its
positive coordinates. It satisfies \cref{simp-new:eq-face-stationarity}.
If the matrix in that system were singular, a nonzero null vector $(h,s)$ would
satisfy $B_{II}h=s\mathbf 1$ and $\mathbf 1^{\mathsf T}h=0$. Symmetry gives

\begin{equation}
s=u^{\mathsf T}B_{II}h=(B_{II}u)^{\mathsf T}h
  =t\mathbf 1^{\mathsf T}h=0.
\end{equation}

Hence $h\ne0$ and $B_{II}h=0$. Moving from $u$ along $h$ preserves both
the coordinate sum and the quadratic value. Because $h$ has both positive
and negative coordinates, we can continue until a coordinate first
reaches $0$: take $\tau=\min_{h_j<0}u_j/(-h_j)$ in $u+\tau h$.
All coordinates remain nonnegative, so the resulting point is a minimizer
on a smaller face. This contradicts the choice of $I$. Thus at least
one global minimizer has a nonsingular stationarity system.

The test now follows. Enumerate every nonempty $I$, solve
\cref{simp-new:eq-face-stationarity} when its matrix is nonsingular,
and discard any solution with a negative coordinate of $u$. If a
retained solution has $t<0$, extend $u$ by zeros and return the resulting
negative witness after checking its value. Otherwise, after all sets
$I$ have been considered, certify nonnegativity: the retained points
include a global minimizer. Singleton supports include the vertices,
and the argument also covers singular quadratics and exact zero minima,
without convexity or genericity assumptions.

\end{proof}

On a simplex with $k$ vertices, the procedure examines $2^k-1$ nonempty
supports and solves the stationarity system whenever its matrix is
nonsingular. In the global verification procedure, $k\leq n+2$ for SP
comparisons and $k\leq n+1$ for approximation comparisons.

To test nonnegativity on a general polytope, we triangulate it and check
each simplex separately. On each piece, a change to barycentric
coordinates puts the polynomial into the form handled by the preceding
lemma. If every piece passes the test, the polynomial is nonnegative
on the whole polytope. Otherwise, a failing test supplies a negative
witness.

\begin{lemma}[Finite quadratic nonnegativity test]\label{simp-new:lem-quadratic-minimization}

Assuming exact arithmetic and exact equality and order comparisons,
the oracle in \cref{simp-new:def-quadratic-oracle} has a finite
implementation, including for lower-dimensional polytopes and singular
quadratics.

\end{lemma}

\begin{proof}

We reduce the input polynomial $q$ on the compact polytope $P$ to
finitely many tests of the preceding form. First, enumerate the vertices
and faces of $P$ by exact linear algebra and inequality checks, and
triangulate $P$ in its affine hull by the construction in the proof of
\cref{quad-search:prop-verification}. Nonnegativity on $P$ is equivalent
to nonnegativity on every simplex in this finite cover.

To apply \cref{simp-new:lem-simplex-nonnegativity} to one such simplex,
we express its points in barycentric coordinates. Let $V$ have the
simplex's $k$ vertices as columns. Then its points are exactly
$x=V\lambda$ with $\lambda\geq0$ and
$\mathbf 1^{\mathsf T}\lambda=1$. This description also applies to
lower-dimensional simplices and to a single point, where $k=1$.

It remains to put the polynomial in the matrix form required by the
simplex test. Write $q(x)=x^{\mathsf T}Hx+g^{\mathsf T}x+c$, with $H$
symmetric, and set $w=V^{\mathsf T}g$. Substitution gives
$q(V\lambda)=\lambda^{\mathsf T}V^{\mathsf T}HV\lambda+w^{\mathsf T}\lambda+c$.
Since the coordinates of $\lambda$ sum to $1$, we can absorb the linear
and constant terms into a quadratic form by setting

\begin{equation}
B=V^{\mathsf T}HV+
  \tfrac12(w\mathbf 1^{\mathsf T}+\mathbf 1w^{\mathsf T})
  +c\mathbf 1\mathbf 1^{\mathsf T}.
\end{equation}

This symmetric matrix satisfies
$q(V\lambda)=\lambda^{\mathsf T}B\lambda$ throughout the standard
simplex. Apply \cref{simp-new:lem-simplex-nonnegativity} to $B$.
A negative witness $\lambda$ gives the negative witness $V\lambda\in P$
for the original polynomial. If every simplex test certifies
nonnegativity, their union is $P$, so $q$ is nonnegative on all of $P$.

\end{proof}

If the polynomial coefficients and polyhedral input data are rational,
the triangulation, stationarity calculations, and any returned negative
witness can all be computed using exact rational arithmetic.
No polynomial-time or bit-complexity bound is asserted here.
The finite procedure implements a single oracle query; it supplies no
convergence guarantee for the sampled search.

\section{Attribution classes, constraints, and separation}

\label{simp-new:app-attribution-separation}

The search in \cref{simp-new:heuristic-search} fixes selectors $F$
and regions $\mathcal T$ and varies their affine mixing functions.
We first justify enlarging these choices, then give coordinates and
finite constraints for the admissible weights. Finally, we express
cost comparisons in those coordinates and show how a failed global
verification gives a separating inequality. The auxiliary subdivisions
used to write these constraints introduce no new optimization variables.
All statements in this section allow real graph and coefficient data.

\subsection{Common enlargement of attribution classes}

\label{simp-new:proof-common-enlargement}

\begin{lemma}[Common enlargement of attribution classes]\label{simp-new:lem-common-enlargement}

Fix $G$ and $n$, and let $(F_1,\mathcal T_1)$ and $(F_2,\mathcal T_2)$
satisfy the PL-attribution assumptions. Set $F_*=F_1\cup F_2$.
For any common simplicial refinement of the two underlying triangulations,
let $\mathcal T_*$ be its family of maximal simplices. Then
$\mathcal M_{\mathcal T_1,F_1}\cup\mathcal M_{\mathcal T_2,F_2}
\subseteq
\mathcal M_{\mathcal T_*,F_*}$.

\end{lemma}

\begin{proof}

Let $\mathcal T_*$ be the family of maximal simplices of any such common
refinement, and fix
$j\in\{1,2\}$ and an attribution $A\in\mathcal A_{\mathcal T_j,F_j}$.
For each maximal simplex $S\in\mathcal T_*$, choose a region
$T\in\mathcal T_j$ containing $S$. For $f\in F_j$, use
$a_{T,f}|_S$ as the new weight on $S$; for $f\in F_*\setminus F_j$,
use weight $0$. These weights are affine, nonnegative, and sum to $1$.
Each new region is contained in a product of graph edges because its
containing original region is. On every new region the mixture equals
$M_A$, so the resulting lotteries agree on overlaps. This defines a PL
attribution in $\mathcal A_{\mathcal T_*,F_*}$ representing $M_A$.
Applying the construction to either value of $j$ proves the inclusion.

\end{proof}

\subsection{Vertex coordinates and interpolation}

\label{simp-new:app-vertex-coordinates}

To optimize over $\mathcal V_{\mathcal T,F}$, we need finitely many
variables describing each regional affine weight. Values at the vertices
of its original simplex provide such coordinates. Barycentric
interpolation then expresses every evaluation used by the overlap,
separation, and SSP constraints in these same variables.

\begin{remark}[Vertex parametrization]\label{simp-new:vertex-parametrization}

For explicit calculations and experiments, one can choose the regional
vertex values as coordinates. The map

\begin{equation}
A\longmapsto
\bigl(a_{T,f}(v)\bigr)_{T\in\mathcal T,\ f\in F,\ v\in\operatorname{Vert}(T)}
\end{equation}

identifies $\mathcal V_{\mathcal T,F}$ linearly with $\mathbb R^m$.
Here $m=|F|\sum_{T\in\mathcal T}(\dim T+1)$, as in
\cref{simp-new:optimization}: each affine weight has one coordinate
per vertex of its simplex.
The associated basis has one family for each $(T,f,v)$: its
$(T,f)$-component is $1$ at $v$ and $0$ at every other vertex of $T$,
and all other components vanish. Shared geometric vertices retain
separate coordinates for each region, since admissibility requires
agreement of the lotteries rather than individual selector weights.
The interpolation formulas below express evaluations in these coordinates.

\end{remark}

For $v\in\operatorname{Vert}(T)$, let $\lambda_{T,v}$ be the unique
affine function on $T$ taking value $1$ at $v$ and $0$ at every other
vertex. These are the barycentric coordinate functions. For all
$\bm{x}\in T$ and $v\in\operatorname{Vert}(T)$,

\begin{equation}
\bm{x}=\sum_{v\in\operatorname{Vert}(T)}\lambda_{T,v}(\bm{x})v,
\qquad
\sum_{v\in\operatorname{Vert}(T)}\lambda_{T,v}(\bm{x})=1,
\qquad \lambda_{T,v}(\bm{x})\geq0.
\end{equation}

Every coefficient family is recovered from its regional vertex values by

\begin{equation}
\label{simp-new:eq-weight-coordinates}
a_{T,f}(\bm{x})
=\sum_{v\in\operatorname{Vert}(T)}
\lambda_{T,v}(\bm{x})a_{T,f}(v),
\quad\text{for all }T\in\mathcal T, f\in F, \bm{x}\in T.
\end{equation}

The families obtained by placing a single $\lambda_{T,v}$ in component
$(T,f)$ and $0$ in all other components are precisely the basis from
\cref{simp-new:vertex-parametrization}. The formula
also expresses evaluations at vertices of an auxiliary subdivision in
the original coordinates. In particular, applying it to
\cref{simp-new:eq-consistency-group-rows} gives an explicit matrix for the
overlap equations without adding variables.

\subsection{Finite equations for overlap consistency}

\label{simp-new:app-attribution-constraints}

Nonnegativity and normalization of the affine weights are equivalent to
finitely many conditions at the regional vertices. Agreement of lotteries on overlaps needs more care: selectors
may coincide only on part of an overlap, so checking the lotteries at its
vertices is insufficient. We group selectors by agreement on each whole
affine piece and impose equations on their total weights.

\restatestatement{simp-new:prop-attribution-polytope}{\claimAttributionPolytope}

\begin{proof}

The finite probability constraints are

\begin{equation}
\label{simp-new:eq-probability-rows}
a_{T,f}(v)\geq0,
\qquad
\sum_{f\in F}a_{T,f}(v)=1,\quad\text{for all }T\in\mathcal T, v\in\operatorname{Vert}(T), f\in F.
\end{equation}

By barycentric interpolation, these vertex inequalities and equalities
are equivalent to nonnegativity and normalization throughout every
region.

We now construct the finite overlap-consistency equations.

Fix $T,U\in\mathcal T$ with nonempty intersection. Subdivide $T\cap U$
into finitely many convex polytopes $Q$ on which every selector is affine
as a map to the Euclidean space containing $G$. Such a subdivision
exists by intersecting the finite affine pieces of the selectors.
The functions $a_{T,f}|_Q$ and $a_{U,f}|_Q$ remain affine.

Define an equivalence relation on $F$ by

\begin{equation}
f\sim_Q g
\quad\Longleftrightarrow\quad
f(\bm{x})=g(\bm{x}),\quad\text{for all }\bm{x}\in Q.
\end{equation}

Since the selectors are affine on $Q$, this relation is determined by
their values at its vertices. For every equivalence class $J\subseteq F$,
impose the equations

\begin{equation}
\label{simp-new:eq-consistency-group-rows}
\sum_{f\in J}a_{T,f}(v)
-\sum_{f\in J}a_{U,f}(v)=0,\quad\text{for all }v\in\operatorname{Vert}(Q).
\end{equation}

Each evaluation here is a linear functional of the original regional
affine functions. A new overlap vertex does not introduce a new independent
weight.

We show that these equations are equivalent to overlap consistency.
Suppose \cref{simp-new:eq-consistency-group-rows} holds. The difference
between the two weight sums for a fixed group $J$ is affine and vanishes
at all vertices of $Q$, so it vanishes throughout $Q$. At any particular
profile, several groups may select the same point. The probability at
that point is the sum of their group weights, which is therefore equal
in the two regional descriptions. This proves consistency on $Q$.

Conversely, suppose the two lotteries agree at every point of $Q$.
For $f\not\sim_Q g$, the equality set of $f$ and $g$ is contained in a
proper affine subspace of $\operatorname{aff}(Q)$, since their affine
difference is not identically $0$. When $\dim Q>0$, the complement of
the finitely many such equality sets is dense in $Q$: a finite union of
proper affine subspaces cannot contain an open subset of its affine
hull. On this dense set, distinct groups select distinct points, so
equality of the lotteries forces equality of every group's total weight.
These affine weight differences are continuous and hence vanish on all
of $Q$, including its vertices. If $Q$ is a point, the groups already
correspond exactly to its distinct selected locations, giving the same
conclusion directly.

Thus the finite equations are equivalent to consistency on every overlap.
Together with \cref{simp-new:eq-probability-rows}, they characterize
$\mathcal A_{\mathcal T,F}$. These finitely many linear equalities and
inequalities define a closed convex set. The probability constraints
require $a_{T,f}(v)\in[0,1]$ for every region $T$, selector $f$, and vertex
$v$ of $T$. Since these finitely many values determine the affine mixing
functions, the set is bounded and hence a compact convex polytope. This
completes the proof of \cref{simp-new:prop-attribution-polytope}.

\end{proof}

Grouping selectors independently at each tested vertex would not give
this criterion. For example, on an overlap line segment with coordinate
$s\in[0,1]$, the affine selectors $f(s)=0$, $g(s)=s$, and $h(s)=1$
allow the regional lotteries $\delta_{g(s)}$ and
$(1-s)\delta_{f(s)}+s\delta_{h(s)}$. They agree at both endpoints
and differ at every interior point. Their three selectors belong to
distinct groups when agreement is tested throughout the line segment, and
the group equations detect the discrepancy.

\subsection{Regional constraints and linearity}

\label{simp-new:app-regional-expressions}

The verifier evaluates costs of one admissible attribution. Search and
separation require the same comparisons as expressions in variable
coefficients. We therefore define regional costs on the full coefficient
space, then identify the inequalities they represent on admissible weights.

For every coefficient family $A\in\mathcal V_{\mathcal T,F}$, define the
regional expression

\begin{equation}
\label{simp-new:eq-regional-cost}
c_T(A;\bm{x},t)
=\sum_{f\in F}a_{T,f}(\bm{x})d_G(f(\bm{x}),t),
\quad\text{for all }\bm{x}\in T, t\in G.
\end{equation}

For $A\in\mathcal A_{\mathcal T,F}$, this equals
$C_{M_A}(\bm{x},t)$ on every region containing $\bm{x}$.
For arbitrary $A\in\mathcal V_{\mathcal T,F}$, the same formula defines a
linear expression in the coefficients without assigning a mechanism to
inadmissible weights. These expressions let us formulate the incentive
and approximation constraints, including separating inequalities, on the
full coefficient space.

\begin{lemma}[Equivalence of regional constraints]\label{simp-new:lem-residual-constraints}

For every admissible attribution $A$ and region $T\in\mathcal T$,
the regional cost satisfies
$c_T(A;\bm{x},t)=\operatorname{cost}(M_A(\bm{x}),t)$ for every
$\bm{x}\in T$ and $t\in G$. Consequently:

\begin{enumerate}
\def\labelenumi{\arabic{enumi}.}
\tightlist
\item
  $M_A$ is SP if and only if
  $c_T(A;\bm{x},x_i)\leq c_U(A;(z,\bm{x}_{-i}),x_i)$
  for every $i\in N$, $\bm{x}\in G^n$, and $z\in G$, and every
  $T,U$ containing $\bm{x}$ and $(z,\bm{x}_{-i})$, respectively.
\item
  For $\alpha\geq1$, $M_A$ is an $\alpha$-approximation if and only if
  $\sum_{i\in N}c_T(A;\bm{x},x_i)\leq\alpha\operatorname{SC}(g(\bm{x}),\bm{x})$
  for every $T\in\mathcal T$, $\bm{x}\in T$, and $g\in\Pi(V(G))$.
\item
  $M_A$ is SSP if and only if
  $c_T(A;(y,\bm{x}_{-i}),x_i)\leq c_U(A;(z,\bm{x}_{-i}),x_i)$
  for every $i\in N$, $\bm{x}\in G^n$, and $y,z\in G$ with
  $y\in[x_i,z]_G$, and every $T,U$ containing
  $(y,\bm{x}_{-i})$ and $(z,\bm{x}_{-i})$, respectively.
\end{enumerate}

\end{lemma}

\begin{proof}

For admissible $A$, the regional lottery represents $M_A(\bm{x})$,
so its expected distance to $t$ equals $c_T(A;\bm{x},t)$. Overlap
consistency makes this equality valid for every applicable regional
choice. Substitution into the incentive comparison gives exactly the SP
inequality when $y=x_i$, and the SSP inequality when $y\in[x_i,z]_G$.
Thus retaining a separate inequality for each regional choice does not
change either requirement.

For the approximation ratio, the regional identity shows that the
left-hand side of the inequality in point 2 equals
$\operatorname{SC}(M_A(\bm{x}),\bm{x})$. By \cref{quad-search:lem-benchmarks}, the least
benchmark cost is $\operatorname{OPT}(\bm{x})$. Since $\alpha\geq1$,
comparison with every benchmark is equivalent to comparison with
$\alpha\operatorname{OPT}(\bm{x})$. This includes unanimous profiles.

Together with admissibility, these equivalences give exactly the
mechanism constraints. Every represented mechanism has an admissible
attribution, so taking infima over these constraints gives the same
value as taking the infimum of approximation ratios over represented
mechanisms with the chosen incentive property.

\end{proof}

At a fixed violating profile, all distances in these expressions are
constants. The following linearity statement lets us use that profile
to separate a candidate from every admissible attribution satisfying the
corresponding requirement.

\begin{lemma}[Linearity of regional expressions]\label{simp-new:lem-regional-linearity}

For fixed regional choices, profiles, and locations, the regional costs
$c_T(A;\bm{x},t)$ and both sides of each SP or SSP inequality in
\cref{simp-new:lem-residual-constraints} are linear functions of
$A\in\mathcal V_{\mathcal T,F}$. Both sides of each approximation
inequality in that lemma are linear functions of
$(A,\alpha)\in\mathcal V_{\mathcal T,F}\times\mathbb R$.
The same holds for the differences between their right- and left-hand sides.

\end{lemma}

\begin{proof}

At fixed profiles and locations, evaluation of a regional weight is
linear in $A$, and every distance factor is constant. The regional costs
are therefore linear in $A$, as are their sums and differences. The
benchmark social cost $\operatorname{SC}(g(\bm{x}),\bm{x})$ is also
constant for the fixed benchmark and profile, so multiplying it by $\alpha$
gives a linear function of $(A,\alpha)$. This proves the claims for both
sides of the inequalities and their differences.

\end{proof}

\subsection{Separation at violating locations}

\label{simp-new:proof-oracle-separation}

Fix $F$ and $\mathcal T$. A failed verification gives locations at
which the candidate violates a cost comparison. By
\cref{simp-new:lem-regional-linearity}, keeping these locations fixed
turns the cost comparison into a linear inequality in the original
coefficients. Together with the global verification procedure, this gives
the following separation result.

\restatestatement{simp-new:prop-separation-reduction}{\claimSeparation}

\begin{proof}

Suppose SP verification returns a violating agent $i$, profile
$\bm{x}_*$, and report $z_*$. Choose original attribution regions
$T,U\in\mathcal T$ containing $\bm{x}_*$ and
$(z_*,\bm{x}_{*,-i})$, respectively. For a variable coefficient family
$A'\in\mathcal V_{\mathcal T,F}$, the inequality

\begin{equation}
c_U(A';(z_*,\bm{x}_{*,-i}),x_{*,i})
-c_T(A';\bm{x}_*,x_{*,i})\geq0
\end{equation}

is linear in $A'$. Every admissible SP attribution satisfies it, while
the queried attribution $A$ violates it.

If approximation verification returns a violating profile
$\bm{x}_*$ and benchmark $g$, choose $T\in\mathcal T$ containing
$\bm{x}_*$. The inequality

\begin{equation}
\alpha'\operatorname{SC}(g(\bm{x}_*),\bm{x}_*)-
\sum_{i\in N}c_T(A';\bm{x}_*,x_{*,i})\geq0
\end{equation}

is linear in $(A',\alpha')$. Every admissible pair whose approximation
ratio is at most its proposed factor satisfies it, while the queried
pair $(A,\alpha)$ violates it.

Both inequalities are defined on the full coefficient space by the
regional formulas. At shared faces, admissibility makes every applicable
formula agree with the corresponding mechanism's expected cost. Thus
boundary witnesses also give valid separating inequalities. Together
with the successful outcomes of \cref{quad-search:prop-verification},
these observations prove \cref{simp-new:prop-separation-reduction}.

\end{proof}

The verifier's subdivision need not be shared by other candidates in this class.

To obtain separation for the full feasible set in
$\mathcal V_{\mathcal T,F}\times\mathbb R$, first check the finite
admissibility constraints and $\alpha\geq1$. A failed inequality already
separates the query. If an equality fails, use its violated direction.
For queries passing these checks, apply the two independent separation
procedures above. This additional admissibility check is not part of
verification for a given admissible attribution.

\section{Proofs for the finite SSP characterization}

\label{finite-ssp:proofs}

\Cref{simp-new:ssp-analysis} reduces SSP to local comparisons and then to
linear constraints on attribution coefficients. We first prove the
locality theorem in the original report and true-location coordinates.
We then turn within-cell comparisons into vertex derivative tests. The
final subsection constructs a compatible subdivision using only the fixed
graph, selectors, and attribution regions, and uses it to prove
\cref{simp-new:ssp-constraints}. Thus the same finite inequalities apply
to every coefficient family in the class.

\subsection{Path subdivision and locality}

We work on $X=G^n\times G$, whose points $(\bm{x},t)$ record a reported
profile and the true location at which cost is evaluated. For an SSP
comparison, only agent $i$'s report moves; the truth $t$ and the other
reports remain fixed. We call a movement from $y$ to $z$ \emph{outward}
when it extends a shortest path from $t$, that is, when
$d_G(t,z)=d_G(t,y)+d_G(y,z)$.
To prove \cref{finite-ssp:thm-locality-ssp}, we
subdivide this movement along a shortest path into portions lying in
individual cells. The following lemma gives the required finite
partition, including when the path lies on cell boundaries.

\begin{lemma}[Partitioning a polygonal path among covering polytopes]\label{finite-ssp:lem-path-subdivision}

Let $k$ be an arbitrary nonnegative integer, and let $Y\subseteq\mathbb R^k$
be covered by a finite family $\mathcal U$ of closed convex polytopes.
Let $\gamma:[0,1]\to Y$ be a continuous
piecewise-affine path with finitely many affine pieces. Then there are

\begin{equation}
0=s_0<s_1<\cdots<s_m=1
\end{equation}

and $U_1,\ldots,U_m\in\mathcal U$ such that

\begin{equation}
\gamma([s_{j-1},s_j])\subseteq U_j,\quad\text{for all }j\in\{1,\ldots,m\}.
\end{equation}

\end{lemma}

\phantomsection\label{finite-ssp:proof-lem-path-subdivision}

\begin{proof}

First partition $[0,1]$ into intervals on which $\gamma$ is
affine. On each such interval $I$, every set
$I\cap\gamma^{-1}(U)$, for $U\in\mathcal U$, is a closed interval,
a singleton, or empty, by continuity and convexity. Refine the partition
using the endpoints of all these nonempty sets. There are finitely many
such endpoints.

For consecutive partition points $a<b$, choose $s\in(a,b)$ and a polytope
$U$ containing $\gamma(s)$. On the affine piece containing $[a,b]$, the
preimage of $U$ is a closed interval containing $s$. Neither of its
endpoints can lie in $(a,b)$, so it contains $[a,b]$. Thus
$\gamma([a,b])\subseteq U$, as required.

\end{proof}

\phantomsection\label{finite-ssp:proof-locality-ssp}

\restatestatement{finite-ssp:thm-locality-ssp}{\claimSSPLocality}

\begin{proof}

If $M$ is SSP, all comparisons within individual cells hold by
restriction. Conversely, suppose they hold, and fix an agent $i$,
a true location $t$, other reports $\bm{x}_{-i}$, and reports $y,z$ with
$d_G(t,z)=d_G(t,y)+d_G(y,z)$. If $y=z$, we are done.
Otherwise, concatenate shortest paths from $t$ to $y$ and from $y$ to
$z$. The distance equality makes their concatenation a shortest path
from $t$ to $z$.

Let $\gamma:[0,1]\to G$ traverse the portion from $y$ to $z$.
Since $G$ is a finite straight-edge graph, this path has a continuous
piecewise-affine parametrization with finitely many pieces. Keeping
the other reports and the truth fixed gives the polygonal path
$s\mapsto((\gamma(s),\bm{x}_{-i}),t)$ in $X$.
By \cref{finite-ssp:lem-path-subdivision}, it has a partition
$0=s_0<\cdots<s_m=1$ such that each portion lies in one cell of
$\mathcal U$. Write $y_j=\gamma(s_j)$. Successive reports lie in order
on the same shortest path from $t$, so
$d_G(t,y_j)=d_G(t,y_{j-1})+d_G(y_{j-1},y_j)$ for every
$j\in\{1,\ldots,m\}$. Thus, the assumed within-cell SSP comparisons give

\begin{equation}
C_M((z,\bm{x}_{-i}),t)-C_M((y,\bm{x}_{-i}),t)
=\sum_{j=1}^{m}\left[
C_M((y_j,\bm{x}_{-i}),t)-C_M((y_{j-1},\bm{x}_{-i}),t)
\right]\geq0.
\end{equation}
This is the required SSP inequality. The argument uses only the values
of $C_M$ at shared endpoints and requires no continuity assumption on
$M$.

\end{proof}

\subsection{Quadratic monotonicity}

Locality reduces SSP to comparisons within individual cells. When a
cell has a quadratic cost formula and a fixed outward direction, the
following lemma reduces those comparisons to tests at its vertices.

Let $H$ be an affine space, with direction space
$\operatorname{lin}(H)=\{x-y:x,y\in H\}$. For a polynomial $q$ on $H$
and $d\in\operatorname{lin}(H)$, write

\begin{equation}
D_dq(x)=\left.\frac{\mathrm d}{\mathrm ds}q(x+sd)\right|_{s=0},
\quad\text{for all }x\in H.
\end{equation}

Monotonicity in direction $d$ means that moving a nonnegative distance
in that direction cannot decrease the function, whenever both endpoints
remain in the domain.

\phantomsection\label{finite-ssp:lem-quadratic-monotonicity-full}

\begin{lemma}[Quadratic monotonicity on a polytope]\label{finite-ssp:lem-quadratic-monotonicity}

Let $P$ be a nonempty compact convex polytope, let
$H=\operatorname{aff}(P)$, and let
$d\in\operatorname{lin}(H)\setminus\{0\}$. For a polynomial
$q:H\to\mathbb R$ given as a finite sum of products of two affine
functions, the following are equivalent:

\begin{enumerate}
\def\labelenumi{\arabic{enumi}.}
\tightlist
\item
  $q(x+hd)-q(x)\geq0$ whenever $h\geq0$ and $x,x+hd\in P$.
\item
  $D_dq(x)\geq0$ for every $x\in\operatorname{relint}(P)$.
\item
  $D_dq(x)\geq0$ for every $x\in P$.
\item\label{finite-ssp:condition-vertex-derivatives}
  $D_dq(v)\geq0$ for every $v\in\operatorname{Vert}(P)$.
\end{enumerate}

The vertex inequalities in \hyperref[finite-ssp:condition-vertex-derivatives]{Condition~\ref*{finite-ssp:condition-vertex-derivatives}}
are linear in the coefficients of $q$.

\end{lemma}

\phantomsection\label{finite-ssp:proof-lem-quadratic-monotonicity}

\begin{proof}

If the first condition holds and $x\in\operatorname{relint}(P)$,
then $x+hd\in P$ for all sufficiently small $h>0$, since $d$ is parallel
to $H$. Dividing by $h$ and taking $h\downarrow0$ gives $D_dq(x)\geq0$,
which proves the second condition.

The relative interior is dense in $P$ and $D_dq$ is continuous, so the
second condition implies the third. Restricting to the vertices gives
the fourth condition.

Finally, assume the fourth condition. The derivative $D_dq$ is affine by the product rule,
so its value at a convex combination of vertices is the same convex
combination of its vertex values. Thus $D_dq$ is nonnegative throughout
$P$. On the line $x+sd$, each factor is affine in $s$.
Expanding their products gives the exact identity

\begin{equation}
q(x+hd)-q(x)=h\,D_dq\left(x+\frac h2d\right).
\end{equation}

If $h\geq0$ and both endpoints belong to $P$, so does their midpoint,
and the right-hand side is nonnegative. This proves the first condition,
including $h=0$.

Differentiation and evaluation are linear in the polynomial coefficients,
so the vertex tests are linear inequalities in those coefficients.

\end{proof}

The vertex tests in \hyperref[finite-ssp:condition-vertex-derivatives]{Condition~\ref*{finite-ssp:condition-vertex-derivatives}} of
\cref{finite-ssp:lem-quadratic-monotonicity} evaluate $D_dq(v)$ using
the polynomial defined on $H=\operatorname{aff}(P)$. They do not require
$v+hd$ to remain in $P$ for small $h>0$.

\subsection{Cell formulas and the finite test}

\label{finite-ssp:cell-formulas}

To apply \cref{finite-ssp:lem-quadratic-monotonicity}, we need cells
on which expected cost is quadratic and outward report movements have a
fixed direction. The compatibility conditions below specify both
requirements. \Cref{finite-ssp:subdivision-construction} constructs a
subdivision satisfying them.

Keep the fixed attribution class from \cref{simp-new:optimization}, its
coefficient space $\mathcal V_{\mathcal T,F}$, and its admissible subset $\mathcal A_{\mathcal T,F}$.
For calculations on the full coefficient space, use the regional
expressions $c_T$ defined in
\cref{simp-new:app-regional-expressions}.
They agree with mechanism costs for an admissible attribution $A$, while $M_A$ is used
only for those admissible families.

Let $K_G$ be the geometric simplicial complex of $G$. Products of its
edges and vertices form a polytopal complex on $X=G^n\times G$, whose
maximal cells are the edge products $P=e_1\times\cdots\times e_n\times e_{n+1}$.
The last factor contains the true location. For a selector $f\in F$,
its expected-cost function is

\begin{equation}
C_f(\bm{x},t)=d_G(f(\bm{x}),t),
\quad\text{for all }\bm{x}\in G^n, t\in G.
\end{equation}

\phantomsection\label{simp-new:ssp-cells}

\begin{definition}[Compatible subdivision for SSP]\label{finite-ssp:def-compatible-subdivision}

A finite simplicial subdivision $\mathcal U$ of the product graph
complex on $X$ is \emph{compatible with $(\mathcal T,F)$} if, for every
maximal simplex $U$, the following conditions hold:

\begin{enumerate}
\def\labelenumi{\arabic{enumi}.}
\item\label{finite-ssp:condition-affine-hull}
  For an edge product $P$ containing $U$,
  $\operatorname{aff}(U)=\operatorname{aff}(P)$.
\item\label{finite-ssp:condition-attribution-region}
  There is an original attribution region $T(U)\in\mathcal T$ with
  $U\subseteq T(U)\times G$.
\item\label{finite-ssp:condition-affine-distance}
  For each $f\in F$, the distance function $C_f$ is affine on $U$.
\item\label{finite-ssp:condition-outward-direction}
  For every $i\in N$, there is a fixed nonzero vector
  $\eta_{i,U}\in\operatorname{lin}(\operatorname{aff}(U))$ changing only
  report coordinate $i$. For any two points
  $((y,\bm{x}_{-i}),t),((z,\bm{x}_{-i}),t)\in U$,

  \begin{equation}
  \begin{aligned}
  d_G(t,z)&=d_G(t,y)+d_G(y,z)\\
  &\quad\Longleftrightarrow\quad
  ((z,\bm{x}_{-i}),t)=((y,\bm{x}_{-i}),t)+h\eta_{i,U}
  \text{ for some }h\geq0.
  \end{aligned}
  \end{equation}
\end{enumerate}

The subdivision, containing attribution regions, and directions are chosen
independently of the coefficient family $A$.

\end{definition}

The equivalence in \hyperref[finite-ssp:condition-outward-direction]{Condition~\ref*{finite-ssp:condition-outward-direction}} of
\cref{finite-ssp:def-compatible-subdivision} is required on the closed cells, including their
boundaries. Thus an outward comparison within $U$ is exactly a
nonnegative displacement in $\eta_{i,U}$ with both endpoints in $U$.
The direction is parallel to the affine hull of $U$; this does not
require that a positive step from every boundary point remain inside $U$.
The construction uses the slopes $+1$ and $-1$ of graph-distance
branches and the distance formula on a straight edge.
\hyperref[finite-ssp:condition-affine-hull]{Condition~\ref*{finite-ssp:condition-affine-hull}} ensures that the directions remain parallel to the
affine hulls of maximal cells after refinement, as proved in
\cref{finite-ssp:subdivision-construction}.

Suppose a subdivision $\mathcal U$ satisfies these compatibility
conditions. We now derive the finite test on each maximal cell.
Write $\mathcal U^{\max}$ for its maximal simplices and
$H_U=\operatorname{aff}(U)$. For $A\in\mathcal V_{\mathcal T,F}$ and $f\in F$, the
functions

\begin{equation}
w_{U,f,A}(\bm{x},t)=a_{T(U),f}(\bm{x}),
\qquad h_{U,f}(\bm{x},t)=C_f(\bm{x},t),
\quad\text{for all }(\bm{x},t)\in U
\end{equation}

are affine. Denote their unique affine extensions to $H_U$ by
$\widehat w_{U,f,A}$ and $\widehat h_{U,f}$. These extensions are unique
because $U$ spans $H_U$. We use them to take derivatives on $H_U$,
including at boundary vertices of $U$. Define

\begin{equation}
q_{U,A}(\bm{x},t)=\sum_{f\in F}
\widehat w_{U,f,A}(\bm{x},t)\widehat h_{U,f}(\bm{x},t),
\quad\text{for all }(\bm{x},t)\in H_U.
\end{equation}

For each agent $i$ and vertex $v$ of $U$, define the derivative test

\begin{equation}
L_{i,U,v}(A)=D_{\eta_{i,U}}q_{U,A}(v).
\end{equation}

\begin{lemma}[Cell costs on the full coefficient space]\label{finite-ssp:lem-cell-costs}

For every $A\in\mathcal V_{\mathcal T,F}$, $q_{U,A}$ is a polynomial of degree at
most $2$ and satisfies

\begin{equation}
q_{U,A}(\bm{x},t)=c_{T(U)}(A;\bm{x},t),
\quad\text{for all }(\bm{x},t)\in U.
\end{equation}

The map $A\mapsto q_{U,A}$ is linear. For admissible $A$, its restriction
to $U$ also equals $C_{M_A}|_U$.

\end{lemma}

\phantomsection\label{finite-ssp:proof-lem-cell-costs}

\begin{proof}

The polynomial is a sum of products of two affine functions.
On $U$, its factors are the regional weights and selector distances,
which gives the displayed regional-cost identity.

The operations taking $A$ to $a_{T(U),f}$, treating this weight as a
function of $(\bm{x},t)$ independent of $t$, restricting to $U$, and
extending affinely to $H_U$ are linear.
The distance factors are fixed. This proves linearity in $A$.
For $A\in\mathcal A_{\mathcal T,F}$, the regional lottery formula identifies the sum
with $C_{M_A}(\bm{x},t)$.

Differentiation and evaluation are linear, so $L_{i,U,v}$ is a linear
functional on $\mathcal V_{\mathcal T,F}$. More explicitly, the product rule gives

\begin{equation}
L_{i,U,v}(A)
=\sum_{f\in F}\left[
\bigl(D_{\eta_{i,U}}\widehat w_{U,f,A}\bigr)
\widehat h_{U,f}(v)
+\widehat w_{U,f,A}(v)
\bigl(D_{\eta_{i,U}}\widehat h_{U,f}\bigr)
\right].
\end{equation}

The derivatives of the affine factors are constants on $H_U$. The
distance factors and the direction are independent of $A$.
Only the original weights $a_{T(U),f}$ enter the formula, so each
inequality uses their coefficients alone. Values at vertices introduced by
$\mathcal U$ are evaluations of these same affine weights, expressed
in the original coordinates by interpolation as in
\cref{simp-new:app-vertex-coordinates}. They are not new optimization
variables.

\end{proof}

\begin{proposition}[Finite equivalence on the full coefficient space]\label{finite-ssp:prop-cell-equivalence}

For every $i\in N$, $U\in\mathcal U^{\max}$, and $A\in\mathcal V_{\mathcal T,F}$,
the following are equivalent:

\begin{enumerate}
\def\labelenumi{\arabic{enumi}.}
\tightlist
\item
  $q_{U,A}((z,\bm{x}_{-i}),t)-q_{U,A}((y,\bm{x}_{-i}),t)\geq0$,
  for all $t,y,z\in G$, $\bm{x}_{-i}\in G^{n-1}$ with both evaluation points
  in $U$ and $d_G(t,z)=d_G(t,y)+d_G(y,z)$.
\item
  $L_{i,U,v}(A)\geq0$, for all $v\in\operatorname{Vert}(U)$.
\end{enumerate}

The inequalities in point 2 are linear in $A$ and use
only the coefficients of the original affine weights
$(a_{T(U),f})_{f\in F}$.

\end{proposition}

\phantomsection\label{finite-ssp:proof-prop-cell-equivalence}

\begin{proof}

\hyperref[finite-ssp:condition-outward-direction]{Condition~\ref*{finite-ssp:condition-outward-direction}} of
\cref{finite-ssp:def-compatible-subdivision} identifies the pairs in the first
condition with nonnegative displacements in direction $\eta_{i,U}$.
Conversely, such a displacement changes only report $i$, and any two
endpoints in $U$ satisfy the required shortest-path equality. Thus the
first condition is exactly monotonicity of $q_{U,A}$ in this direction
on $U$.

By \cref{finite-ssp:lem-cell-costs}, $q_{U,A}$ is a sum of affine products.
The direction is nonzero and parallel to $H_U$, so
\cref{finite-ssp:lem-quadratic-monotonicity} makes this monotonicity
equivalent to nonnegativity of $D_{\eta_{i,U}}q_{U,A}$ at every vertex
of $U$. These are precisely the tests in the second condition.

Finally, differentiation and evaluation preserve the linear dependence
of $q_{U,A}$ on the weights $a_{T(U),f}$. The product-rule formula in
the preceding proof expresses each test using only the coefficients of
those weights. This cellwise equivalence holds on the full coefficient
space; admissibility is needed later to identify these formulas with
one mechanism across cells.

\end{proof}

\subsection{Global finite characterization}

\label{finite-ssp:global-characterization}
\label{finite-ssp:subdivision-construction}

To obtain one finite SSP characterization for the fixed attribution class,
we need a compatible subdivision that is independent of the coefficient
family $A$. We first construct such a subdivision and then combine the
cellwise tests with locality to prove the global characterization.

We construct three subdivisions: one keeps each cell inside an attribution
region, one makes every selector distance affine, and one fixes the
outward report directions. We then take a common refinement.
The following fact ensures that the refined maximal cells retain the
affine hulls needed for the directional test. Here a cell is
\emph{full-dimensional} if it has dimension $n+1$, the dimension of an
edge product in $X$, regardless of the dimension of the Euclidean space
containing $X$.

\begin{lemma}[Full affine hulls under subdivision]\label{finite-ssp:lem-full-dimension}

Every maximal simplex $U$ of any finite simplicial subdivision of the
product graph complex on $X$ has dimension $n+1$. If $U$ lies in the
edge product $P$, then

\begin{equation}
\operatorname{aff}(U)=\operatorname{aff}(P).
\end{equation}

\end{lemma}

\phantomsection\label{finite-ssp:proof-lem-full-dimension}

\begin{proof}

Every vertex of the graph is incident to an edge.
Hence the maximal edge products have dimension $n+1$ and cover $X$.
Fix a finite simplicial subdivision of the product graph complex and
restrict it to one maximal edge product $P$.

Every simplex of dimension below $n+1$ lies in a proper affine
subspace of $\operatorname{aff}(P)$. A finite union of these subspaces
cannot contain a relatively open subset of $P$. Thus points of
$\operatorname{relint}(P)$ outside the lower-dimensional simplices
are dense in $P$, and all belong to simplices of dimension $n+1$.
The union of these full-dimensional simplices is finite and closed, so
it covers $P$. Applying this to every maximal edge product shows that
the full-dimensional simplices cover $X$.

For any simplex $S$ in the subdivision, choose
$x\in\operatorname{relint}(S)$. Some full-dimensional simplex $U$
contains $x$. The intersection $S\cap U$ is a face of $S$ containing a
relative-interior point, so it equals $S$. Therefore $S\subseteq U$.
Every maximal simplex has dimension $n+1$. If $U\subseteq P$, then
$\operatorname{aff}(U)=\operatorname{aff}(P)$ because their dimensions
agree.

\end{proof}

We first construct a subdivision respecting the attribution regions.
Taking products of attribution regions with true-location edges fixes a
regional weight formula on each cell of $X$.

\begin{lemma}[A subdivision respecting attribution regions]\label{finite-ssp:lem-carrier-subdivision}

There is a finite simplicial subdivision $\mathcal R^{\mathrm{attr}}$
of the product graph complex on $X$ such that, for every simplex $U$,
$U\subseteq T\times G$ for some $T\in\mathcal T$.
It can be chosen independently of $A$.

\end{lemma}

\phantomsection\label{finite-ssp:proof-lem-carrier-subdivision}

\begin{proof}

Take the products of the attribution triangulation on report space
with the edges and vertices of $K_G$ in the true-location coordinate.
These products form a finite polytopal subdivision of the product graph
complex on $X$.
Each projects into an original attribution simplex, hence into some
maximal region $T\in\mathcal T$. A compatible triangulation of this
polytopal complex preserves that containment. The construction uses
$\mathcal T$ and the graph, but no coefficient family $A$.

\end{proof}

We next use \cref{simp-new:lem-pl-calculus} to make all selector
distances affine on the same subdivision.

\begin{lemma}[A subdivision with affine selector distances]\label{finite-ssp:lem-distance-subdivision}

There is a finite simplicial subdivision $\mathcal R^{\mathrm{cost}}$
of the product graph complex on $X$ such that every $C_f$ is affine on
every simplex. It can be chosen independently of $A$.

\end{lemma}

\phantomsection\label{finite-ssp:proof-lem-distance-subdivision}

\begin{proof}

For each $f\in F$, the map $(\bm{x},t)\mapsto(f(\bm{x}),t)$ is PL.
Composing it with graph distance shows that $C_f$ is PL by
\cref{simp-new:lem-pl-distance,simp-new:lem-pl-pq-calculus}.
By \cref{simp-new:lem-pl-calculus}, these finitely many functions admit
a common triangulation on which they are affine, refining the product
graph complex. Only the fixed selectors and graph enter this construction.

\end{proof}

For outward directions we use the distances
$\rho_i(\bm{x},t)=d_G(t,x_i)$ from truth to report. Their signed unit
slopes determine which direction along a report edge extends a shortest
path from the truth.

\begin{lemma}[A subdivision with fixed outward report directions]\label{finite-ssp:lem-outward-direction}

There is a finite simplicial subdivision $\mathcal R^{\mathrm{ord}}$
of the product graph complex on $X$ such that, on each maximal simplex
$U$, nonzero directions $\eta_{i,U}$ satisfy
\hyperref[finite-ssp:condition-outward-direction]{Condition~\ref*{finite-ssp:condition-outward-direction}} of
\cref{finite-ssp:def-compatible-subdivision} for every agent $i$.

This subdivision and its directions depend only on the graph and the
number of agents, and are independent of $\mathcal T$, $F$, and $A$.

\end{lemma}

\phantomsection\label{finite-ssp:proof-lem-outward-direction}

\begin{proof}

Orient each graph edge and write $s_e$ for its arclength coordinate.
Use \cref{simp-new:lem-distance-branches}
to fix a finite subdivision $\mathcal D$ of $G\times G$, refining its
product graph complex, such that on each cell $Q\subseteq e\times e'$
there is a branch

\begin{equation}
d_G(x,y)=b_Q+\varepsilon_Qs_e(x)+\sigma_Qs_{e'}(y),
\qquad \varepsilon_Q,\sigma_Q\in\{-1,1\}.
\end{equation}

Put $\psi_i(\bm{x},t)=(t,x_i)$ and $\rho_i=d_G\circ\psi_i$. For each
product graph cell $P$ and each $Q\in\mathcal D$, take
$P\cap\psi_i^{-1}(Q)$. Each nonempty intersection is a polytope:
substituting $\psi_i(\bm{x},t)$ into the affine constraints defining $Q$
gives additional affine constraints on $P$.
Within $P$, the inverse images of common target faces are common
faces of the resulting polytopes. The constructions on different
product cells also agree on their shared faces.
Thus these intersections and their faces form a finite polytopal
subdivision. Take a common refinement for all agents and a compatible
triangulation, denoted by $\mathcal R^{\mathrm{ord}}$.

Let $U$ be one of its maximal simplices, in the edge product
$P=e_1\times\cdots\times e_n\times e_{n+1}$. By
\cref{finite-ssp:lem-full-dimension}, $\operatorname{aff}(U)=\operatorname{aff}(P)$.
The projection $\psi_i$ retains the independently varying true-location
and $i$th report coordinates, so it has rank $2$ on this affine space.
Consequently, $\psi_i(U)$ is two-dimensional. It lies in a cell $Q_{i,U}$ of $\mathcal D$ within
$e_{n+1}\times e_i$, which must therefore also be two-dimensional.
Its report-coordinate slope $\sigma_{i,U}\in\{-1,1\}$ is the actual
slope of the distance on that cell.

Let $\bm{e}_{i,U}$ be the vector increasing the arclength coordinate on $e_i$
at unit speed and fixing all other coordinates. It is parallel to
$\operatorname{aff}(U)$. For the affine extension $\widehat\rho_{i,U}$
of $\rho_i|_U$, define

\begin{equation}
\eta_{i,U}=\sigma_{i,U}\bm{e}_{i,U},\qquad
D_{\bm{e}_{i,U}}\widehat\rho_{i,U}=\sigma_{i,U},\qquad
D_{\eta_{i,U}}\widehat\rho_{i,U}=1.
\end{equation}

Take two points $((y,\bm{x}_{-i}),t),((z,\bm{x}_{-i}),t)\in U$.
They differ only in report coordinate $i$, so their difference is
$h\eta_{i,U}$ for a unique $h\in\mathbb R$. Affineness gives
$d_G(t,z)-d_G(t,y)=h$.
For two points on the same straight graph edge, their graph distance
is their arclength separation: the straight segment attains their
Euclidean distance, and no graph path can be shorter. Hence
$d_G(y,z)=|h|$, and the shortest-path equality in
\hyperref[finite-ssp:condition-outward-direction]{Condition~\ref*{finite-ssp:condition-outward-direction}} of
\cref{finite-ssp:def-compatible-subdivision} is
$h=|h|$, equivalent to $h\geq0$.
The direction changes only report $i$, as required. These formulas
hold on the closed cell, including its boundary. The construction uses
only $G$ and the set of coordinates.

\end{proof}

Containment in an attribution region, affine formulas, and the equivalence
in \hyperref[finite-ssp:condition-outward-direction]{Condition~\ref*{finite-ssp:condition-outward-direction}} of
\cref{finite-ssp:def-compatible-subdivision} remain valid on smaller cells. By
\cref{finite-ssp:lem-full-dimension}, a refined maximal cell has the
same affine hull as the maximal cell containing it, so its inherited
directions remain parallel to that hull. This need not hold for a
lower-dimensional face.

\begin{lemma}[A common compatible subdivision]\label{finite-ssp:lem-subdivision-existence}

Any finite common simplicial refinement of
$\mathcal R^{\mathrm{attr}}$, $\mathcal R^{\mathrm{cost}}$, and
$\mathcal R^{\mathrm{ord}}$ is compatible with $(\mathcal T,F)$.
Such a refinement exists and can be chosen independently of $A$.
Every further finite simplicial subdivision of a compatible subdivision
is compatible, with inherited attribution regions and directions on its
maximal cells.

\end{lemma}

\phantomsection\label{finite-ssp:proof-lem-subdivision-existence}

\begin{proof}

The three finite simplicial subdivisions admit a finite common
polytopal refinement and a compatible triangulation. Let $\mathcal U$
be any common simplicial refinement. By
\cref{finite-ssp:lem-full-dimension}, each maximal $U$ spans the edge
product containing it, giving the first compatibility condition.

Choose maximal parent simplices $U_{\mathrm{attr}}$,
$U_{\mathrm{cost}}$, and $U_{\mathrm{ord}}$ containing $U$ in the
three subdivisions. The region $T$ chosen for $U_{\mathrm{attr}}$
also satisfies $U\subseteq T\times G$, and the affine distance formulas
on $U_{\mathrm{cost}}$ restrict to $U$. These give the second and third
conditions.

The equivalence in \hyperref[finite-ssp:condition-outward-direction]{Condition~\ref*{finite-ssp:condition-outward-direction}} of
\cref{finite-ssp:def-compatible-subdivision} on $U_{\mathrm{ord}}$ restricts to pairs
in $U$. Both $U$ and $U_{\mathrm{ord}}$ have dimension $n+1$, and
$U\subseteq U_{\mathrm{ord}}$, so their affine hulls coincide. The
inherited nonzero vector $\eta_{i,U_{\mathrm{ord}}}$ is therefore
parallel to $\operatorname{aff}(U)$, giving the fourth condition.
All choices are independent of $A$.

The same argument applies to a further finite simplicial subdivision
of any compatible subdivision: its maximal cells have dimension $n+1$
and inherit the chosen attribution region, affine formulas, and directions
from containing maximal cells. This proves the final
assertion. Directions on a maximal cell need not be parallel to the
affine hulls of its lower-dimensional faces.

\end{proof}

The compatible subdivision and the cellwise equivalence now yield the
global characterization.

\restatestatement{simp-new:ssp-constraints}{\claimFiniteSSP}

\begin{proof}

Choose a compatible subdivision $\mathcal U$ as constructed in
\cref{finite-ssp:lem-subdivision-existence}. The subdivision, containing attribution
regions, and outward directions depend only on $G$, $F$, and $\mathcal T$.
For an admissible attribution $A$, \cref{finite-ssp:lem-cell-costs}
identifies each $q_{U,A}$ with the mechanism cost on $U$.
By \cref{finite-ssp:prop-cell-equivalence}, the derivative tests are
equivalent to all SSP comparisons within that cell. The maximal cells
and their faces form a finite subdivision of $X$, so
\cref{finite-ssp:thm-locality-ssp} makes these local comparisons equivalent
to global SSP. Therefore

\begin{equation}
\label{finite-ssp:eq-global-tests}
\begin{aligned}
M_A\text{ is SSP}
&\quad\Longleftrightarrow\quad L_{i,U,v}(A)\geq0,\\
&\qquad\text{for all }i\in N, U\in\mathcal U^{\max}, v\in\operatorname{Vert}(U).
\end{aligned}
\end{equation}

Admissibility makes the cell polynomials agree in value on shared
faces. Their derivatives can differ there: when a vertex belongs to
several maximal cells, every incident maximal cell contributes its own
test for every agent. These tests concern the polynomial on the cell's
affine hull and do not require a feasible outward step from that vertex.

Each test is linear and uses only the coefficients of the weights
$a_{T(U),f}$ on the original region. The auxiliary subdivision introduces no new attribution
variables. Intersecting the compact attribution polytope with these
finitely many closed linear half-spaces gives the compact polytope
$\mathcal A_{\mathcal T,F}^{\mathrm{SSP}}$.

\end{proof}

\section{Justification of the sampled bounds}

\label{quad-search:sampled-bounds}

The finite SSP constraints enforce the incentive requirement globally,
but the LP in \cref{simp-new:eq-sampled-program} tests approximation
only at sampled profiles. We justify the two uses of this LP in
\cref{simp-new:ssp-optimization}: its exact optimum is a lower bound
on the best global ratio within the fixed SSP class, and global
verification supplies an upper bound for the chosen candidate. Their
difference bounds the candidate's suboptimality within that class.

Fix the attribution class and assume that the sampled LP
\cref{simp-new:eq-sampled-program} has an exactly feasible optimal solution
$(A,\alpha_{\mathcal P})$. Its admissibility and SSP constraints ensure
that $M_A$ is globally SSP. Write

\begin{equation}
\alpha^*_{\mathrm{SSP}}
=\inf_{A'\in\mathcal A_{\mathcal T,F}^{\mathrm{SSP}}}
  \operatorname{apx}(M_{A'}).
\end{equation}

Every globally feasible pair in this class is feasible for the sampled
LP. Taking infima gives
$\alpha_{\mathcal P}\leq\alpha^*_{\mathrm{SSP}}$, while membership of
$A$ in the class gives
$\alpha^*_{\mathrm{SSP}}\leq\operatorname{apx}(M_A)$.
Thus any globally verified approximation factor $U$ for $M_A$ satisfies

\begin{equation}
\alpha_{\mathcal P}\leq\alpha^*_{\mathrm{SSP}}
\leq\operatorname{apx}(M_A)\leq U,
\qquad
0\leq\operatorname{apx}(M_A)-\alpha^*_{\mathrm{SSP}}
\leq U-\alpha_{\mathcal P}.
\end{equation}

To obtain $U$, use the exact global approximation verifier in
\cref{quad-search:prop-verification}, whose queries require only a
nonnegativity verdict or a negative witness. If verification at the proposed
threshold $\beta\geq1$ fails, its violating comparison proves
$\operatorname{apx}(M_A)>\beta$. If it succeeds, then
$\operatorname{apx}(M_A)\leq\beta$. In this case, test $\alpha_{\mathcal P}$.
Acceptance forces equality throughout the first three terms of the
displayed chain, so $M_A$ is optimal in the fixed SSP class and we can
take $U=\alpha_{\mathcal P}$.

Otherwise, $\alpha_{\mathcal P}<\operatorname{apx}(M_A)\leq\beta$.
Bisect this interval, retaining a rejected factor as the lower endpoint
and an accepted factor as the upper endpoint. After $k$ bisection tests
the interval has width $(\beta-\alpha_{\mathcal P})/2^k$, so finitely
many tests reduce its width to at most any prescribed $\varepsilon>0$.
The retained upper endpoint $U\leq\beta$ then satisfies

\begin{equation}
U-\varepsilon\leq\operatorname{apx}(M_A)\leq U.
\end{equation}

Bounds are also available without an exact sampled optimum. By weak
duality, any feasible sampled dual solution with value $L$ gives
$L\leq\alpha^*_{\mathrm{SSP}}$. Together with an exactly feasible SSP
candidate and its globally verified factor $U$, this gives the same
suboptimality bound with $U-L$ in place of $U-\alpha_{\mathcal P}$.
A violating approximation comparison can be added to the sampled LP:
every globally feasible pair must satisfy it, so the lower-bound
interpretation is preserved while the attribution class stays fixed.
All SSP constraints are retained throughout. Sampled feasibility alone gives
no global approximation guarantee.

\section{The finite-domain reduction}

\label{lower-exp:proof-output-reduction}

\Cref{simp-new:lower-bounds} obtains unrestricted lower bounds by
restricting reports to a finite set $D$ while preserving the costs of
continuous-output mechanisms. We prove \cref{lower-exp:lem-output-reduction}
by constructing a compatible output set and redistributing each outcome
onto its subdivision endpoints. Preserving costs at every true location
in $D$, including at profiles reached by deviations, transfers SP. Agreement
of the finite and continuous social optima then transfers approximation.

\restatestatement{lower-exp:lem-output-reduction}{\claimFiniteDomainReduction}

\begin{proof}

For each $t\in D$, \cref{simp-new:lem-pl-distance} implies that
$y\mapsto d_G(t,y)$ has finitely many affine pieces. Taking the union
of their breakpoints with $D$ and $V(G)$ gives a finite set $B$
compatible with $D$. This construction resolves only distances from
points in $D$. Additional points in $B$ need not be treated as true
locations.

Now let $B$ be any set compatible with $D$. On an edge of the subdivision
with endpoints $u,v\in B$, write $s$ for the arclength from $u$ to $y$ and
$\ell$ for the length of this edge. Set $\lambda=s/\ell\in[0,1]$.
Affineness gives

\begin{equation}
d_G(t,y)=(1-\lambda)d_G(t,u)+\lambda d_G(t,v),\quad\text{for all }t\in D.
\end{equation}

Replace an outcome at $y$ by
$(1-\lambda)\delta_u+\lambda\delta_v$, leaving points of $B$ unchanged.
This redistribution is independent of the report profile. Apply it
to each atom of the finite-support output lottery.

For any mechanism $M:D^n\to\Delta(G)$, the resulting mechanism
$M_{D,B}:D^n\to\Delta(B)$ satisfies, by linearity of expectation,

\begin{equation}
C_{M_{D,B}}(\bm{x},t)=C_M(\bm{x},t),\quad\text{for all }\bm{x}\in D^n, t\in D.
\end{equation}

For any $\bm{x}\in D^n$, agent $i$ and alternative report $z\in D$,
apply this identity at true location $t=x_i$ to both $\bm{x}$ and
$(z,\bm{x}_{-i})$. The two mechanisms therefore have the same SP
comparisons within $D$. In particular, restricting an SP mechanism on
$G$ to reports in $D$ and applying this redistribution gives an SP
mechanism $M_{D,B}$.

\phantomsection\label{lower-exp:proof-approximation}\label{lower-exp:cor-approximation}
It remains to show that the redistribution preserves approximation
guarantees on $D^n$. For any $\bm{x}\in D^n$, the finite-benchmark lemma,
\cref{quad-search:lem-benchmarks}, gives an optimal outcome among the
reports and graph vertices. Write $\operatorname{OPT}_B(\bm{x})$
for the minimum social cost when outputs are restricted to $B$.
All reports and graph vertices belong to $B$, so

\begin{equation}
\operatorname{OPT}_B(\bm{x})
=\operatorname{OPT}_G(\bm{x})
=\min_{b\in B}\sum_{i=1}^n d_G(x_i,b).
\end{equation}

Summing the preserved values $C_M(\bm{x},x_i)$ over the agents
shows that $M$ and $M_{D,B}$ have the same social cost at $\bm{x}$.
Since the finite and continuous optima agree, the two mechanisms have
the same approximation guarantees on $D^n$. Thus every approximation
factor for an SP mechanism $M$ on $G$ is also valid for $M_{D,B}$,
proving \cref{lower-exp:lem-output-reduction}.
Taking the infimum over SP mechanisms on $G$ gives

\begin{equation}
\operatorname{infapx}(\mathcal M_{D,B}^{\mathrm{SP}})
\leq\operatorname{infapx}(\mathcal M_G^{\mathrm{SP}}).
\end{equation}

\end{proof}

\section{The finite LP and dual certificates}

\label{lower-exp:app-dual}

The preceding reduction lets us prove lower bounds for continuous SP
mechanisms by proving them for a finite problem. We now turn this
principle into certificates that can be checked directly. For the finite
LP, we follow Meir's construction \cite[Section 4.1]{meir2019strategyproof},
extending it to allow different report and output sets. We then derive its dual:
a feasible dual solution weights the mechanism constraints to give a
lower bound on every feasible approximation factor. This bound transfers
to the continuous graph, even if the dual solution is not optimal.
The formulation below makes the certificate inequalities and their
lower-bound value explicit, establishing
\cref{lower-exp:prop-lower-certificate}.

For finite report and output sets $D\subseteq B\subseteq G$, with
$B$ compatible with $D$, let $p_{\bm{x},b}$ be the probability of output
$b\in B$ at profile $\bm{x}\in D^n$. The program is

\begin{equation}
\label{lower-exp:eq-primal}
\begin{aligned}
\text{minimize}\quad &\alpha\\
\text{subject to}\quad
&\alpha\geq1,\qquad p_{\bm{x},b}\geq0,
&&\text{for all }\bm{x}\in D^n, b\in B,\\
&\sum_{b\in B}p_{\bm{x},b}=1,
&&\text{for all }\bm{x}\in D^n,\\
&\sum_{b\in B}p_{\bm{x},b}\sum_{i=1}^n d_G(x_i,b)
   \leq\alpha\operatorname{OPT}_B(\bm{x}),
&&\text{for all }\bm{x}\in D^n,\\
&\sum_{b\in B}d_G(x_i,b)p_{\bm{x},b}
   \leq\sum_{b\in B}d_G(x_i,b)p_{(z,\bm{x}_{-i}),b},
\\ &\hspace{1em}\text{for all }\bm{x}\in D^n, i\in\{1,\ldots,n\}, z\in D.
\end{aligned}
\end{equation}

Every lottery on $B$ is specified by its finitely many probabilities.
The LP imposes exactly normalization, nonnegativity, all SP comparisons
within $D$, and the approximation inequalities for every profile in
$D^n$. Its optimum is therefore $\operatorname{infapx}(\mathcal M_{D,B}^{\mathrm{SP}})$.
Combining the reduction with weak duality proves
\cref{lower-exp:prop-lower-certificate}.

Let $\mathcal I$ index all incentive inequalities of the finite program:
$r=(\bm{x},i,z)$ ranges over $\bm{x}\in D^n$,
$i\in\{1,\ldots,n\}$ and $z\in D$. Write $\bm{x}^-(r)=\bm{x}$ and
$\bm{x}^+(r)=(z,\bm{x}_{-i})$ for its source and deviated profiles, and
$t_r=x_i$ for the original true location.
Assign nonnegative multipliers $\lambda_{\bm{x}}$ to the approximation
inequalities and $\beta_r$ to the incentive inequalities. Assign
a multiplier $\gamma_{\bm{x}}\in\mathbb R$, with no sign restriction,
to each normalization equality. Define $\omega$ and, for every
$\bm{x}\in D^n$ and $b\in B$, the expression $D_{\bm{x}}(b)$ by

\begin{equation}
\begin{aligned}
\omega&=\sum_{\bm{x}\in D^n}\lambda_{\bm{x}}
   \operatorname{OPT}_B(\bm{x}),\\
D_{\bm{x}}(b)
&=\lambda_{\bm{x}}\sum_{i=1}^n d_G(x_i,b)
 +\sum_{r:\bm{x}^-(r)=\bm{x}}\beta_r d_G(t_r,b)
 -\sum_{r:\bm{x}^+(r)=\bm{x}}\beta_r d_G(t_r,b).
\end{aligned}
\end{equation}

Substitute $\alpha=1+\tau$, with $\tau\geq0$, into the primal.
Its dual is

\begin{equation}
\label{lower-exp:eq-dual}
\begin{aligned}
\text{maximize}\quad
&1+\sum_{\bm{x}\in D^n}\gamma_{\bm{x}}-\omega\\
\text{subject to}\quad
&\lambda_{\bm{x}}\geq0,\quad\text{for all }\bm{x}\in D^n,\\
&\beta_r\geq0,\quad\text{for all }r\in\mathcal I,\\
&\omega\leq1,\\
&D_{\bm{x}}(b)\geq\gamma_{\bm{x}},
\quad\text{for all }\bm{x}\in D^n, b\in B.
\end{aligned}
\end{equation}

For weak duality, let $p,\alpha$ be primal feasible and
$\lambda,\beta,\gamma$ dual feasible. Weighting and summing the
approximation and incentive inequalities, and using normalization, gives

\begin{equation}
\sum_{\bm{x}}\gamma_{\bm{x}}
\leq\sum_{\bm{x},b}p_{\bm{x},b}D_{\bm{x}}(b)
\leq\alpha\omega.
\end{equation}

Since $\alpha\geq1$ and $\omega\leq1$,
$1+\sum_{\bm{x}}\gamma_{\bm{x}}-\omega
\leq1+(\alpha-1)\omega\leq\alpha$, proving the certificate bound.

The multipliers have the same roles as the social-cost and incentive
weights in the circle certificates of Chan, Lin and Wang
\cite[Section 4, Lemmas 10 and 11]{chan2026randomized}.
Their formulation uses representative profiles and averages weighted
cost expressions over profile symmetries.

The reduction is valid for real graph and location data. Rational edge
lengths and arclength coordinates for $D$ and $B$ give a rational LP.
Normalization to $\omega=1$ is not required. A restricted-output
dual certifies only its restricted program; the unrestricted lower
bound requires the full output set justified above.

\section{The bipartite specialization}

\label{lower-exp:proof-bipartite}

Applying the finite LP requires a compatible output set for the chosen
reports. Even if all reports are graph vertices, compatibility can
require additional output locations inside edges. We show that for
bipartite graphs with equal edge lengths, the vertex set itself is
compatible: distances from a report vertex are affine along every edge.
This gives a finite program using only vertices for both reports and
outputs, whose dual certificates still bound the unrestricted continuous
problem.

\restatestatement{lower-exp:cor-bipartite}{\claimBipartiteReduction}

\begin{proof}

Let $D=B=V(G)$ and let $\ell$ be the common edge length.
For $t\in D$ and adjacent vertices $u,v\in B$,

\begin{equation}
|d_G(t,u)-d_G(t,v)|\leq\ell.
\end{equation}

Both distances are integer multiples of $\ell$. Bipartiteness gives
opposite parities for their lengths in edges: $u$ and $v$ lie in
opposite parts, and every edge switches parts. The two distances
therefore differ by exactly $\ell$ in absolute value.
For a point $y_s$ at arclength
$s$ from $u$ on this edge, every route from $t$ enters its interior
through an endpoint. Thus

\begin{equation}
d_G(t,y_s)
=\min\{d_G(t,u)+s,\ d_G(t,v)+\ell-s\}.
\end{equation}

If $d_G(t,v)=d_G(t,u)+\ell$, the first expression is minimal for
all $s\in[0,\ell]$; if the reverse distance equality holds, the
second expression is minimal. In either case,

\begin{equation}
d_G(t,y_s)
=\left(1-\frac{s}{\ell}\right)d_G(t,u)
 +\frac{s}{\ell}d_G(t,v).
\end{equation}

This proves that $B=V(G)$ is compatible with $D=V(G)$. The
finite-domain reduction therefore applies, and weak duality gives
\cref{lower-exp:cor-bipartite}.

\end{proof}

The equal-length assumption matters: an abstract bipartite graph with
unequal edge lengths may have interior distance breakpoints. The
bipartite assumption also ensures the condition for the stated vertex
set. For example, on a unit-edge triangle the distance from one vertex
to the midpoint of the opposite edge is $3/2$, whereas its distance to
each endpoint is $1$.

\section{Comparison with Meir's support reduction}

\label{lower-exp:app-correction}

In the proof sketch of his support reduction,
Meir \cite[Proposition 6, p. 26]{meir2019strategyproof} proposes
redistributing outcomes onto the current reports and their antipodes.
On a circle of circumference $1$, the antipode of $x$ is
$x+1/2\pmod 1$, the point halfway around the circle. Thus the proposed
output set at profile $\bm{x}$ is

\begin{equation}
\left\{x_i,\ x_i+\tfrac12 \pmod 1:i=1,\ldots,n\right\}.
\end{equation}

An outcome between consecutive points of this set is replaced by a
lottery on those endpoints, with probabilities preserving its expected
position along the intervening arc.

This redistribution preserves costs evaluated at the current reports.
An SP comparison, however, evaluates the lottery at a deviating profile
using the agent's original true location. That location need not occur
among the new reports, and its distance breakpoints may be absent from
the new output set. Redistributing outcomes can therefore change this
cost and create a profitable deviation.

Our finite-domain reduction (\cref{lower-exp:lem-output-reduction})
retains the endpoint redistribution but uses the same output set $B$
at every profile. On the circle, $B$ contains the graph vertices,
every allowed true location in $D$, and its antipode.
This satisfies the compatibility condition
in \cref{lower-exp:def-compatible-set} and preserves the original true
location's cost on both sides of every SP comparison. The following
example shows why the profile-dependent redistribution needs this
correction.

\subsection{A redistribution that destroys strategyproofness}

\label{lower-exp:redistribution-example}

Consider a circle of circumference $1$, with coordinates modulo $1$,
and the constant mechanism $M(\bm{x})=\delta_0$. This mechanism
is SP. Apply the proposed redistribution at each profile, replacing
an outcome by a lottery over adjacent peaks or antipodes of the
current reports.

At $\bm{x}=(1/2,1/4,1/4)$, the outcome $0$ is an antipode of the
first report and remains unchanged. If the first agent reports $1/4$,
the new profile is $\bm{y}=(1/4,1/4,1/4)$. Its peaks and antipodes
are $1/4$ and $3/4$, and redistribution replaces $0$ by
$\tfrac12\delta_{1/4}+\tfrac12\delta_{3/4}$. For the true location
$1/2$, the two costs are

\begin{equation}
d_G(1/2,0)=\frac12,
\qquad
\frac12d_G(1/2,1/4)+\frac12d_G(1/2,3/4)=\frac14.
\end{equation}

The transformed mechanism is therefore manipulable. This counterexample
concerns preservation of SP by the proposed redistribution; it does not by
itself disprove Meir's support-existence claim
\cite[Proposition 6]{meir2019strategyproof}.

\section{Additional experimental results and verification records}

\label{lower-exp:app-results}

This section describes the certificates supporting the exact bounds in
\cref{simp-new:verified-experiments}, the numerical refinement study, and
the smaller-grid exact cross-checks.

\subsection{Verified mechanisms and unrestricted lower certificates}

\label{lower-exp:app-certificates}
\label{lower-exp:app-symmetry}

In the LLM-based search, lower-bound certificate files were limited to
$50$ MiB, and each upper-bound verification attempt was allowed at most
$12$ hours, including staging. These resource limits constrained the
certificates we could obtain and validate, and therefore the reported
bounds.

\Cref{lower-exp:table-verified-bounds} lists all $16$ selected
configurations: $13$ paired bounds and $3$ lower-only cases.
The upper certificates establish SP and approximation guarantees for the
specified mechanisms; the lower certificates apply to all SP mechanisms
on the corresponding continuous graph with the stated number of agents.
Endpoints are paired only when the metric graphs and agent counts match,
allowing uniform scaling and degree-two subdivision. Decimals are rounded
outward to at most $12$ decimal places: lower bounds downward and upper
bounds upward. Exact rational bounds are retained with the certificates.
A dash means that no upper certificate is included for that configuration.

Here $C_k$ with a pendant edge is a unit-edge $k$-cycle with one additional
unit edge attached at a cycle vertex. Its pendant length is $1/k$ of the
circumference. A theta graph has two hubs joined by three internally
disjoint paths, with the indicated relative lengths. Equal-edge $K_{2,k}$
has $k$ equal-length hub-to-hub paths, and $K_{3,3}$ has all nine edges
of equal length. The domino consists of two unit-edge $4$-cycles sharing
one edge, and the bowtie of two equal $4$-cycles sharing one vertex.

\begin{longtable}{@{}lrrr@{}}
\caption{Bounds on $\operatorname{infapx}(\mathcal M_G^{\mathrm{SP}})$ for continuous graphs.}\label{lower-exp:table-verified-bounds}\\
\toprule
Continuous graph & Agents & Lower bound & Upper bound \\
\midrule
\endfirsthead
\toprule
Continuous graph & Agents & Lower bound & Upper bound \\
\midrule
\endhead
\bottomrule
\endlastfoot
Cycle & $3$ & $1.0591173691$ & $1.10747$ \\
Cycle & $4$ & $1.08915743224$ & $1.25025$ \\
\addlinespace[0.4em]
$C_4$ with pendant edge & $3$ & $1.04973428268$ & $1.13212$ \\
$C_6$ with pendant edge & $3$ & $1.050939365103$ & $1.19092$ \\
\addlinespace[0.4em]
Equal-path theta & $3$ & $1.078216584909$ & $1.2028$ \\
Theta $(1:1:2)$ & $3$ & $1.065238837183$ & $1.25055$ \\
Theta $(1:2:2)$ & $3$ & $1.075635097787$ & $1.21725$ \\
\addlinespace[0.4em]
Equal-edge $K_{2,4}$ & $3$ & $1.077807650022$ & $1.23605$ \\
Equal-edge $K_{2,4}$ & $4$ & $1.118056448043$ & --- \\
Equal-edge $K_{2,5}$ & $3$ & $1.078119703605$ & $1.248$ \\
Equal-edge $K_{2,5}$ & $5$ & $1.133587351026$ & --- \\
Equal-edge $K_{2,6}$ & $3$ & $1.077986717685$ & $1.25885$ \\
Equal-edge $K_{2,6}$ & $6$ & $1.140204823651$ & --- \\
Equal-edge $K_{3,3}$ & $3$ & $1.078138407691$ & $1.24495$ \\
\addlinespace[0.4em]
Domino & $3$ & $1.063643013134$ & $1.2511$ \\
Bowtie & $3$ & $1.04470368845$ & $1.25$ \\
\end{longtable}

The cycle upper certificate uses the three report selectors on a
four-edge realization of the circle, with $23{,}400$ full-dimensional
attribution simplices across its $64$ edge-product cubes. The $q=4$ search
in Appendix~\ref{lower-exp:app-legacy} uses a different triangulation, with
six simplices per cube and $384$ in total. The theta certificate uses report
selectors and $17$ constant selectors on a subdivision with $11$ vertices
and $12$ edges. Its three hub-to-hub paths have equal length, so subdivision
and uniform rescaling identify it with the equal-edge $K_{2,3}$ in the
main text, preserving SP and approximation ratios. Both candidates passed
complete exact verification. These acceptances do not assert SSP or
optimality within a search class.

For the cycle lower certificate, all $240$ mesh vertices serve as reports
and outputs. For $K_{2,6}$, subdividing each edge into $16$ equal edges
gives $188$ vertices and $192$ edges, and all vertices again serve as
reports and outputs.

\paragraph{Upper-bound verification.}

The upper-bound checker receives proposed simplicial covers of the
products of unit edges, each identified with a cube, and validates them.
It checks that every simplex is full-dimensional and contained in its
cube. A supplied separation tree specifies an affine function $h$ at each
internal node. The hyperplane $h=0$ defines the cut, with branches
corresponding to $h\leq0$ and $h\geq0$. A simplex goes only to the negative
branch if $h\leq0$ at all its vertices, only to the positive branch if
$h\geq0$ at all its vertices, and to both branches otherwise. Each cut is
evaluated on the original simplex vertices. This establishes disjoint
interiors: an interior point
common to two cells and off all cut hyperplanes would route both cells
to the same leaf, contradicting the requirement that each leaf contain
exactly one named cell.

The checker also requires the sum of exact simplex volumes to equal the
cube's volume. Together with disjoint interiors, this gives full volume
to their union. The finite union is closed, so a missing point, even on
the cube boundary, would have a neighborhood of positive cube volume
outside the union. Thus the whole closed cube is covered. Volume equality
alone would not suffice, since overlaps could compensate for gaps.

Every comparison polynomial must represent the actual cost difference
throughout its closed cell. The checker validates containment in the
referenced attribution regions and the selected affine distance formulas
on the whole cell, including boundaries. Auxiliary comparison covers may
be nonconforming, since they only supply domains for these tests. The
attribution itself must form a conforming triangulation. Within each
edge-product cube, the checker pairs every internal facet with an
identical facet of exactly one simplex on the opposite side. Each facet
on the cube boundary belongs to exactly one simplex in that cube. On
every shared face of the edge-product cubes, including lower-dimensional
intersections, the checker requires all incident cubes to induce the same
triangulation. The attribution must also satisfy probability validity and
equality of lotteries across faces, as expressed by the constraints in
\cref{simp-new:app-attribution-constraints}.
Admissibility is a hypothesis of \cref{quad-search:prop-verification}.
The implementation checks it before testing comparisons with the exact
nonnegativity procedure in \cref{simp-new:app-quadratic-minimization}.

\paragraph{Lower-bound verification.}

The lower checker represents a profile by its multiset of report
vertices, identifying profiles that differ only in the agents' labels.
This reduces the finite program without changing its optimum.
To see why, average a mechanism over all permutations of the agents,
choosing the permutation independently of the reports. Each SP inequality
for the average is an average of valid inequalities for the corresponding
agents and permuted profiles. Social cost and its optimum are invariant
under permutations, so this averaging also preserves any approximation
guarantee. The resulting mechanism is anonymous. Thus restricting the
finite program to anonymous mechanisms leaves its optimum unchanged.

The checker reconstructs every single-agent deviation from a vertex
present in that multiset to any graph vertex, evaluates both lotteries
at the original true vertex, and allows every graph vertex as an output.
The dual inequalities and weak-duality argument in
\cref{lower-exp:app-dual} apply with these anonymous profile indices.
The checker accepts no supplied quotient by graph symmetries: any
symmetry reduction used in the search must already have been converted
into a multiplier map for this explicit anonymous program, which is
checked directly. Its correctness does not depend on the producer's
graph-symmetry reduction or lifting procedure.

Feasibility is checked by the exact multiplier signs, the inequality
$\omega\leq1$, and every inequality
$D_{\bm{x}}(b)\geq\gamma_{\bm{x}}$ of \cref{lower-exp:eq-dual}.
The claimed objective is recomputed. A sparse certificate assigns $0$
to omitted multipliers.
Permutation averaging and \cref{lower-exp:cor-bipartite} transfer these
bounds to unrestricted continuous mechanisms.

\subsection{Numerical SSP search and smaller-grid cross-checks}

\label{lower-exp:app-legacy}

The cycle study introduced in \cref{simp-new:cycle-search-class} provides
a concrete application of the paper's heuristic-search framework to three
agents on a continuous cycle. We use the finite SSP constraints to optimize
over PL attributions and examine how finer representation meshes affect
approximation performance. The search minimizes a sampled approximation
factor, and bisection subsequently estimates the candidate's global
approximation ratio. For the smaller meshes, rational reconstruction
and exact verification yield certified mechanisms close to the numerical
candidates.

\paragraph{Search class and numerical checks.}

We use the sampled LP from \cref{simp-new:ssp-optimization}, together with
constraints imposing anonymity, rotations by whole mesh intervals, and
reflections. Anonymity permutes the reports and their report-selector
labels together. Rotations and reflections transform the reports and
facility locations together, preserving agent identities. The implementation
uses ordered report profiles and rotation-normalized variables, together
with linear equalities enforcing the remaining rotation and reflection
conditions. The LP retains all attribution and SSP constraints, so an
exactly feasible solution represents a globally SSP mechanism. Only the
approximation comparisons are sampled: the objective value returned by
the LP solver need not be an upper bound on the mechanism's approximation
ratio over the full continuous domain.

For the representation in \cref{simp-new:cycle-search-class}, measure
arclength in mesh intervals, so the circumference is $q$.
Approximation sampling subdivides each interval into $s$ equal parts and
uses every unordered triple of the resulting $sq$ locations, including
repeated reports. The refinement study fixes $s=2$.

The linear programming solver CBC, called through PuLP, first minimizes
the sampled approximation factor.
A secondary solve minimizes total expected social cost over the mesh
profiles $(0,u,v)$, where $u,v$ are integers with $0\leq u\leq v<q$.
Each profile is counted once, and the sampled factor may exceed the
primary optimum by at most $10^{-6}$.
This secondary objective favors lower costs across the grid and,
empirically, helped narrow the gap between the sampled factor and the
continuous approximation ratio subsequently estimated by bisection.

After optimization, we recheck the saved candidate using the derivative
inequalities from the finite SSP characterization
(\cref{simp-new:ssp-constraints}). For an exactly feasible LP solution,
these checks are mathematically redundant; here they test the
implementation and its floating-point output. To estimate the global
approximation ratio, we use bisection with numerical evaluations of the
continuous quadratic approximation comparisons from the global
verification procedure (\cref{simp-new:pl-properties}). The resulting
interval has width at most $10^{-5}$. Optimization and checking use
floating-point arithmetic, so this implements the mathematical search
and verification framework numerically without providing exact
certificates.

\paragraph{Refinement results.}

\begin{figure}[htbp]
\centering
\includegraphics[width=0.9\linewidth]{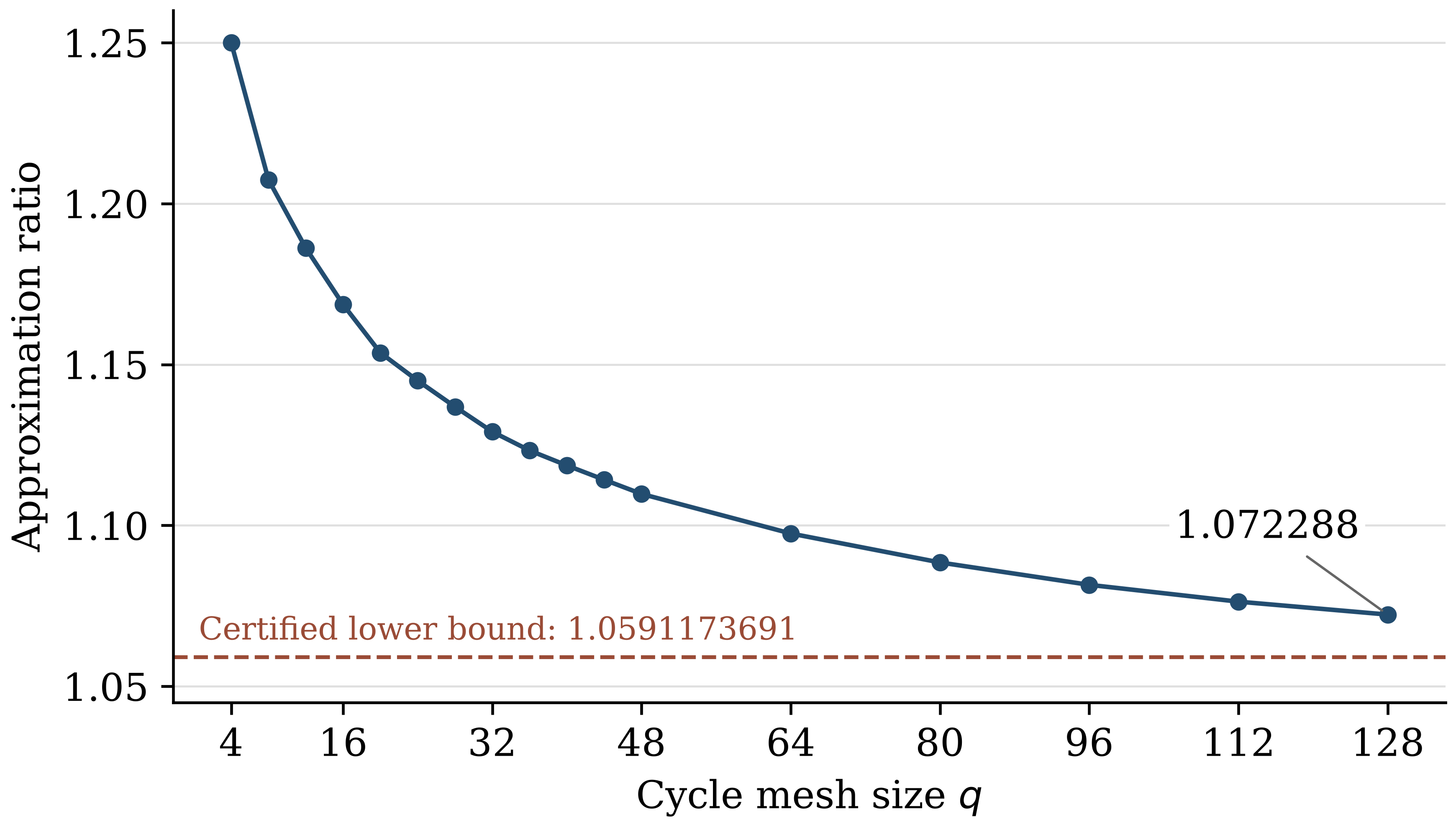}
\caption{Numerical continuous approximation upper endpoints for the $17$
saved cycle mechanisms ($s=2$). Lines connect tested mesh sizes.
The floating-point bisection intervals have width at most $10^{-5}$.
The dashed horizontal line marks the certified unrestricted lower bound
$1.0591173691$ for three agents on the continuous cycle.}
\label{lower-exp:fig-cycle-refinement}
\end{figure}

\Cref{lower-exp:fig-cycle-refinement} shows the numerical approximation
upper bounds for all $17$ saved cycle mechanisms. Each candidate thus
provides a numerical upper bound on the optimal approximation ratio
attainable by an SP mechanism for three agents on the continuous cycle.
Each completed the SSP check
without a reported counterexample. At the coarsest mesh, $q=4$, the search
recovers Meir's PCD mechanism \cite{meir2019strategyproof}, whose
approximation ratio is $5/4$. At the finest mesh, $q=128$, the
best-performing candidate has a numerical approximation interval of
$[1.072278,1.072288]$, with endpoints rounded outward to six decimal
places. Both endpoints concern this candidate's ratio. The lower endpoint
does not bound the unrestricted design problem from below.

On an AMD Ryzen~9 9950X processor, the $q=128$ run took approximately
$24.5$ hours for construction and $3.8$ hours for checking, with peak
resident memory of $19.5$ and $29.2$ GiB, respectively. These measurements
give resource estimates, not controlled benchmarks: background load and
runtime settings varied across the study.

\paragraph{Exact recovery of smaller-grid candidates.}

Our goal is to recover exactly verified SP mechanisms from the numerical
candidates while preserving their approximation performance. For
$q=4,8,12,16$, rational reconstruction achieves this with only a small
change to the saved coefficients, retaining the triangulations and report
selectors. The reconstructed mechanisms have accepted exact upper
factors
\[
\begin{array}{c|cccc}
q & 4 & 8 & 12 & 16\\ \hline
U & \dfrac54 & \dfrac{120738663}{100000000}
  & \dfrac{118617969}{100000000} & \dfrac{116861459}{100000000}
\end{array}.
\]
These four factors do not improve the headline cycle bound $1.10747$;
their role is to connect the numerical search with exact checking.

The recovered coefficients satisfy all constraints of the search LP
exactly. By the finite linear characterization
(\cref{simp-new:ssp-constraints}), this implies that the recovered
mechanisms are SSP. The general exact verifier independently certifies
strategyproofness and global approximation upper bounds no larger than
the corresponding numerical upper bounds. The independent verifier checks
SP and approximation; the SSP conclusion follows from exact feasibility
of the search LP. These results concern only the four recovered mechanisms
and do not certify the numerical candidates at larger mesh sizes.

Recovery changes the worst-case cost ratio by less than
$1.04\times10^{-7}$. To see this, let $p_i$ and $\widehat p_i$ denote the
saved and recovered coefficients, respectively, and define
\[
\delta=\max_{v}\sum_{i=1}^{3}|\widehat p_i(v)-p_i(v)|,
\]
where $v$ ranges over regional vertices. Affine interpolation extends
this bound to every profile. Since locating at any report has social cost
at most $2\operatorname{OPT}$ for three agents, the change in worst-case
cost ratio is at most $2\delta<1.04\times10^{-7}$. Here the saved decimals
are interpreted as exact rationals. This comparison does not assume that
the saved numerical coefficients define an exactly valid or SP mechanism.

Finally, exact evaluation at a selected profile bounds each recovered
mechanism's approximation ratio from below. These lower bounds differ
from the certified upper bounds by less than $7\times10^{-9}$. For $q=4$,
both equal $5/4$. They concern the individual recovered mechanisms, not
the optimum of the unrestricted design problem.

\section{A counterexample to quarter-QCD}

\label{simp-new:app-qcd}

The computational comparison in \cref{simp-new:verified-experiments}
excludes quarter-QCD as an SP mechanism with factor $7/6$. The following
exact calculation gives both an improving deviation and a profile
exceeding that factor for the published rule.

\begin{remark}[Quarter-QCD is not strategyproof and exceeds the claimed bound]

Let $M$ be the $\tfrac14$-QCD mechanism of Meir
\cite[Definition 2]{meir2019strategyproof}. For three distinct reports
on a circle of circumference $1$, the arc opposite a report joins
the other two reports without passing through that report.
Each report's weight is the larger of the squared length of its
opposite arc and $1/16$. Dividing by the sum of the three weights
gives its selection probability. Scaling all lengths by $8$ multiplies
both quantities by $64$, giving the equivalent rule on a circle of
circumference $8$ with threshold $4$.

Use arclength coordinates modulo $8$ and consider the profiles
$\bm{x}=(0,3,4)$ and $\bm{y}=(1,3,4)$. The rule gives

\begin{equation}
M(\bm{x})=\frac{4\delta_0+16\delta_3+9\delta_4}{29},
\qquad
M(\bm{y})=\frac{4\delta_1+25\delta_3+4\delta_4}{33}.
\end{equation}

The first agent, whose true location is $0$, benefits from reporting $1$:

\begin{equation}
\operatorname{cost}(M(\bm{x}),0)
-\operatorname{cost}(M(\bm{y}),0)
=\frac{84}{29}-\frac{95}{33}
=\frac{17}{957}>0.
\end{equation}

At $\bm{x}$, the social costs of locations $0,3,4$ are $7,4,5$,
respectively. The triangle inequality gives a lower bound of
$d_G(0,4)=4$ for every location's social cost, and location $3$
attains it. Thus $\operatorname{OPT}(\bm{x})=4$ and

\begin{equation}
\frac{\operatorname{SC}(M(\bm{x}),\bm{x})}
     {\operatorname{OPT}(\bm{x})}
=\frac{4\cdot7+16\cdot4+9\cdot5}{29\cdot4}
=\frac{137}{116}
=\frac76+\frac{5}{348}>\frac76.
\end{equation}

These calculations contradict the SP and $7/6$
approximation claims for this definition
\cite[Theorem 2 and Proposition 3]{meir2019strategyproof}.

\end{remark}

\section{Deterministic characterization}

\label{simp-new:app-deterministic}

We prove \cref{simp-new:thm-deterministic-representation} and relate
deterministic attributions to general PL attributions. We first establish
the attribution lemmas, then construct the representation on a line
segment and extend it to general graphs.

\restatestatement{simp-new:thm-deterministic-representation}{\claimDeterministicCharacterization}

\subsection{Proofs of the attribution lemmas}

\begin{lemma}[Deterministic attributions with selectors in $\Pi(W)$ imply SSP]\label{simp-new:lem-deterministic-attribution-ssp}

For every finite $W\subseteq G$ and every
deterministic PL $(\Pi(W),\mathcal T)$-attribution $A$, the mechanism $M_A$
is SSP.

\end{lemma}

\begin{proof}\label{simp-new:proof-deterministic-attribution-ssp}

Let $A$ be a deterministic PL $(\Pi(W),\mathcal T)$-attribution.
On each region $T$,
the attribution assigns weight $1$ to a fixed selector $f_T\in\Pi(W)$.
Fix a true profile
$\bm{x}$, an agent $i$, and reports $y,z$ with $y\in[x_i,z]_G$
such that $(y,\bm{x}_{-i})$ and $(z,\bm{x}_{-i})$ belong to $T$.
If $f_T=\pi_i$, the two costs under $M_A$ are $d_G(x_i,y)$ and
$d_G(x_i,z)$, respectively, and the former is no larger. If the selector
is another agent's peak or a constant, the two outcomes, and hence the
two costs, are equal.

The products $T\times e$, where $T$ ranges over attribution regions
and $e$ over graph edges, are the maximal cells of a finite subdivision
of $G^n\times G$. The preceding argument proves every SSP comparison
whose endpoints lie in one such cell. By
\cref{finite-ssp:thm-locality-ssp}, these local comparisons imply that
$M_A$ is SSP.

\end{proof}

\begin{lemma}[Deterministic representations]\label{simp-new:lem-deterministic-selector}

Every PL attribution representing a deterministic mechanism can be
replaced by a deterministic attribution with the same selectors and regions.

\end{lemma}

Thus restricting to deterministic attributions preserves every
deterministic mechanism admitting a PL attribution, with the same selectors
and regions. In particular, condition (3) of the characterization theorem
is equivalent to admitting a PL $(\Pi(W),\mathcal T)$-attribution
without requiring it to be deterministic.

\begin{proof}\label{simp-new:proof-deterministic-selector}

Let $A=(a_{T,f})$ be a PL $(F,\mathcal T)$-attribution and suppose that
$M_A$ is deterministic. We construct a deterministic attribution $A'$
representing the same mechanism with the same selectors and regions.

Fix a region $T$. A nonnegative affine weight that is not identically
$0$ on $T$ is positive at some vertex, and therefore at every convex
combination of the vertices with all coefficients positive.
At every such point, all selectors whose weights are not identically
$0$ must coincide, because their mixture is a point mass.
These points are dense in $T$, and the selectors are continuous, so
those selectors coincide throughout $T$.

Normalization ensures that at least one weight is not identically $0$.
Choose one corresponding selector $f_T$. At every point of $T$, the
remaining weights either vanish or select the same location as $f_T$,
so $f_T$ represents $M_A$ throughout $T$. Define $A'=(a'_{T,f})$ by
assigning weight $1$ to $f_T$ on each region and $0$ to all other
selectors. The resulting regional lotteries agree on intersections
because they all equal $M_A$ there. Thus $A'$ is a deterministic PL
$(F,\mathcal T)$-attribution with $M_{A'}=M_A$.

\end{proof}

\subsection{Representation on a line segment}

\label{simp-new:proof-segment-representation}

\begin{lemma}[Representation on a line segment]\label{simp-new:lem-segment-representation}

Let $G=[0,L]$, with $L>0$. Every deterministic SP onto mechanism
$M:G^n\to G$ admits a deterministic PL
$(\Pi(W),\mathcal T)$-attribution for some finite
$W\subseteq G$ and some finite triangulation whose maximal simplices
form a family $\mathcal T$ satisfying the PL attribution assumptions.

\end{lemma}

\begin{proof}

By the characterization stated by Schummer and Vohra
\cite[Section 3, Proposition 1]{schummer2002strategy}, $M$ is a
generalized median voter scheme. Hence there are thresholds
$t_S\in[0,L]$, indexed by $S\subseteq N$, such that

\begin{equation}
t_\varnothing=0,\qquad t_N=L,\qquad
t_S\le t_U,\quad\text{for all }S\subseteq U\subseteq N,
\end{equation}

and

\begin{equation}
M(\bm{x})=
\max_{S\subseteq N}
\min\bigl(\{x_i:i\in S\}\cup\{t_S\}\bigr),
\quad\text{for all }\bm{x}\in G^n.
\end{equation}

Let $W=\{t_S:S\subseteq N\}$. Subdivide the products of graph
edges by the hyperplanes $x_i=x_j$ and $x_i=t_S$. On each
resulting closed cell, the reports and thresholds have a fixed weak
ordering. Thus each minimum selects one fixed argument throughout
the cell, and the outer maximum selects one of these arguments.
Ties do not matter because the tied arguments have the same value.
The mechanism therefore agrees with a single selector on the cell,
either a report selector or a constant selector.

Choose a finite triangulation refining these cells, with the triangulations
agreeing on shared faces, and let $\mathcal T$ consist of its maximal simplices.
Let $A$ be the family assigning weight $1$ to the selected component
on each simplex, with all other weights $0$. Each simplex remains
contained in a product of graph edges, and the weights are constant,
hence affine.
On the intersections of simplices, the selected locations agree because both
regional descriptions evaluate $M$. Thus $A$ is a deterministic PL
$(\Pi(W),\mathcal T)$-attribution with $M_A=M$.

\end{proof}

The construction shows how the mechanism itself determines suitable
selectors and regions: its thresholds supply the constants, and
its comparisons supply the subdivision. The choices of $W$ and
$\mathcal T$ may depend on the deterministic SP onto mechanism $M$
being represented.

\subsection{Representation on general graphs}

\label{simp-new:app-general-representation}

The following extends the line segment representation lemma to every
graph under our standing assumptions.

\begin{lemma}[Representation on general graphs]\label{simp-new:lem-general-graph-representation}

Every deterministic SP onto mechanism $M:G^n\to G$ admits a deterministic PL
$(\Pi(W),\mathcal T)$-attribution for some finite
$W\subseteq G$ and some finite triangulation whose maximal simplices
form a family $\mathcal T$ satisfying the PL attribution assumptions.

\end{lemma}

\begin{proof}

Let $M$ be a deterministic SP onto mechanism. We use Schummer and
Vohra's characterizations \cite[Theorems 1 and 4]{schummer2002strategy}.

\textbf{Trees.}

Suppose first that $G$ is a tree. For a path $\gamma$ joining two leaves,
let $\operatorname{proj}_\gamma:G\to\gamma$ be the nearest-point
projection. It is the identity on edges of $\gamma$ and is constant
at an attachment vertex on every other edge. In particular, it is PL.

Schummer and Vohra's tree characterization supplies, for each such
path, a generalized median voter scheme $M_\gamma:\gamma^n\to\gamma$.
In arclength coordinates, its formula is a finite expression of minima
and maxima of reports and fixed thresholds. Writing
$(\operatorname{proj}_\gamma)^n$ for coordinatewise projection and

\begin{equation}
\widehat M_\gamma
=M_\gamma\circ(\operatorname{proj}_\gamma)^n,
\end{equation}

the characterization gives

\begin{equation}
M(\bm{x})\in\gamma
\quad\Longrightarrow\quad
M(\bm{x})=\widehat M_\gamma(\bm{x}),\quad\text{for all }\bm{x}\in G^n.
\end{equation}

Let $W$ contain all graph vertices and all threshold locations used
by these finitely many path schemes. On a product of graph edges,
each projected report is either the original report or a constant
vertex. Its arclength coordinate on $\gamma$ is therefore affine.
Subdivide the product by the equalities between these coordinates
and between coordinates and thresholds, for every leaf-to-leaf path.
On each resulting closed convex region $T$, the order of the
arguments in every min/max formula is fixed. Consequently each
$\widehat M_\gamma|_T$ is one fixed selector from $\Pi(W)$.
If that selector is nonconstant, its image lies in the graph edge
assigned to the corresponding input coordinate.

It remains to show that $M|_T$ is itself one fixed selector, despite
the conditional form of the displayed characterization. Any two
distinct points in a finite tree lie on a common leaf-to-leaf path.
Apply this to two distinct points of $M(T)$. For a path $\gamma$
containing them, $\widehat M_\gamma|_T$ takes both values, so it
is nonconstant and those values lie in a single graph edge.

In a tree, a set with this pairwise property lies in one graph edge.
Indeed, fix distinct points $x,y$ in an edge $e$. A point outside $e$
sharing an edge with each of them would force $x,y$ to be the endpoints
of $e$ and produce a three-edge cycle. Thus, if $M(T)$ is not a
singleton, it lies in one edge and hence in one leaf-to-leaf path
$\gamma$. A singleton also lies on such a path. The characterization
now gives $M|_T=\widehat M_\gamma|_T$, proving the required
selector description.

Take a compatible finite triangulation of the regions and let
$\mathcal T$ consist of its maximal simplices.
Let $A$ assign weight $1$ to the selected member of $\Pi(W)$ on each
region $T\in\mathcal T$ and $0$ to all other selectors. The regional lotteries agree
on intersections because they all equal $M$ there. Thus $A$ is a
deterministic PL $(\Pi(W),\mathcal T)$-attribution with $M_A=M$ on a tree.

\textbf{Graphs containing cycles.}

Let $C_G$ be the smallest connected subgraph of $G$ containing every
cycle, including the connecting paths between cycles. The remaining
parts of $G$ form finitely many attached trees, denoted $H_j$, each
meeting $C_G$ at one vertex $v_j$. Branches with the same attachment
vertex are grouped together. The nearest-point projection
$\operatorname{proj}_{H_j}:G\to H_j$ is the identity on $H_j$
and constant at $v_j$ outside it, so it is affine on each graph edge.

Schummer and Vohra's characterization for graphs with cycles gives
an agent $i\in N$ and deterministic SP onto tree mechanisms
$M_j:H_j^n\to H_j$. For all $\bm{x}\in G^n$, they satisfy

\begin{equation}
x_i\in C_G\quad\Longrightarrow\quad M(\bm{x})=x_i,
\end{equation}

and, for every attached tree $H_j$,

\begin{equation}
x_i\in H_j\quad\Longrightarrow\quad
M(\bm{x})
=M_j\bigl((\operatorname{proj}_{H_j})^n(\bm{x})\bigr)
\in[x_i,v_j]_G.
\end{equation}

The latter interval is the unique tree path. At $x_i=v_j$, both
formulas therefore give $v_j$.

Apply the tree construction to each $M_j$, and let $W$ contain all
its fixed selector locations together with the vertices of $G$.
On each product of graph edges, intersect the product with the inverse
images of the tree regions under $(\operatorname{proj}_{H_j})^n$,
and take a common finite refinement. These intersections are convex
polytopes because the projection is affine on that product.
A report selector after projection is either an original
report or the constant attachment point; a constant selector remains
constant. Thus each composed tree mechanism is a fixed member of
$\Pi(W)$ on every refined region.

The input edge of agent $i$ lies either in $C_G$ or in one attached
tree. The displayed characterization consequently makes $M$ one
fixed member of $\Pi(W)$ throughout each refined region.
Choose a compatible triangulation and let $\mathcal T$ consist of its
maximal simplices. Let $A$ assign the selected component weight $1$ on
each region $T\in\mathcal T$, with all other weights $0$.
The formulas agree at attachment points and all other overlaps
because they represent $M$. If there are no attached trees, the
mechanism is simply $\pi_i$, and $A$ assigns it weight $1$ on every
region $T\in\mathcal T$. In every case, $A$ is a deterministic PL
$(\Pi(W),\mathcal T)$-attribution with $M_A=M$.

\end{proof}

\begin{proof}[Proof of \cref{simp-new:thm-deterministic-representation}]

The general-graph representation lemma supplies a deterministic PL
$(\Pi(W),\mathcal T)$-attribution $A$ with $M_A=M$, establishing
$(1)\Rightarrow(3)$ in
\cref{simp-new:thm-deterministic-representation}.
For a deterministic mechanism satisfying (3), choose a deterministic PL
$(\Pi(W),\mathcal T)$-attribution $A$ with $M_A=M$.
Then \cref{simp-new:lem-deterministic-attribution-ssp} implies that $M$ is SSP.
Thus $(3)\Rightarrow(2)$, while $(2)\Rightarrow(1)$ follows from
the definition of SSP.

\end{proof}

\section*{AI methodology}
\addcontentsline{toc}{section}{AI methodology}

\paragraph{Preparation of this disclosure.}
This disclosure was drafted by AI and refined through discussion and
correction by the authors. It draws on Git history, attribution notes,
retained conversations, and summaries from the manuscript and local research
environments, supplemented by the authors' account of earlier work and
handwritten notes. These records are incomplete, and first recorded
appearance was not treated as establishing who originated an idea. The
transcripts and full working environment are not being released with this
disclosure because they have not undergone review for potentially
confidential information.

Documentation evolved from a human-written main draft with selected AI
assistance to separate AI-developed notes and then systematic interaction
records. By August 2026, AI-generated attribution notes were collected during
development and consolidated before manuscript commits. In October, we moved
to retaining available transcripts and preparing summaries at manuscript
checkpoints, distinguishing conversation dates from summary-generation dates.

\paragraph{Human and AI contributions.}
The first author led the mechanism design research agenda and
developed the initial attribution concept and early strong-strategyproofness
ideas. The original numerical cycle implementation was developed in 2025,
predominantly by a human author, with reviewed AI-generated snippets.

AI involvement is documented from 2025 and increased during the project.
ChatGPT, GitHub Copilot, and Gemini helped explore mathematical foundations,
identify literature, and develop supporting arguments. Codex, including
GPT-6 Astra, subsequently contributed substantial mathematical development
across the theoretical sections and appendices: formulations, constructions,
proofs, and verification procedures. Some supporting arguments were
independently reproved by AI; others refined existing author proofs. AI also
helped identify and prove folklore lemmas in piecewise-linear topology and
polynomial optimization, including tests for nonnegativity of quadratic
polynomials on polytopes, supporting the exploration of possible proof
strategies.
The authors supplied research questions and mathematical ideas, checked and
challenged proposed arguments, and selected the results and assumptions to
retain.

AI assistance with writing was extensive throughout the main body and
appendices. Codex generated drafts ranging from whole sections to individual
passages, including exposition of ideas developed earlier by the authors.
The authors reviewed and revised these drafts, directing their mathematical
content, organization, and emphasis.

For the computational study, Codex with GPT-6 Astra implemented the
upper-bound verifier from a specification developed through human--AI
discussion. Claude developed the lower-bound verifier through an iterative
human--AI workflow. AI also assisted with reviewing the code for the
numerical cycle experiments and reconciling earlier experiment records.

Claude Opus 5.5 operated the autonomous search loop for candidate mechanisms
and lower-bound certificates, selecting tasks, implementing procedures, and
assessing results across more than $100$ experiments. The authors
occasionally supplied ideas or changed priorities, directly or through
human-prompted Codex sessions, and assessed progress qualitatively each day.

Certified bounds required acceptance by separate exact-arithmetic verifier
programs. Their guarantees depend on the mathematical reductions and verifier
implementations, rather than the research agent's assessment. The authors
retain responsibility for the paper's correctness, originality, and
integrity, including its references.

\clearpage
\phantomsection
\addcontentsline{toc}{section}{\refname}
\bibliographystyle{alpha}
\bibliography{ijcai24}
\end{document}